\documentclass[aps,twocolumn,
floatfix,footinbib,longbibliography]{revtex4-2}
\usepackage{hyperref}
\hypersetup{
  breaklinks=true, 
  citecolor=blue,
  colorlinks=true, 
  linkcolor=blue,
  urlcolor=blue}
\usepackage{float}
\usepackage{epsfig}
\usepackage{graphicx}
\usepackage{amsmath}
\usepackage{amsfonts}
\usepackage{amssymb}
\usepackage{bm}
\usepackage{bbold}
\usepackage{epsfig}
\usepackage{graphicx}
\usepackage{color}
\usepackage{pictex}
\usepackage{placeins} 
\usepackage{ulem}

\usepackage{mathtools}
\usepackage{extarrows}
\usepackage{lipsum}
\usepackage{footmisc}

\usepackage{braket}
\usepackage{xcolor}

\begin{document}

\title{Stability of Floquet sidebands and quantum coherence in 1D strongly interacting spinless fermions}
\author{Karun Gadge}

\affiliation{Institute for Theoretical Physics, Georg-August-University G\"{o}ttingen, Friedrich-Hund-Platz 1, D-37077 G\"{o}ttingen, Germany}
\author{Salvatore R. Manmana}

\affiliation{Institute for Theoretical Physics, Georg-August-University G\"{o}ttingen, Friedrich-Hund-Platz 1, D-37077 G\"{o}ttingen, Germany}


%
%
\begin{abstract}
For strongly correlated quantum systems, fundamental questions about the formation and stability of Floquet-Bloch sidebands (FBs) upon periodic driving remain unresolved.
Here, we investigate the impact of electron-electron interactions and perturbations in the coherence of the driving on the lifetime of FBs by directly 
computing
time-dependent single-particle spectral functions using exact diagonalization (ED) and matrix product states (MPS). 
We 
study interacting metallic and correlated insulating phases in a chain of correlated spinless fermions. 
At high-frequency driving we obtain clearly separated, long-lived FBs of the full many-body excitation continuum. 
However, if there is significant overlap of the features, which is more probable in the low-frequency regime, the interactions lead to strong heating, which results in a significant loss of quantum coherence and of the FBs. 
Similar suppression of FBs is obtained in the presence of noise.
The emerging picture is further elucidated by the behavior of real-space single-particle propagators, of the energy gain, and of the momentum distribution function, which is related to a quantum Fisher information that is directly accessible by spectroscopic measurements. 
\end{abstract}
\maketitle

\section{Introduction}\label{sec:Intro}
A promising direction of research for realizing interesting quantum states of matter is Floquet engineering, where the interaction of the light field with electrons in materials~\cite{Basov2017,OkaRev2019,delaTorre2021,Bao2021,ValmispildNatPhotonics2024,PRL_Li2020,Reviewmurakami2023} is used to tailor the band structure. 
For example, such periodically driven systems are predicted to offer a tunable platform to realize Fractional Chern insulators (FCIs) \cite{Grushin2014,Jotzu2014}, engineered topological states \cite{ Mikami2016,Lindner2011,McIver2019, Dehghani2015HC, Dehghani2015TI,Oka2009}, and coherent excitations in experiments with ultracold gases \cite{ReviewBloch2008,Jotzu2014,Wintersperger2020,Weitenberg2021,Lei2024,DiCarli2023} (for a review see Ref.~\onlinecite{Eckardt2017}).
A hallmark of such periodically driven crystalline lattice structures is the emergence of Floquet-Bloch sidebands (FBs).
These FBs are replica of the original band structure and appear at energies which are integer multiples of the driving frequency.
They can directly be detected in pump–probe experiments by using time- and angle-resolved photoemission spectroscopy (trARPES) \cite{Sentef2015,Wang2013,Mahmood2016,DeGiovannini2016,Gierz2021,Marco2024,Choi2025,Bielinski2025,fragkos2024}, which gives insight into the time-dependent single-particle spectral function. 
The direct observation of FBs has been reported, e.g., 
for the topological insulator Bi$_2$Se$_3$~\cite{Wang2013,Mahmood2016}, for the semiconductor material WSe$_2$~\cite{Gierz2021}, and recently for mono-layer graphene~\cite{Choi2025,Marco2024}
and for the topological antiferromagnet MnBi$_2$Te$_4$~\cite{Bielinski2025}. 
These experimental findings are for weakly interacting systems.  
Model calculations and experimental work shows that it is non-trivial to understand the Floquet physics even for weak interactions \cite{Gierz2021,Schler2020,Schler2022}.
Theoretical studies of interacting systems have been performed using cluster perturbation theory, or in infinite spatial dimensions using Floquet-DMFT and out-of-equilibrium DMFT
~\cite{Murakami2018,Tsuji2008,Aoki2014,Puviani2016}.
However, in 1D and 2D strongly interacting systems the interplay of Floquet driving and many-body effects on the spectral functions and on the stability of FBs is still a topic of ongoing investigations~\cite{Gierz2021}.
\begin{figure*}[t]
    \centering
    \includegraphics[width=0.90\linewidth]{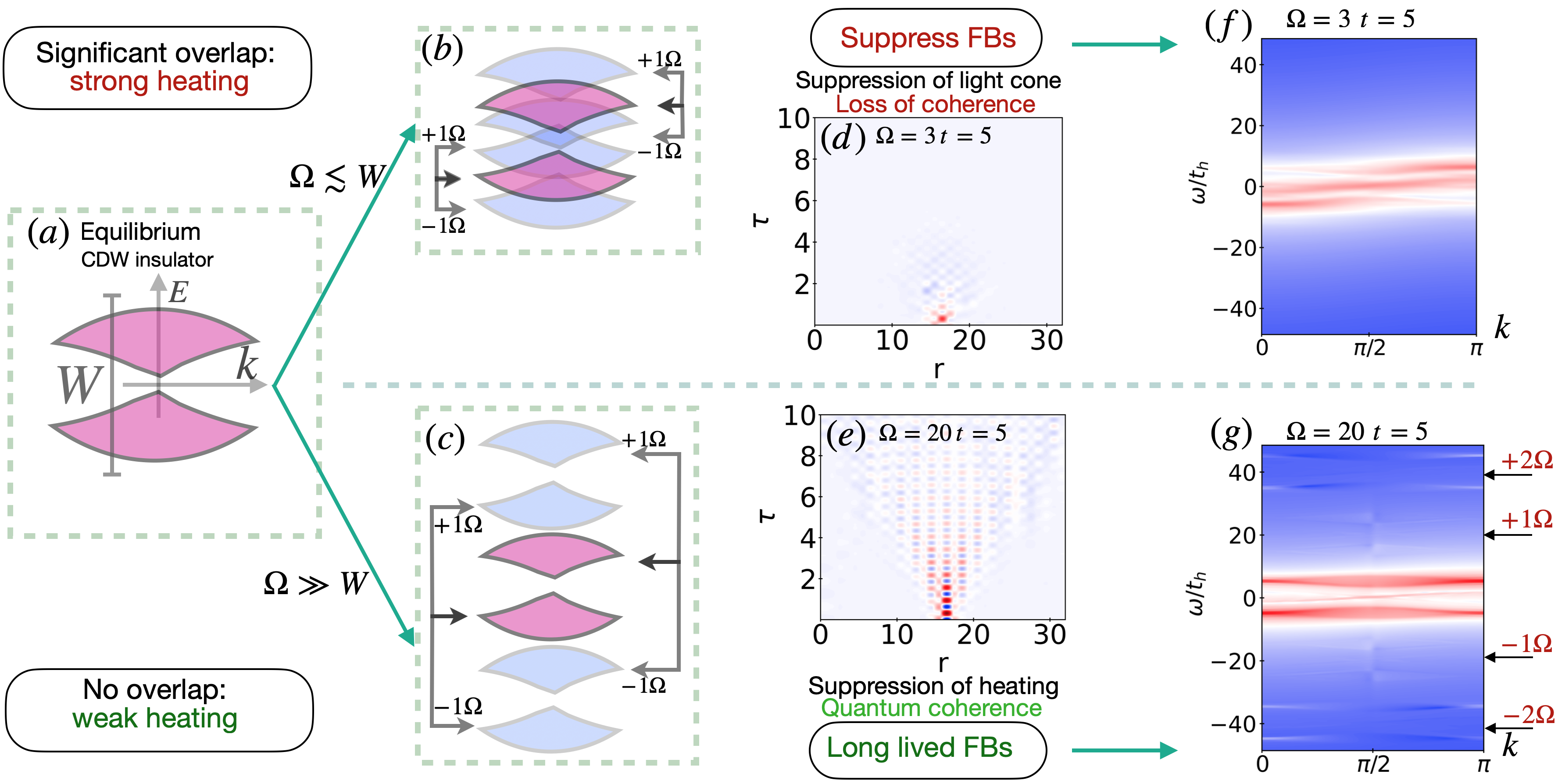} 
    \caption{Summary of the two extreme cases of our findings.
    (a) Sketch of a typical equilibrium result for the retarded spectral function $A^{\rm ret}_k(\omega)$ in the charge density wave (CDW) insulating phase at half filling for model \eqref{eq:Peierls_Hamiltonian}. 
    We denote by $W$ the \textit{spectral width} (see Sec.~\ref{sec:results}), in which the spectral function has populated continua.
    (b) and (c) 
    illustrate the general expectation for FBs in such a strongly correlated system: 
    In both sketches, the magenta region illustrates the zeroth Floquet sector, which is located at the position of the original equilibrium spectral function. Note that, due to the driving, in comparison to the equilibrium result the shape can be modified~\cite{Dunlap1986}, and additional effects (e.g., in-gap bands~\cite{Osterkorn2023}) can come into appearance. 
    This spectral function in the zeroth Floquet sector then is replicated at energies, which are multiples of the driving frequency $\pm n\Omega$.
    Here, we only sketch the 1st replicas at $\pm \Omega$ (light blue).
    In (b) we sketch the situation, when the driving frequency $\Omega < W$. This leads to substantial overlap of the different Floquet sectors.
    In contrast to this, in (c) we sketch the situation if $\Omega > W$, so that there is no overlap between the Floquet sectors. 
    (d) and (e) illustrate the effect of the driving on correlations in real space. They display the time evolution of the single particle propagator Eq.~\eqref{eq:propagator} in real space in the CDW phase for driving frequencies $\Omega = 3.0$ and $\Omega = 20.0$, respectively, at waiting time $t=5$, indicating the stability of quantum coherence in both regimes.
    (f) and (g) show examples for the resulting time dependent spectral function $A^{\rm ret}_k(t,\omega)$ [Eq.~\eqref{eq:Spectral_function}] in the CDW state at low and high frequency driving, $\Omega = 3.0$ (no FBs visible) and $\Omega=20$ (FBs visible at $\pm n\Omega$ indicated by the arrows), respectively, at waiting time $t=5$.
    All results are at $V/t_h=5$.
    } 
    \label{fig:Sketch}
\end{figure*}

Floquet systems have also been explored in the context of many-body localization (MBL) and thermalization~\cite{Ponte2015, Lazarides2015, Sierant2023, Zhang2016,Decker2020}.
 In particular, by fine-tuning drive parameters or suppressing hopping terms, one can induce localization even in clean systems (dynamical localization~\cite{Bukov2015,Dunlap1986}), revealing rich regimes of Floquet MBL and prethermal dynamics.
Other important effects of periodic driving are energy absorption~\cite{Kennes2018,Tsuji2008,Aoki2014}, dressing of electrons~\cite{Dunlap1986}, light induced gaps~\cite{Broers2021,Broers2022}, metal-insulator transitions~\cite{Tsuji2008,Aoki2014,Puviani2016}, and 
in strongly correlated systems it can also induce in-gap features in the presence of symmetry broken ground states~\cite{Osterkorn2023}.
A major challenge in interacting Floquet systems is the significant heating~\cite{Decker2020,Tsuji2024,Bukov2016,Zhao2022,Abanin2015,Haldar2018,Fleckenstein2021,Peronaci2018,Machado2019}, which drives the system toward an infinite-temperature state.
As a result, one expects Floquet effects to play a role on some transient time scale, before the heating wins~\cite{Tsuji2024}.

The scope of this paper is to investigate how for strongly correlated electron systems the interactions, the driving parameters, and noise influence the Floquet physics.
We do so by directly studying the time evolution of the retarded time-dependent spectral function $A_k^{\rm ret}(t,\omega)$, of real space correlations, and of the momentum distribution function $\langle n_k\rangle (t)$, which relates to a Quantum Fisher Information (QFI)~\cite{Hauke2016}.
All results are obtained with time-dependent matrix product states (MPS)~\cite{Schollwck2011,Paeckel2019} and exact diagonalization (ED)~\cite{Noack2005,Sandvik2010,Weinberg2019} techniques. 
While most of the calculations can be done using ED, it is useful to compare the results for the spectral functions with MPS results, since we can treat larger systems and hence reach a better resolution in $k$-space there. 
Studying larger systems will also be beneficial when exploring the dynamics of real-space correlations (see Sec.~\ref{subsec:CorrN_kE_t}).
We focus on one-dimensional (1D) systems, as MPS methods are particularly efficient in this setting.
Since correlation effects are strongest in 1D, we believe that studying them in this context will provide a generic understanding that is likely to hold in higher dimensions as well.
This, in turn, will help us understand how interactions affect the stability of FBs in correlated systems in general.
We complement this by studying coherence in real space correlations and in the momentum distribution function, which is a simple QFI, in strongly interacting Floquet-driven systems.
This is motivated by recent inelastic neutron scattering studies on quantum magnets, which analyze the time evolution of correlation functions in real space and of the QFI~\cite{Tennant2022,Scheie2021,ScheieReview2025,Assaad2024}. 
Further work relating the single-particle spectral function~\cite{Assaad2024,Malla2024} in- and out-of-equilibrium~\cite{Hales2023,Baykusheva2023} to QFIs is ongoing. 

A summary showing the two extreme cases with and without FBs is sketched in Fig.~\ref{fig:Sketch}.
We find in the high frequency regime that the FBs are stable for long times even in the presence of strong interactions, and are only suppressed when adding noise to the drive.
In the low frequency regime, they are suppressed if there is a large overlap between the FBs and the original spectral function.
In general, our results for clean monochromatic driving indicate that for interacting systems the stability of the FBs depends on this overlap.

This paper is organized as follows.
In Sec.~\ref{sec:Model}, we introduce the model used in our study.
Our main findings are presented in Sec.~\ref{sec:results},
where we discuss the system's behavior under a low-frequency drive (Sec.~\ref{subsec:LowFreq}), 
a high-frequency drive (Sec.~\ref{subsec:HighFreq}),
the stability of the FBs against noise (Sec.~\ref{subsec:NoiseCase}),
and the evolution of real-space correlations, the momentum distribution function, and heating (Sec.~\ref{subsec:CorrN_kE_t}). 
We summarize our conclusions and provide an outlook in Sec.~\ref{sec:conclusions}. 
Several technical details and complementary results are provided in the appendices.
Details on the numerical methods are provided in 
Apps.~\ref{app_details_of_calculation} and~\ref{app_window_fucntion}, with 
App.~\ref{app_window_fucntion} discussing details of the Fourier transform used, in particular the choice of a window function.
Further supporting results for the main text and an error analysis is provided in 
Apps.~\ref{app_v1p5_LL_wt},~\ref{app_real_space_corr_errorestimate},~\ref{app_E_nk_errorestimate}, and~\ref{app_crosssectional_more_k}.

\section{Model}\label{sec:Model}
We consider a system of periodically driven one-dimensional interacting spinless fermions ($tV$-chain),
\begin{equation}\label{eq:Peierls_Hamiltonian}
\begin{aligned}
H(t)= &  - t_h \sum_j  \left( e^{iA_V(t)} c_j^{\dagger} c^{\phantom{\dagger}}_{j+1}+\text { H.c. } \right)
 +V\left(n_j n_{j+1}\right) .
\end{aligned}
\end{equation}
Here, $c_j^{(\dagger)}$ is the annihilation (creation) operator for a spinless fermion at site $j$, $n_j=c_j^{\dagger} c^{\phantom{\dagger}}_{j}$ is the operator of the local density on site $j$, $t_h$ the hopping strength between nearest neighboring lattice sites, and $V$ the strength of the interaction between particles sitting on neighboring sites.
The periodic driving is realized as Peierls' substitution~\cite{Peierls1933} via a time-dependent vector potential $A_V(t)=A_0 \sin(\Omega t)$, which is activated at time $t=0$.
Unless stated otherwise, we use open boundary conditions (OBC) and the hopping parameter is set to $t_h \equiv 1$ throughout, as well as $\hbar \equiv 1$.
Without the driving (i.e., for $A_0=0$), the model Eq.~\eqref{eq:Peierls_Hamiltonian} at half filling exhibits a Berezinskii-Kosterlitz-Thouless-like (BKT) phase transition from a Luttinger liquid (LL)~\cite{giamarchi2004quantum}
to a strongly correlated charge density wave (CDW) insulator at $V/t_h=2$, as obtained from Bethe ansatz after mapping to an XXZ spin-chain via Jorgan-Wigner transform~\cite{DesCloizeaux1966}.
In order to suppress artifacts due to the OBC, we apply a pinning-field $H_{\rm pin} = \mu n_j$ at one of the edge sites, which enforces the numerics to converge to one of the two possible CDW ground states~\cite{Osterkorn2023,PhdConstantin}.
Throughout the paper, the initial state is prepared as a ground state of the system at temperature $T=0$.
We compute the non-equilibrium generalization of the spectral function by performing a Fourier transform (FT) of the single-particle Green’s function,
\begin{align}\label{eq:ret_greens_function}
G_{\alpha \beta}^{\mathrm{ret}}\left(t, t^{\prime}\right) & 
& =-i \theta\left(t-t^{\prime}\right)\left\{
\left\langle c_\alpha(t) , \, c_\beta^{\dagger}\left(t^{\prime}\right)\right\rangle
\right\}\,. 
\end{align}
The FT to $\omega$-space is not unique~\cite{Kalthoff2018}.
Typical choices are Wigner coordinates, or relative-time coordinates.
The qualitative behavior typically does not depend on the particular choice of these coordinates~\cite{Kalthoff2018,NghiemPRL2017,Nghiem2020}.
For our numerical approach, relative-time coordinates $\tau = t' - t$ for the FT are more convenient, 
since in this way one can reduce the number of MPS states that need to be stored in the course of the calculation.
The protocol is:

\begin{align}\label{eq:proto}
\ket{\psi} \quad&\xrightarrow[\text{evolve in time variable } t]{t=0: \text{ turn on driving}}|\psi(t)\rangle \nonumber \\
\ket{\psi(t)}&\xrightarrow[\text{compute Green's fct.}]{\text{evolve in } \tau} G_{\alpha,\beta}(t, t+\tau) \nonumber \\
G_{\alpha,\beta}(t, t+\tau)&\xrightarrow[\tau \rightarrow \omega]{FT} A_k(t,\omega)\,,
\end{align}

where we interpret the evolution time $t$ as a waiting time, since we let evolve the state for time $t$ under the influence of the  periodically driven time dependent Hamiltonian before we compute the Green's function ~\cite{PhdConstantin,Osterkorn2023}.
The evolution in the time variable $\tau$ is also done with the time-dependent Hamiltonian, and then the retarded single-particle spectral function is obtained via a FT over $\tau$, 

\begin{equation}\label{eq:Spectral_function}
\begin{aligned}
& A_k^{\mathrm{ret}}(t, \omega)=-\operatorname{Im} \frac{1}{\sqrt{2 \pi}} \int_{-\infty}^{\infty} \mathrm{d} \tau 
e^{i \omega \tau} W(\tau) 
G_{k k}^{\mathrm{ret}}(t, t+\tau) \, .
\end{aligned}
\end{equation}
The function $W(\tau)$ is a suitable window function (see 
App.~\ref{app_window_fucntion}).
Note that even at waiting time $t=0$ the results differ from the ground state behavior, since the Fourier-transform to $\omega$-space is done on results, which were obtained with the time-dependent Hamiltonian~\eqref{eq:Peierls_Hamiltonian}.
The retarded spectral function Eq.~\eqref{eq:Spectral_function} can then be interpreted as the spectral function measured at waiting time $t$, with the driving being turned on at $t=0$. 
In contrast to the equilibrium case, $A^{\rm ret}_k(t,\omega)$ can have negative values, see, e.g., Refs.~\onlinecite{UhrigPRL2019,Blum2025}, which, however, in some strongly interacting systems have been found to be absent at later times~\cite{Osterkorn2023,Blum2025}. 
Here, in the context of the existence of FBs, we focus our discussion on the absolute value $|A^{\rm ret}_k(t,\omega)|$.  
We use OBC for our MPS results, since MPS are more efficient there~\cite{Schollwck2011}.
For the results at long times, we need to use Lanczos time evolution~\cite{Manmana2005}, for which we use PBC and smaller system sizes. 
The results with OBC and PBC show the same behavior, so that in the main part of the paper we discuss them side by side, where needed.
For details of the MPS calculations and an error estimate, 
see Apps.~\ref{app_details_of_calculation}, \ref{app_real_space_corr_errorestimate}, and \ref{app_E_nk_errorestimate}.
\section{Results}\label{sec:results}

\begin{figure}[t]
    \centering
    \includegraphics[width=0.95\linewidth]{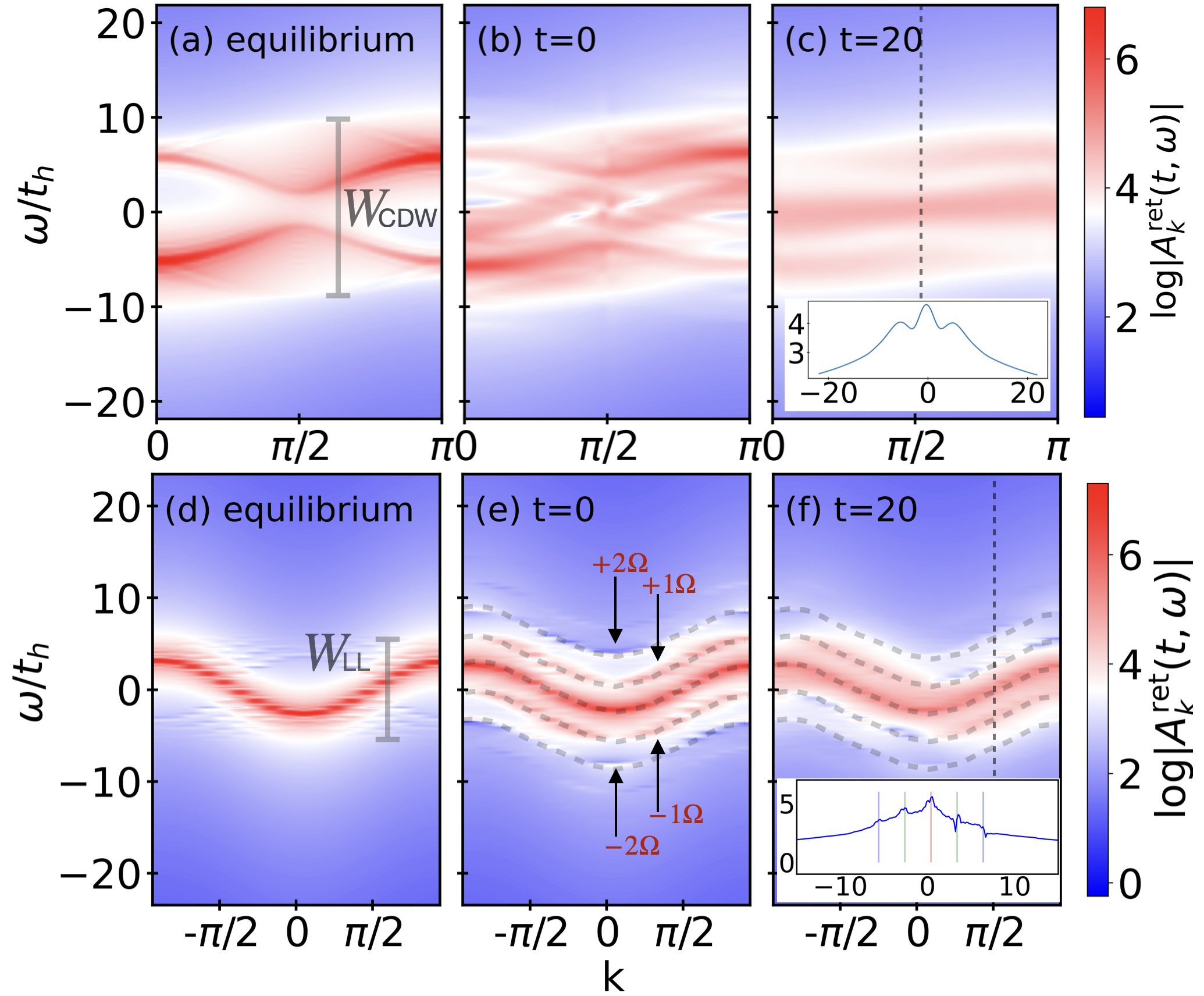} 
    \caption{Retarded single-particle spectral function $A^{\rm ret}_k(t,\omega)$ [Eq.~\eqref{eq:Spectral_function}] for the $tV$-chain~\eqref{eq:Peierls_Hamiltonian}. (a)-(c):  Results in the CDW phase for $V/t_h=5$ for $L=32$ and OBC at half filling obtained with MPS. (a) Results at equilibrium.
    $W_{\rm CDW} \approx 18$ estimates the spectral width, in which $A^{\rm ret}_k(\omega) > \epsilon$, where $\epsilon$ is $2\%$ of the peak value of $A^{\rm ret}_k(t,\omega)$ (see main text). 
    (b) Driven case with $\Omega = 3$ and $A_0=1.0$ at waiting time $t=0$. (c) The same driven case at waiting time $t=20$. The inset in (c) shows the result at fixed $k=\pi/2$ for $t=20$. 
    (d)-(f):  Results in the LL phase $V/t_h=1.5$ for $L=18$ and PBC at half filling obtained with ED. (d) Results at equilibrium, $W_{\rm LL} \approx 10$ denotes the spectral width, as before. (e) and (f) nonequilibrium results for the same driving parameters as in (b) and (c).
     In (e) the first two side bands are indicated with arrows.
    The inset in (f) shows the results at fixed $k=\pi/2$ for $t=20$, vertical lines indicate the position of the side bands. Dashed lines in (e) and (f) are guides to the eye for locating the side bands.}
\label{fig:Low_freq_Eq_and_driven}
\end{figure}
We present in Figs.~\ref{fig:Low_freq_Eq_and_driven} and~\ref{fig:High_freq_drive} our main results.
In Fig.~\ref{fig:Low_freq_Eq_and_driven}, we investigate the dynamics at low-frequency driving for two distinct phases of the system: the CDW insulator (top row) and the LL phase (bottom row).
For the high frequency drive, we focus on the CDW insulator, since here interaction effects would be strongest, see Fig.~\ref{fig:High_freq_drive} for our results.

In order to better distinguish between the low- and high-frequency regime, we introduce the spectral width $W$ in the ground-state retarded spectral function at equilibrium, $A^{\rm ret, \, eq}_k(\omega)$.
In the context of our discussion, we are most interested in the influence of correlation effects on the properties of the FBs.
On general grounds, for noninteracting systems we expect well-defined band structures and no continua in $A^{\rm ret,\,eq}_k(\omega)$.
Hence, the presence of continua in $A^{\rm ret,\,eq}_k(\omega)$ is indicative for correlation effects.
This is further confirmed by mean-field calculations (Hartree-Fock), which cannot reproduce the excitation continua discussed here in the tV-chain, see Ref.~\onlinecite{Osterkorn2023}. 
In our estimate of $W$, we therefore focus on the extend in $\omega$-direction covered by such continua. 
In Fig.~\ref{fig:Low_freq_Eq_and_driven}(a) we denote by $W_{\rm CDW}$ the spectral width in the CDW phase; since for these parameters the most prominent aspects of the spectral function are continua, $W_{\rm CDW}$ essentially covers the entire region in which $|A^{\rm ret,\,eq}_k(\omega)| > \epsilon$, with $\epsilon$ a suitable threshold value. 
Note that the support of $A^{\rm ret,\,eq}_k(\omega)$ can extend over all $\omega$-values~\cite{Pereira2009}.
In general, the continua may decay as a power law and hence have finite weight over a wide range of $\omega$.
However, for the following discussion it is only important to distinguish highly populated regions from weakly populated regions, so that choosing $\epsilon$ small enough suffices.
Here, we choose $\epsilon$ such that $|A^{\rm ret,\,eq}_k(\omega)|$ is larger than $2\%$ of the peak value, leading to a value $W_{\rm CDW} \approx 18$.
In the LL case shown in Fig.~\ref{fig:Low_freq_Eq_and_driven}(d), the situation is different.
Here, $A^{\rm ret,\,eq}_k(\omega)$ is dominated by a $\cos$-like feature, plus additional weak excitation continua~\cite{Pereira2009}.
Using again a threshold value of $2\%$ of the peak value, we obtain a spectral width $W_{\rm LL} \approx 10$.
Based on these considerations, we identify $\Omega/t_h < W$ as a low-frequency drive and $\Omega/t_h > W$ as a high-frequency drive.

In more detail: in Fig.~\ref{fig:Low_freq_Eq_and_driven}(a) we present MPS results for $A_k^{\rm ret,\,eq}(\omega)$ deep in the CDW-regime at $V/t_h=5$ for $L=32$ at half filling.
Two continua of holon excitations, separated by a gap, are observed within the energy range $\omega \in [-10,10]$, consistent with previous studies \cite{Osterkorn2023}.
The gap size $\Delta$ can be calculated using the Bethe ansatz, which for $V/t_h=5$ obtains $\Delta/t_h \approx 1.576$~\cite{DesCloizeaux1966}.
This is in agreement with the gap size seen in Fig.~\ref{fig:Low_freq_Eq_and_driven}(a), which is affected by finite-size effects and the limited spectral resolution of our methods.
Fig.~\ref{fig:Low_freq_Eq_and_driven}(d) shows ED results for $A_k^{\rm ret, \,eq}(\omega)$ for $L=18$ at half filling for $V/t_h=1.5$, i.e. in the LL phase, which shows the aforementioned cosine-like dispersion with a buildup of a weakly populated excitation continuum without gap.
This weakly populated region is observed within the energy range $\omega \in [-5,5]$, leading to our estimated value of $W_{\rm LL}$.

The equilibrium results shown in Figs.~\ref{fig:Low_freq_Eq_and_driven}(a) and (d) serve as the starting point for our investigation. 
A key aspect to consider is the potential overlap of the Floquet replicas with the original GS spectral function:
in case of a substantial overlap,  
interference or scattering events seem likely, which should lead to heating.
This is expected for $\Omega \lesssim W$. 
 
In driven systems, heating can be suppressed by various mechanisms.
For example, in disordered systems, Floquet many body localization (MBL) has been discussed as a possibility to avoid heating up of the system to infinite temperatures, at least on transient time scales~\cite{Ponte2015,Ponte2015-2,Moessner2017,Bordia2017,Sierant2023}. 
Without disorder, it has been reported in the high frequency case that heating is suppressed due to an emergent effective static Hamiltonian. 
A transient prethermal state forms, which can be long lived, before the system finally heats up to infinite temperatures~\cite{Bukov2016,Abanin2015,Peronaci2018,Machado2019,Abanin2017}.
Indeed, our findings for high frequency driving discussed in Sec.~\ref{subsec:HighFreq} indicate this behavior.
In such systems, the driving frequency is typically compared to the intrinsic energy scales of the system~\cite{Abanin2015}, and in some cases, to the noninteracting bandwidth~\cite{Bukov2016}.
At low frequencies, the picture of an effective static Hamiltonian, which is obtained by perturbative ansatzes, may loose its validity.
This makes the question of heating in a generic many-body system a more difficult one due to
non-universal heating rates in low-to-intermediate driving frequencies~\cite{Fleckenstein2021} and may also depend on the presence of high-energy continuum states~\cite{Abanin2015}.

Our findings indicate that the suppression of heating is related to the overlap of the Floquet replica, and also to the actual weight of the spectral function and its replica in the overlapping regions.
We use the spectral width $W$ as defined above in order to estimate the role of correlation effects.
This allows us to determine 
the relevant energy scale against which the driving frequency is compared, 
as illustrated in Fig.~\ref{fig:Sketch}(a).
However, it is not clear a priori whether in the case of overlapping Floquet replicas a transient time scale remains, on which FBs can form.
This can be contrasted to non-interacting systems, where  we do not expect heating due to the absence of scattering terms, even in the presence of a complicated band structure (e.g., multiple bands).

\subsection{Low frequency drive}\label{subsec:LowFreq} 
Here, we set the driving amplitude $A_0=1.0$ and 
driving frequency $\Omega/t_h=3.0$,
 which is smaller than
$W_{\rm CDW}  \approx 18$ and smaller than $W_{\rm LL} \approx 10$.
The results in the CDW phase with $V/t_h=5.0$ are shown in Figs.~\ref{fig:Low_freq_Eq_and_driven}(a)-(c).
One would expect replicas of the entire equilibrium spectral function, which, due to the small value of $\Omega/t_h$, would significantly overlap with the original signal and with each other.
In Fig.~\ref{fig:Low_freq_Eq_and_driven}(b) we observe a significant spectral weight within the gap region, but no signature of FBs.
The in-gap feature at $t=0$ exhibits a rich structure, which is reminiscent
to the avoided-crossing scenario expected in non-interacting two-band models \cite{Tsuji2008, Aoki2014, Oka2009, Park2014, Schler2020}.
However, as shown in Fig.~\ref{fig:Low_freq_Eq_and_driven}(c) for $t=20.0$,
these features 
vanish at later times, and a broad continuum is obtained, with a stronger weight inside the original gap region.
This is further illustrated in the inset of the Fig.~\ref{fig:Low_freq_Eq_and_driven}(c) for momenta $k=\pi/2$ (dashed line). Further results at different momenta $k$ show similar behavior and are shown for $t=0$ and $t=20$ in  App.~\ref{app_crosssectional_more_k}.
This melting of the gap features hints towards strong heating effects due to the chosen drive parameters, see further below.
This indicates that 
once the replicas overlap in this strongly interacting system, heating prevails and the spectral function does not show FBs.

In comparison  to the gapped CDW phase, in the gapless LL phase also at $\Omega/t_h = 3$ we obtain FBs, which persist also at intermediate times, see Figs.~\ref{fig:Low_freq_Eq_and_driven}(d)-(f)
 (results are obtained with Lanczos time evolution, as the low-frequency case exhibits higher entanglement buildup in MPS, leading to lower accuracy. See Apps.~\ref{app_E_nk_errorestimate} and~\ref{app_real_space_corr_errorestimate} for further discussion): 
Fig.~\ref{fig:Low_freq_Eq_and_driven}(d) shows the spectral function in equilibrium, which follows a cosine-like dispersion with a buildup of a weakly populated excitation continuum without gap. 
In Fig.~\ref{fig:Low_freq_Eq_and_driven}(e),  
we observe an overlap of this weakly populated excitation continuum and the FBs at $t = 0$.
The FBs are clearly visible, even at $t = 20$ in Fig.~\ref{fig:Low_freq_Eq_and_driven}(f).
This is further illustrated in the inset of Fig.~\ref{fig:Low_freq_Eq_and_driven}(f) for $k=\pi/2$.
However,
the overlap of the FBs and the original spectral function now leads to a weaker signal of the FBs compared to the $t = 0$ case. 
In addition, the intensity of the FBs is time-dependent, and at longer waiting times $t \gtrsim 50/t_h$ they will be suppressed (see App.~\ref{app_v1p5_LL_wt}). 
Hence, our results indicate that in both, the interacting LL and in the CDW phases, overlapping bands are the main cause for the suppression of FBs 
due to the enhancement of scattering and the resulting heating. 
However, if the overlap is weak as in our results for the LL case, the FBs persist on a transient time scale, which for our parameters is $t_{\rm transient} \sim 50/t_h$ (see App.~\ref{app_v1p5_LL_wt} for additional intermediate and longer waiting time instances).
\subsection{High frequency drive}\label{subsec:HighFreq}
\begin{figure}[t]
    \centering
    \includegraphics[width=0.89\linewidth]{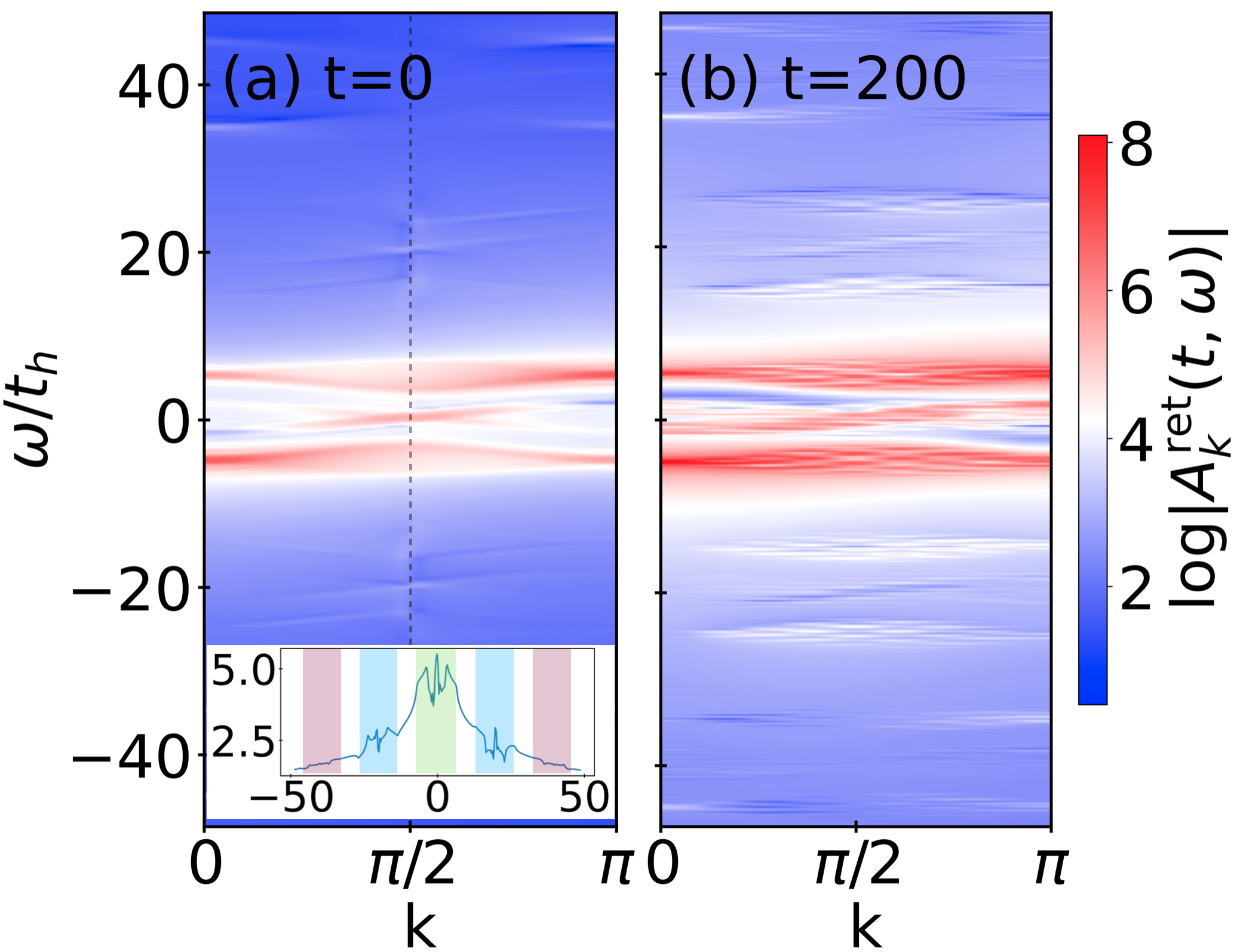}
    \caption{
    The same as in Fig.~\ref{fig:Low_freq_Eq_and_driven} for $V/t_h=5$, but at high-frequency driving $\Omega=20$ and driving amplitude $A_0=1.6$. (a)  
    shows MPS results for $L=32$ and OBC, while the long-time results in (b)
    are obtained using Lanczos time evolution with $L=18$ sites and PBC. 
    The inset in (a) shows the results at fixed $k=\pi/2$; the green highlighted region indicates the spectral function in the 0th Floquet sector, the blue shaded regions indicate the 1st FBs, and the magenta shaded regions indicate the 2nd FBs.
    }
    \label{fig:High_freq_drive}
\end{figure}

In Fig.~\ref{fig:High_freq_drive} we present results for the same set up, but with $A_0=1.6$, $\Omega/t_h = 20 >W_{\rm CDW}$, and $V/t_h=5.0$.
At waiting time $t=0.0$ in Fig.~\ref{fig:High_freq_drive}(a), we observe clear FBs at integer multiples of $\pm \Omega$.
These FBs exhibit the full excitation continuum with renormalized bandwidth (since $t_h^{\rm eff} = J_0(A_0)t_h$~\cite{Dunlap1986,Kennes2018,Tsuji2024}) and the emergence of an in-gap band,
see Ref.~\onlinecite{Osterkorn2023}.
This shows that also in the strongly interacting case the FBs are replicas of the full spectral function of the 0th Floquet sector, including all features.
This is further illustrated in the inset of Fig.~\ref{fig:High_freq_drive}(a) for $k=\pi/2$ (dashed line); see App.~\ref{app_crosssectional_more_k} for further cross sections at other values of $k$.
The question arises, if these FBs are stable in time, or if scattering between the particles eventually will destroy coherence, leading to a suppression of the FBs in $A^{\rm ret}_k(t,\omega)$ at later times. 
A simple estimate from perturbation theory suggests scattering becomes visible on time scales $\sim 1/V^2$.
Hence, a suppression of the FBs at waiting times $t \lesssim 1$ would follow.
However, in Fig.~\ref{fig:High_freq_drive}(b), we reach longer waiting times using the Lanczos time evolution, we have to restrict to system sizes of $L=18$ (PBC).
Our results indicate that even at these substantially longer times of $t=200$,
the FBs remain stable.
In contrast to low frequency drive where heating suppresses FBs already at short times $t \sim 5$ (see above).
This demonstrates that high-frequency driving can yield FBs even in strongly interacting quantum many-body systems, exhibiting stability over the extended timescales we examined. 

\subsection{Stability of the FBs against noise}\label{subsec:NoiseCase}
\begin{figure}[t]
    \centering
    \includegraphics[width=\linewidth]{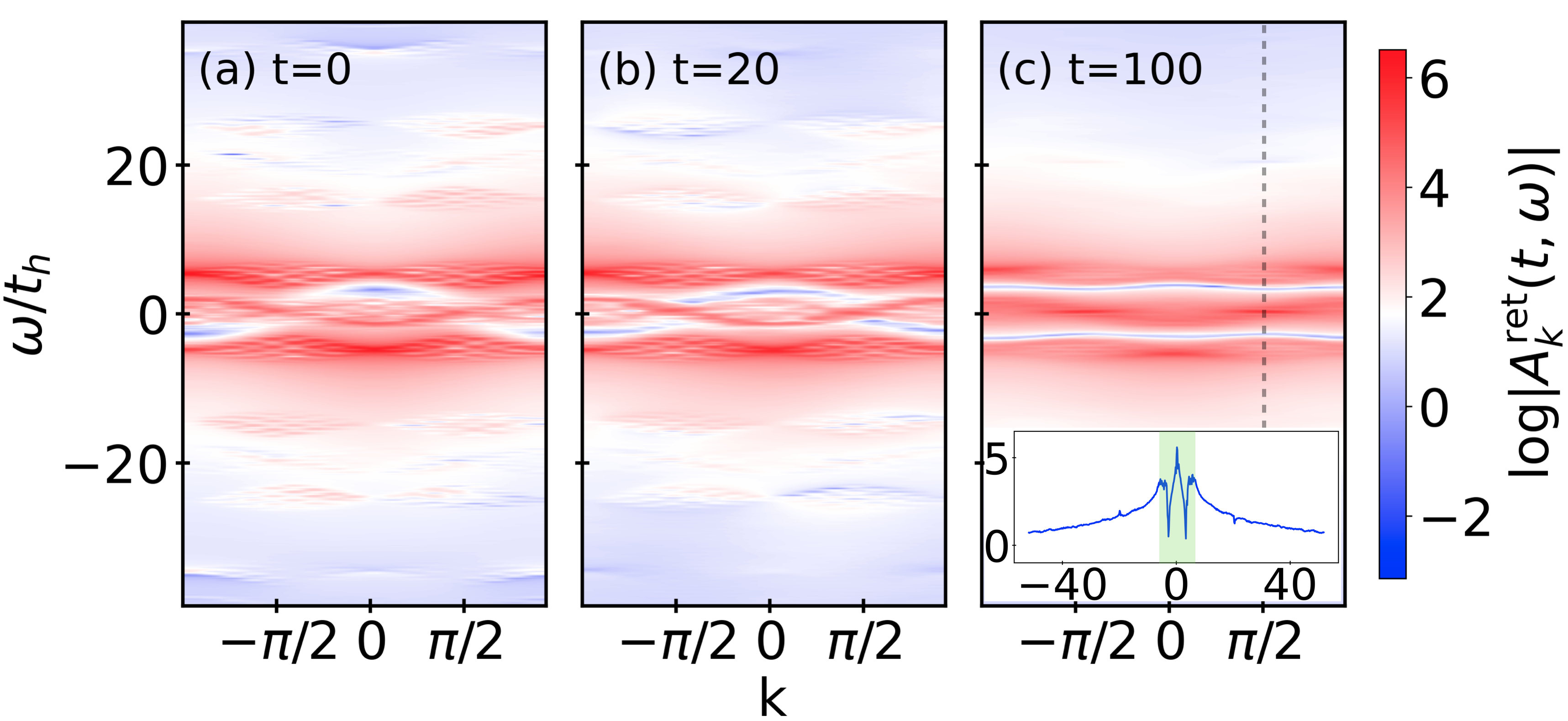} %
    \caption{The same as in Fig.~\ref{fig:High_freq_drive}, but in the presence of time-dependent noise in the driving, where $\Omega = 20 \pm \Omega_{\rm noise}$ and $\Omega_{\rm noise} \in [0,0.1]$. 
    The results are obtained using Lanczos time evolution with $L=18$ and PBC.
    Panels (a)-(c) show $A_k^{\rm ret}(t,\omega)$ at waiting times $t=0$, $t=20$, and $t=100$, respectively.
    The inset in (c) shows results at fixed $k=\pi/2$ at $t=100$.
    All results show an average over 8 realizations of the random noise.
    }
    \label{fig:noise_in_omega}
\end{figure}

We now test these findings by explicitly breaking the coherence when adding incoherent noise to the drive and ask for the stability of the FBs as a function of time.
In order to reach longer times, we again apply the Lanczos time evolution method for $L = 18$ (PBC) at half-filling, $V/t_h = 5.0$, $A_0 = 1.6$, and a frequency $\Omega = 20 \pm \Omega_{noise}$, where $\Omega_{noise}$$\in [0,0.1]$ represents the random time-dependent noise in the driving frequency, such that $A_V(t) = A_0 \sin((\Omega \pm \Omega_{noise})t)$.
We discretize the time variable such that $t= n dt$, with $n \in \mathbb{N}_0$, and then compute $A^{\rm ret}_k(t,\omega)$ when applying $A_V(t) = A_0 \sin((\Omega \pm \Omega_{noise}) n dt)$.
Note that the results depend on the value of $dt$, as discussed in  App.~\ref{app_E_nk_errorestimate}.
Here, we present results for $dt=0.005$ in Fig.~\ref{fig:noise_in_omega}.  
Figures~\ref{fig:noise_in_omega}(a) and (b)
demonstrate that, at small to intermediate waiting times, despite the introduction of noise, FBs are stable.
However, at $t=20$ in Fig.~\ref{fig:noise_in_omega}(b), we observe a noticeable suppression of the second-order FBs.
In Fig.~\ref{fig:noise_in_omega}(c), for waiting time $t=100$, this suppression becomes even more pronounced, with the cross-section at $k = \pi/2$ shown in the inset of Fig.~\ref{fig:noise_in_omega}(c) clearly showing the suppression of all FBs.
Additionally, we notice a higher intensity inside the gap region, similar to what we found for  low frequency driving.
\subsection{Real space correlations, momentum distribution function, and heating}\label{subsec:CorrN_kE_t}
\begin{figure}[t]
    \centering
    \includegraphics[width=\linewidth]{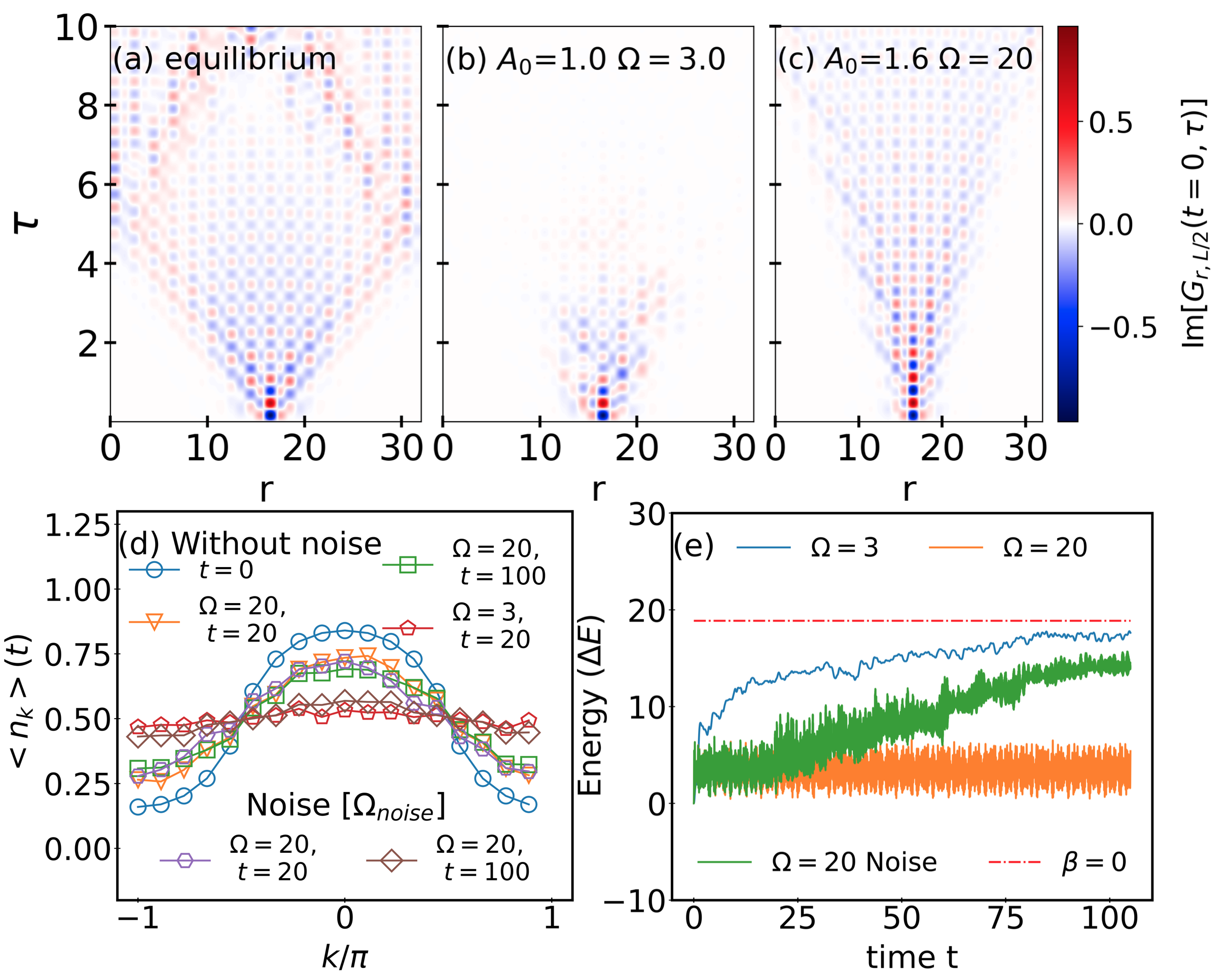}
    \caption{Results at $V/t_h=5$: (a)-(c) show the imaginary part of 
    $G_{r,L/2}(t,\tau)$ [Eq.~\eqref{eq:propagator}] (MPS, $L=32$, OBC)  
    for the equilibrium case, and for the driving parameters of Figs.~\ref{fig:Low_freq_Eq_and_driven} and~\ref{fig:High_freq_drive} at waiting time $t=0$.
    (d) shows $\langle n_k\rangle(t)$ 
    at $\Omega=3$
    and at $\Omega=20$, with and without noise for $L=18$ and PBC.
    (e) shows typical time evolution of the energy gain compared to the initial state (ground state energy $E_{\rm GS}$), $\Delta E(t) = \langle H(t) \rangle - E_{\rm GS}$ for different driving frequencies for $L=14$ and PBC. 
    }
    \label{fig:light_cone_energy_gain}
\end{figure}

We complement our results for $A_k^{\rm ret}(t,\omega)$ by studying the time evolution of correlation functions in real space and of the 
momentum distribution function $\langle n_k\rangle(t)$, 
accompanied by the behavior of the energy as a function of time.
This helps us to monitor the heating, which is intrinsic to driven interacting systems~\cite{Abanin2015,Bukov2016,Haldar2018,Fleckenstein2021,Peronaci2018,Machado2019}.  
In the following we focus on the case $V/t_h=5.0$.
We apply our low and high frequency driving protocols in Figs.~\ref{fig:light_cone_energy_gain}(a)-(c) to the imaginary part of the real-space single-particle propagator
\begin{equation}
G_{r,L/2}(t,\tau) = \bra{\psi(t)} \left[ c^\dagger_r (\tau), c_{L/2} \right] \ket{\psi(t)} \, .
\label{eq:propagator}
\end{equation}
Fig.~\ref{fig:light_cone_energy_gain}(a) shows the typical light-cone-like behavior expected for gapped ground states~\cite{Lieb1972,Calabrese2006,Calabrese2007,Bravyi2006,Laeuchli2008,Manmana2009,Cheneau2012,Kalthoff2019}.
Additionally, one observes 
a `checkerboard-like' pattern, which is similar to the one observed
for spinons in spin correlation function in neutron scattering experiments~\cite{Tennant2022}.
Here, we expect this should translate to the presence of holon excitations.
In stark contrast to the equilibrium results,
we see in Fig.~\ref{fig:light_cone_energy_gain}(b) that for low frequency driving already at $t=0$ the correlation function is quickly suppressed as a function of time.
This demonstrates that in this case coherence is quickly lost, which suggests the absence of FBs in $A_k^{\rm ret}(t,\omega)$.
However, for the high-frequency driving case shown in Fig.~\ref{fig:light_cone_energy_gain}(c), again a light-cone is visible, 
which is narrow due to the reduced hopping $t_h^{\rm eff} = J_0(A_0) t_h$, which leads to a lower propagation velocity of the perturbations in the system.
In the equilibrium case one observes `wakes'~\cite{Tennant2022}, which are created inside the light cone. 
In the driven case, they are more pronounced at later times, and every second one seems to be suppressed. 
Also, the checkerboard pattern is very similar to the one at equilibrium, but with slightly slower oscillation frequency in time. 
This is also obtained at later waiting times (see App.~\ref{app_real_space_corr_errorestimate}). 
These findings illustrate that coherence is maintained, suggesting that here long-lived FBs are possible, as indeed found in Fig.~\ref{fig:High_freq_drive}.

Coherence effects can also be studied 
in the momentum distribution function $\langle n_k \rangle(t) = \langle c_k^\dagger c_k^{\phantom{\dagger}} \rangle(t)$, with $c_k^{(\dagger)} = 1/\sqrt{L}\sum_{r=1}^L e^{(-)ik\cdot r} c_r^{(\dagger)}$, 
which is accessible via time-of-flight experiments in cold-gases setups~\cite{ReviewBloch2008}. 
In Fig.~\ref{fig:light_cone_energy_gain}(d) we show several snapshots of $\langle n_k\rangle(t)$ for $\Omega = 3.0$ and $\Omega = 20.0$ with and without noise (ED results for $L=18$ and $V/t_h=5$). 
At equilibrium, we obtain the typical result for a strongly correlated insulator. 
This behavior persists in the high frequency regime up to the longest times studied by us, here $t=100$. 
However, when turning on the noise, or in the low frequency regime,  
all features are lost and $\langle n_k \rangle(t) \approx const$ for all values of $k$.
In the low frequency case, this happens already at short times $t \sim 5$, while in the case with noise it happens only at later times $t \gtrsim 30$. 
This featureless result for $\langle n_k\rangle$ is expected at very high or infinite temperatures, further showing that in these cases coherence gets lost.

The findings for $G_{r,L/2}(t,\tau)$ and for $\langle n_k \rangle(t)$ can be related to heating effects by comparing with the behavior of the energy gain as a function of time $\Delta E(t) = \langle H(t) \rangle - E_{\rm GS}$ (with $E_{\rm GS}$ the ground state energy), see Fig.~\ref{fig:light_cone_energy_gain}(e).
This has been discussed extensively in the literature~\cite{Abanin2015,Bukov2016,Haldar2018,Fleckenstein2021,Peronaci2018,Machado2019}. 
Further results for $G_{r,L/2}(t,\tau)$, $\langle n_k \rangle(t)$, and $E(t)$ for longer times and at later waiting times, as well as a discussion of the numerical errors, can be found in Apps.~\ref{app_real_space_corr_errorestimate} and~\ref{app_E_nk_errorestimate}. 
The results for coherent driving with $\Omega=3$ and $\Omega=20$ differ clearly at long times. 
While for $\Omega=20$ 
the energy oscillates around a value, which is clearly not at infinite $T$, 
for $\Omega=3$ it approaches this limit quickly.
This seems in agreement with the expected suppression of heating at high driving frequencies~\cite{Abanin2015}. 
When adding noise, we find intermediate behavior: while at times $t \lesssim 25$ the results for the coherent driving are reproduced, at later times the result at infinite temperatures is approached. 
This is in agreement with the findings of Ref.~\onlinecite{Fleckenstein2021_Noise}. 
It is interesting that in the presence of noise a clean transient time window exists, in which the behavior is essentially the one of the coherent driving, as can also be seen in Figs.~\ref{fig:noise_in_omega}(a) and (b).

\section{Conclusions and Outlook}\label{sec:conclusions}
We directly studied the formation of FBs in the presence of strong correlations in the time-dependent spectral functions for different interaction strengths and driving parameters using MPS and ED methods.
Since in the 1D systems studied by us correlation effects are strongest, our results
indicate that, in the absence of noise, Floquet engineering should be possible also in strongly correlated systems and for suitable parameters also for low-frequency driving on a transient time scale.
This should also hold for strongly correlated materials at low enough temperatures, e.g., cuprates~\cite{Dagotto1994}: As long as the coherent driving frequency is larger than the typical spectral width $W$ and noise is suppressed, 
if the overlap of the FBs and the original spectral function is weak, 
the FBs persist on a transient time scale, which depends on details of the system.
In contrast, for large overlap, we find fast loss of coherence and very strong heating, which leads to an immediate suppression of the FBs.
For high frequency driving, (i.e. $\Omega \gg W$), no such overlap is possible, and FBs are long-lived even in the presence of strong interactions.
However, incoherent noise in the driving also here leads to a transient timescale after which they are suppressed.
This is further corroborated by the behavior of the single-particle propagator $G_{r,L/2}(t,\tau)$ and the momentum distribution function $\langle n_k\rangle(t)$, which is related to a QFI $
F_Q(k,t)=\int d\omega A^{\rm ret}_k(t,\omega) = \langle n_k \rangle(t)$ 
when expressing the QFI using the simple entanglement witness operator $\hat{O}=\sum_i (a_i c_i^\dagger + h.c.)$~\cite{Hauke2016}.
This deserves further investigation in order to explore the possibility to relate time-dependent single-particle spectral functions with other QFIs.~\cite{Malla2024}.
Future investigations of FBs in strongly correlated systems should address a more realistic modeling of such materials and include the effect of finite temperatures and of phonons on the stability of the FBs and hence the possibility to realize Floquet engineering in these materials.

\section*{Acknowledgements}
We acknowledge interesting discussions with Theo Costi, Alan Tennant, Manuel Buriks, Fabian Heidrich-Meisner, Anatoli Polkovnikov, Michael Sch\"uler, Maksymilian \'Sroda, and
the Journal Club of the Stefan Mathias group (in particular Marco Merboldt, Marcel Reutzel, and Stefan Mathias).
The calculations were performed on the goegrid cluster in G\"ottingen.
We acknowledge financial support by Deutsche Forschungsgemeinschaft (DFG, German Research
Foundation) Grants No. 436382789, and No. 493420525, via large equipment grants (GOEGrid), and by the Deutsche Forschungsgemeinschaft (DFG, German Research Foundation)
- 217133147/SFB 1073, project B03.
\section*{Data and code availability}
All the scripts and data used to prepare the manuscript are available on Zenodo upon request~\cite{gadge_2025_15115513}.

\appendix

\section{Details of the calculations}\label{app_details_of_calculation}
We compute  
\begin{equation}
    G_{k k}^{\rm ret}\left(t, \tau\right):=-i \theta(\tau)\left\langle\left\{c_k\left(t+\tau\right), c_k^{\dagger}\left(t\right)\right\}\right\rangle \,,
    \label{eq:G_eq}
\end{equation}
which is the retarded Green's function.
For the real-to-$k$-space transform, in the case of PBC we apply a standard Fourier transform, $c_k^{(\dagger)}=1/\sqrt{L} \sum_j e^{(-)ijk} c_j^{(\dagger)}$ with momenta $k=2 \pi/L \cdot \{-L/2, \ldots , \, L/2-1\}$.
For OBC, we apply the sine transform with quasi-momenta $k\in \pi/L+1 \cdot \{1,...,\,L \}$ \cite{Benthien2004,Tiegel2016,Osterkorn2023,Khler2020},
\begin{equation}
    c_k=\sqrt{\frac{2}{L+1}}\sum_{i}\sin(k\cdot i)c_i\,.
    \label{eq:sin-transform}
\end{equation}
This leads for OBC to a nontrivial time dependence of $\langle n_k\rangle(t)$ even for noninteracting systems, as can be seen as follows.
Consider the noninteracting Hamiltonian with Peierls' substitution, 
\begin{equation}\label{eq:Peierls_Hamiltonian_nonint}
\begin{aligned}
H(t)= &  - t_h \sum_{j=1}^{N-1}  \left( e^{iA_V(t)} c_j^{\dagger} c_{j+1}+\text { H.c. } \right)\,.
\end{aligned}
\end{equation}
Using the sine transform, the Hamiltonian is diagonalized, but has time-dependent eigenvalues,
\begin{equation}\label{eq:Peierls_Hamiltonian_nonint1}
\begin{aligned}
H(t)= &  - t_h \sum_{k}  \left[ \cos(k+A_V(t)) + \cos(k-A_V(t)) \right] c_k^{\dagger} c^{\phantom{\dagger}}_{k}\,,
\end{aligned}
\end{equation}
which implies that $\langle n_k \rangle(t)$ is time-dependent.
(At $t=0$, this simplyfies to the usual  expression $H(t=0) =   - 2 t_h \sum_{k}   \cos(k) c_k^{\dagger} c^{\phantom{\dagger}}_{k}$.)

The MPS calculations have been carried out using the SymMPS toolkit (developed by Sebastian Paeckel and Thomas Köhler), which is freely available at \onlinecite{SymMPS}.
For comprehensive reviews we refer to the literature \cite{Paeckel2019,Schollwck2011}.
We follow the steps:\\
\begin{equation} \label{MPS-tevol}
\begin{aligned}
|\psi(t)\rangle & =U_{\mathrm{TDVP}}(t, 0)|\mathrm{\psi_0}\rangle, \\
\left|\psi_l(t)\right\rangle & =c_l|\psi(t)\rangle, \\
\left|\psi_l(t+\tau)\right\rangle & =U_{\mathrm{TDVP}}(t+\tau, t)\left|\psi_l(t)\right\rangle,
\end{aligned}
\end{equation}
with the time-dependent time evolution operator $U_{TDVP}(t,t')$, which is implemented using a $2-$site time-dependent variational principle (TDVP)~\cite{Paeckel2019}.
For time-dependent Hamiltonians, there are better approximations (e.g., commutator free expansions, see Ref.~\onlinecite{ALVERMANN2011}); however, if the value of $dt$ is chosen small enough, we obtain a good accuracy also with this simpler approach. 
We repeat similar calculations for the operator $c_l^{\dagger}$. 
Using the states~\eqref{MPS-tevol} we calculate the quantities
\begin{equation}
C_{m l}(t, \tau)=\left\langle\psi(t+\tau)\left|c_m^{\dagger}\right| \psi_l(t+\tau)\right\rangle \, . \\
\end{equation} 
We work in units of the energy in which the hopping parameter $t_h \equiv 1$, and $\hbar \equiv 1$ throughout the simulations, which leads to discrete time steps $dt$ in these units.
Typically, we set $dt = 0.005$, which is chosen to be one order of magnitude smaller than the highest frequency scale $\sim 1/\Omega$ (with $\Omega=20$ for the high frequency drive).
We vary the bond dimension between $400$ and $800$.
For $\Omega = 20$ for $dt=0.005$ without noise we observe that the typical discarded weight at the end of the time evolutions is $ \lesssim 10^{-8}$, which is a rather small value, indicating a high precision of the results.
This is further corroborated by comparing the results for the time evolution of the energy with different values of $dt$ as shown in Fig.~\ref{fig:dt_vs_Et}, where excellent agreement is found (see also the discussion further below).
In other cases, in particular for $\Omega = 3$, the discarded weight grows much faster, and we reach rather large maximal values $\lesssim 10^{-4}$ at the end of the time evolutions treated.
It is very expensive to perform the calculations at substantially higher bond-dimension, so that we refrain from doing so. 
However, we check for the time-evolution of the energy when varying $dt$, and find that also in these cases the results for different values of $dt$ are in excellent agreement, see Fig.~\ref{fig:dt_vs_Et} and the discussion further below. 

We also analyzed the errors in the time-evolution of the real-space correlations, where we obtain similar trends (high accuracy for $\Omega=20$, lower accuracy in the low frequency case).
However, at waiting time $t=0$ in all cases the errors are smaller than the signal size at small times, and for $\Omega=3$ of comparable value with the value of the correlation functions, which is close to zero there.
This confirms the statement about fast loss of coherence in the low frequency case, as discussed in the main text.
For a more detailled discussion of the error in the correlation functions, see below.
To reach longer times we use the Lanczos time evolution method where the computation of one time step is achieved by projecting the time-evolution operator onto a Lanczos basis for which $\ket{\psi(t)}$ is the initial state of the iteration procedure. 
In this way, also for time-dependent Hamiltonians, we compute~\cite{Manmana2005}
\begin{equation}
e^{-i dt  \hat{H}(t)}|\psi(t)\rangle \approx \mathbf{V}_n(t) e^{-i dt  \mathbf{T}_n(t)} \mathbf{V}_n^{+}(t)|\psi(t)\rangle \, ,
\end{equation}
where the matrix $\mathbf{V}_n(t)$ contains the Lanczos vectors after $n$ iterations, and $\mathbf{T}_n(t)$ is the tridiagonal matrix representation of $H(t)$ in this basis.
As for the MPS calculation, for time-dependent Hamiltonians there are better approximations (see, e.g., Ref.~\onlinecite{ALVERMANN2011}); however, if the value of $dt$ is chosen small enough, we obtain a good accuracy also with this simpler approach. 
The Lanczos simulations were done using the QuSpin package, see Ref.~\onlinecite{Weinberg2019}.

\section{Window Function}\label{app_window_fucntion}

\begin{figure}[t]
    \centering
    \includegraphics[width=0.40\textwidth]{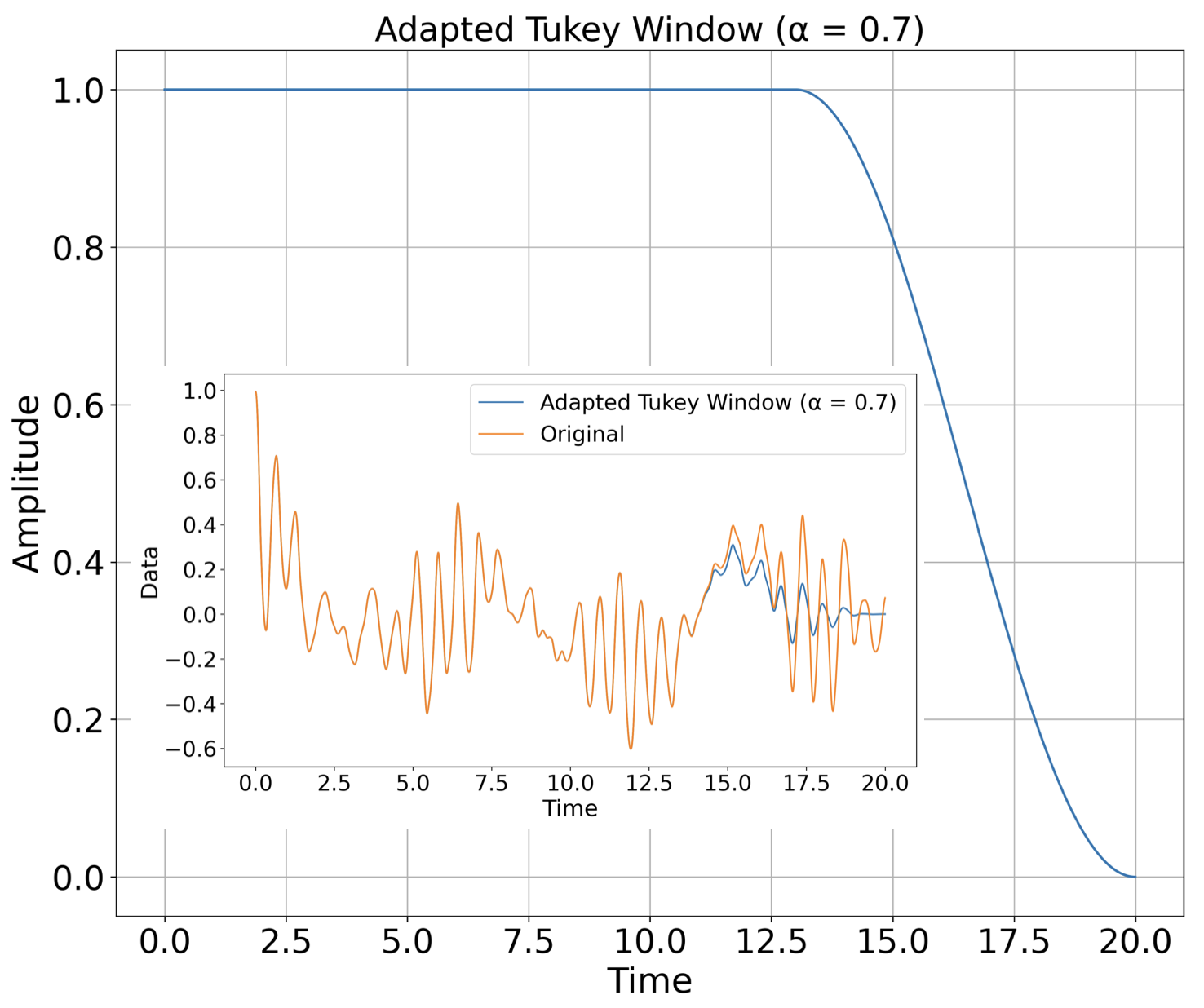}
    \caption{Damping of the data when applying the adapted Tukey window~\eqref{eq:tukeyEQ}.
    The main Figure shows the amplitude of the windowing function for $\alpha=0.7$ and the impact on the raw data is shown in the inset.
    }
    \label{fig:tukey}
\end{figure}
The retarded single-particle spectral function  is computed via
\begin{equation}
A_k^{\rm ret, \, eq}(\omega) = -\operatorname{Im} \frac{1}{\sqrt{2 \pi}} \int_{-\infty}^{\infty} \mathrm{d} \tau  
e^{i \omega \tau}
W(\tau)
G_{k k}^{\mathrm{ret,eq}}(t, \tau) \, 
\label{eq:Aret_GS}
\end{equation}
where $W(\tau)$ is a windowing function.
Typically, we use an adapted Tukey-window~\cite{Bloomfield2000} with the parameter $\alpha=0.7$, where
\begin{equation}
W(\tau)=\frac{1}{2}\left[1-\cos \left(\frac{2 \pi \tau}{\alpha N}\right)\right], \,\,\mbox{for } 0 \leq n<\frac{\alpha N}{2} \, .
\label{eq:tukeyEQ}
\end{equation}
In Fig.~\ref{fig:tukey}, we show an example of the window function in Eq.~\eqref{eq:tukeyEQ} for $\alpha = 0.7$.
We find that with this choice even weak FBs can be resolved in the spectral functions shown in the main text. 

\section{$A_k^{\mathrm{ret}}(t, \omega)$ for $V/t_h=1.5$}\label{app_v1p5_LL_wt}
\begin{figure*}[t]
    \centering
    \includegraphics[width=0.7\textwidth]{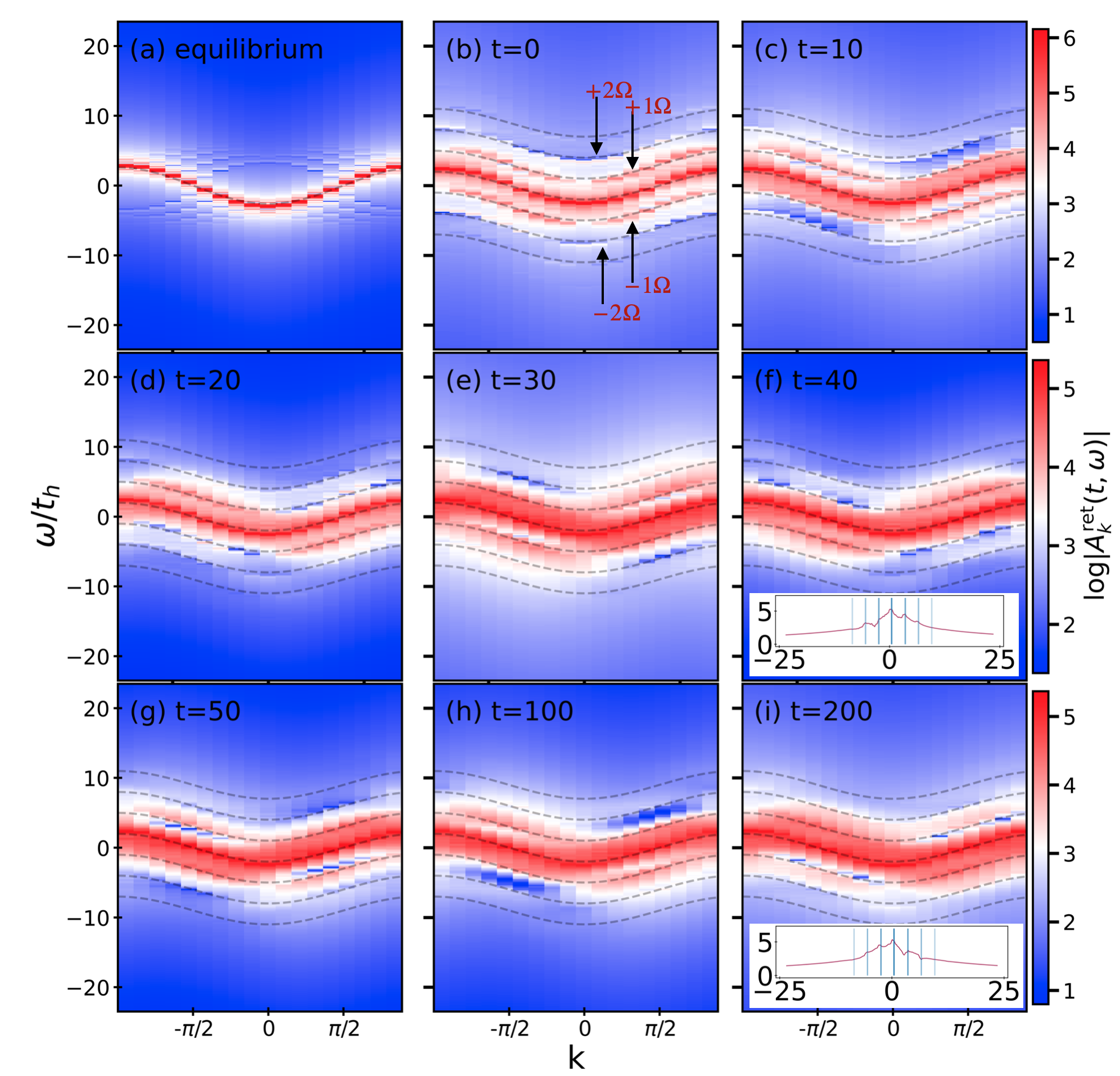} 
    \caption{Retarded time-dependent single-particle spectral function $A^{\rm ret}_k(t,\omega)$  for   $L=18$, $V/t_h = 1.5$, and PBC at half filling obtained with Lanczos time evolution. (a) shows the equilibrium case and (b) to (i) show different waiting times $t$ indicated on the plots for driving parameters $A=1.0$ and $\Omega/t_h=3.0$.
    Dashed lines are guides to the eye and depict a cosine-like band $\epsilon (k) = -t_{LL}\cos(k)$ and its expected Floquet replica for $t_{LL}=2.6$ and for the driven case $t_{eff}= t_{LL} J_0(A_0)$, with $J_0$ the 0th Bessel function.
    Arrows in (b) show the locations of the first two ($\pm n \Omega$) side bands.   
    Insets in (f) and (i) show the cross-section plots for $k=\pi/2$. 
    }
    \label{fig:Low_freq_sf_LL}
\end{figure*}

\begin{figure}[t]
    \centering
    \includegraphics[width=0.95\linewidth]{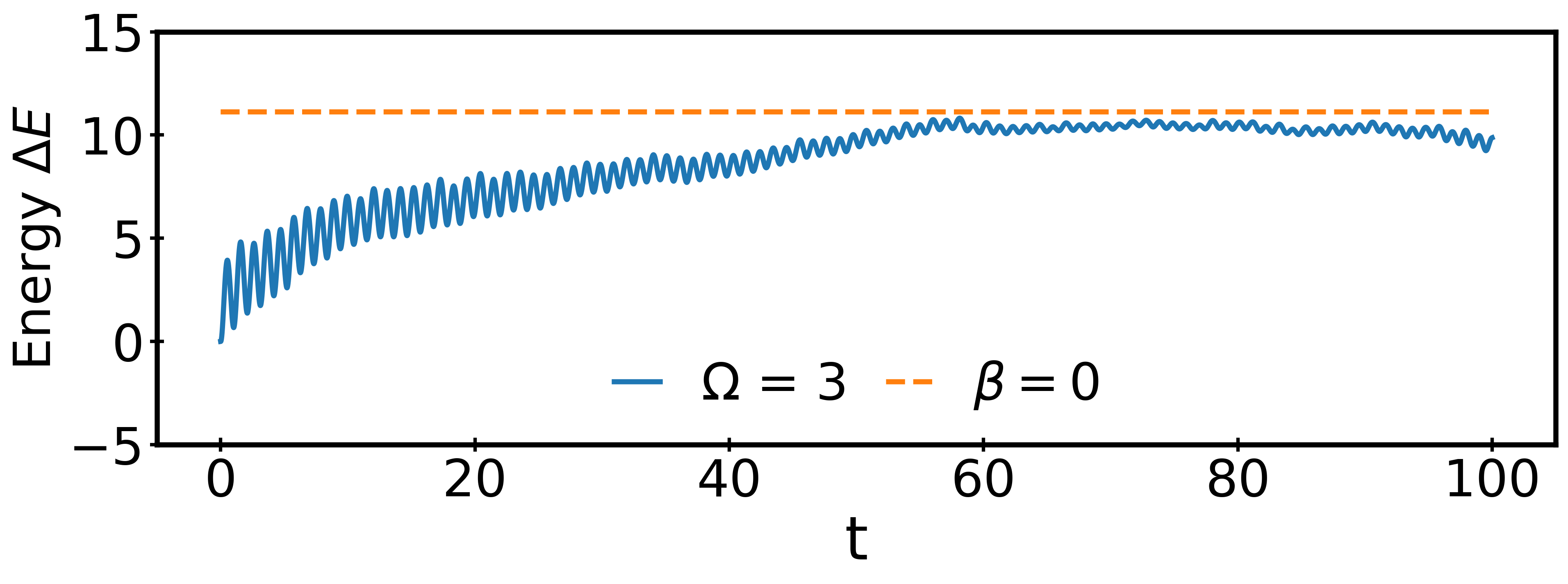}
    \caption{ED results for the energy gain compared to the initial state (ground state energy $E_{\rm GS}$), $\Delta E(t) = \langle H(t) \rangle - E_{\rm GS}$ for the driven case with $\Omega =3.0$ for $L=14$ and $V/t_h=1.5$ at half filling and with PBC.
    The dashed line indicates the result at infinite temperature.
    }
    \label{fig:Et_for_LL_low_freq}
\end{figure}
We show in Fig.~\ref{fig:Low_freq_sf_LL} the spectral function $A^{\rm ret}_k(t,\omega)$ for short to long waiting times, for $L=18$, $V/t_h = 1.5$, and PBC at half filling, obtained using Lanczos time evolution. 
In Fig.~\ref{fig:Low_freq_sf_LL}(b), we indicate the Floquet bands (FBs) with arrows to show the locations of the $\pm1 \Omega$ and $\pm2 \Omega$ sidebands. Dashed lines are included in all the figures as guides to the eye. 
We observe clear FBs for a driving frequency of $\Omega = 3.0$ and $A = 1.0$ for waiting times $t \lesssim 50$.
For the chosen driving frequency, overlapping FBs appear.
For waiting times $t > 50$, we observe a significant suppression of FBs and strong spectral intensities in the $0$th Floquet sector, indicating heating effects. 
This is further illustrated by the insets in Fig.~\ref{fig:Low_freq_sf_LL}, which show the intensity of the FBs at the indicated waiting times and
further supported by Fig.~\ref{fig:Et_for_LL_low_freq}, where we plot the energy gain as a function of time to monitor the heating.

\section{Real space correlations}\label{app_real_space_corr_errorestimate}
\begin{figure*}[t]
    \centering
    \includegraphics[width=0.7\textwidth]{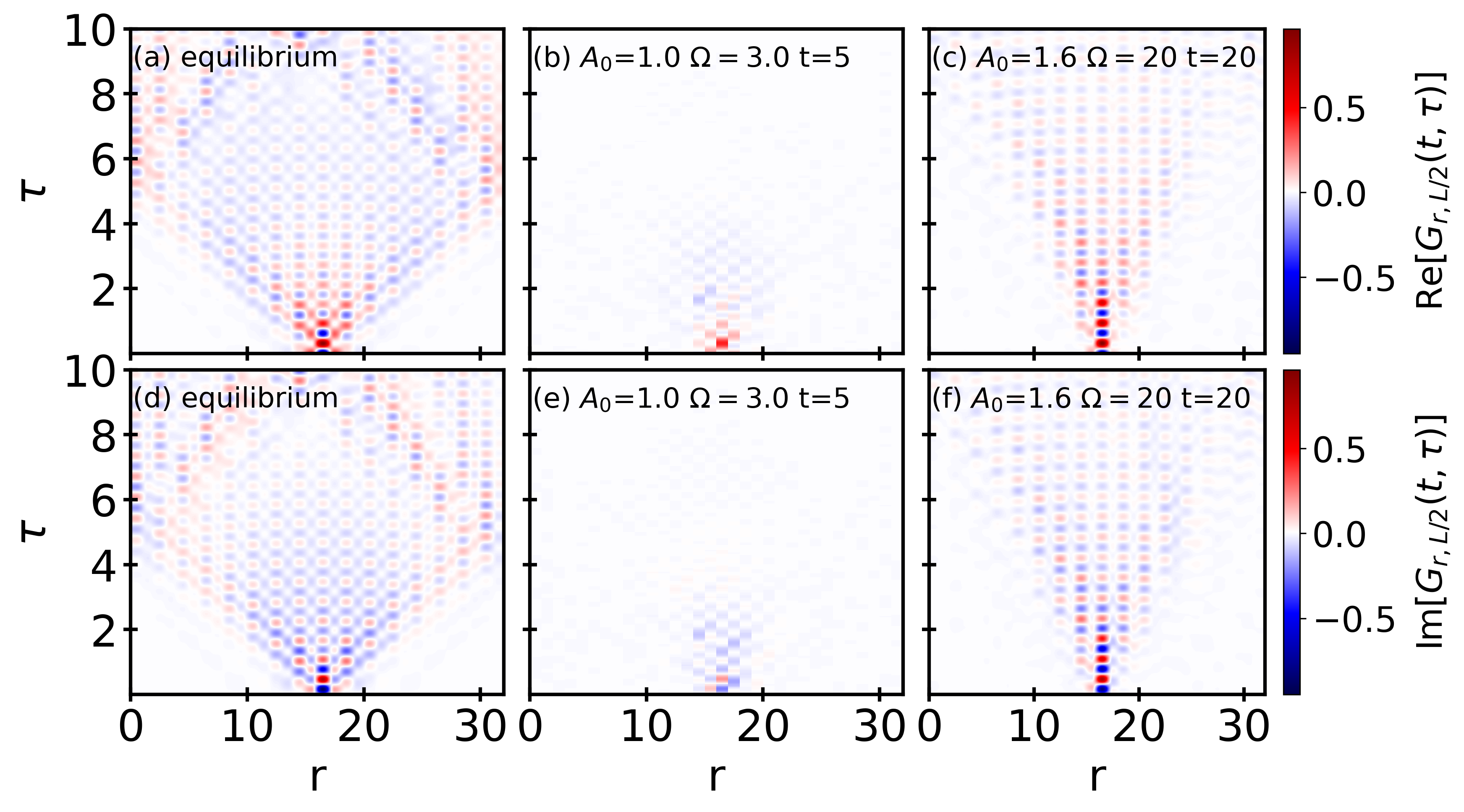}
    \caption{MPS results for $L=32$, $V/t_h=5.0$ at half filling and OBC for the real part (a)-(c) and for the imaginary part (e)-(f) of the real-space correlation function 
    $G_{r,L/2}(t,\tau) = \bra{\psi(t)} \left[ c^\dagger_r (\tau), c_{L/2} \right] \ket{\psi(t)}$
    at equilibrium and for the driven case for waiting times $t=5.0$ and $20.0$.}
    \label{fig:light_cone_t0}
\end{figure*}

We show the real and imaginary parts of the real-space correlation function 
\[
G_{r,L/2}(t,\tau) = \bra{\psi(t)} \left[ c^\dagger_r (\tau), c_{L/2} \right] \ket{\psi(t)}
\]
in Fig.~\ref{fig:light_cone_t0} for $L = 32$ with OBC for $V/t_h = 5.0$ at waiting time $t = 5.0$ for $\Omega =3.0$ and at waiting time $t = 20.0$ for $\Omega =20.0$.
The results are presented at equilibrium in panels (a) and (d), for $\Omega = 3.0$ in panels (b) and (e), and for $\Omega = 20.0$ in panels (c) and (f). 
In equilibrium, we observe a light-cone-like spread of correlations as the system is perturbed at the central site. 
This behavior changes for $\Omega = 3.0$, as also pointed out in the main text.
Already at $t = 0$, we observed a suppression of correlation propagation as from the main text Fig~\ref{fig:light_cone_energy_gain}; here, at $t=5.0$ we observe even stronger suppression.
For $\Omega = 20.0$, even at $t=20.0$, the light-cone-like spread is clearly visible, exhibiting coherence features also at later waiting times.

\begin{figure}[t]
    \centering
    \includegraphics[width=0.47\textwidth]{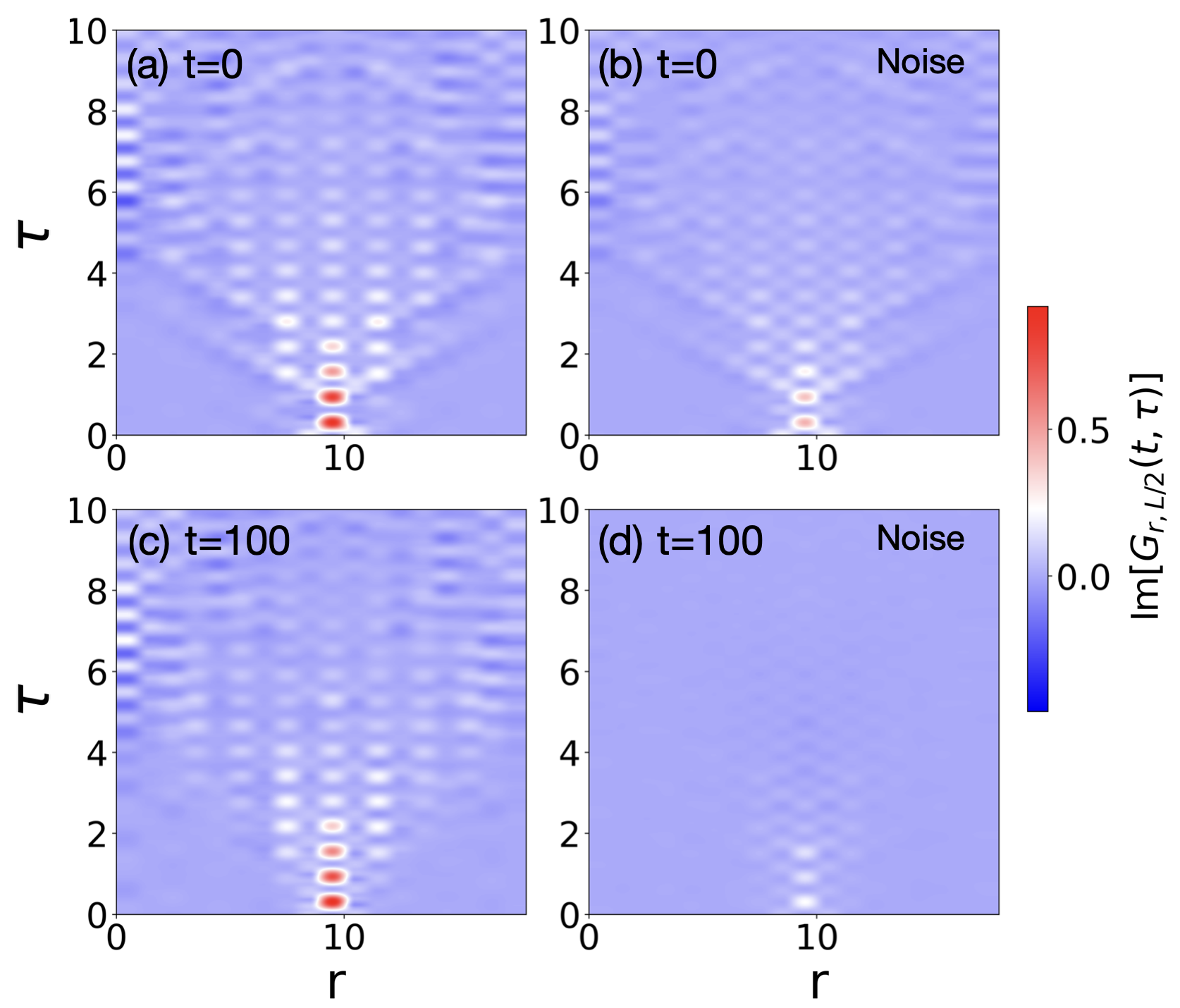}
    \caption{Lanczos results for $L=18$ and PBC for the imaginary part of the real-space correlation function $G_{r,L/2}(t,\tau) = \bra{\psi(t)} \left[ c^\dagger_r (\tau), c_{L/2} \right] \ket{\psi(t)}$, with $A_0=1.6$ and $\Omega = 20.0$. 
    (a) and (c) show results for waiting time $t=0$ and $t=100$ without noise.
    (b) and (d) show results for waiting time $t=0$ and $t=100$ with noise.
    }
    \label{fig:light_cone_pure_noise}
\end{figure}

Furthermore, we compare the imaginary part of $G_{r,L/2}(t,\tau)$ for cases with and without noise in Fig.~\ref{fig:light_cone_pure_noise} at longer waiting times.
In Fig.~\ref{fig:light_cone_pure_noise}, we show results for $L = 18$ (PBC), $V/t_h = 5.0$ at half filling.
Panels (a) and (b) correspond to the case without noise, while panels (c) and (d) represent the noise case at waiting times $t = 0$ and $t = 100$, respectively. 
As is evident from the noise case, at longer waiting times the system loses coherence, as seen in the vanishing imaginary part of the correlation $G_{r,L/2}(t,\tau)$ in Fig.~\ref{fig:light_cone_pure_noise}(d).

\begin{figure}[t]
    \centering
    \includegraphics[width=0.47\textwidth]{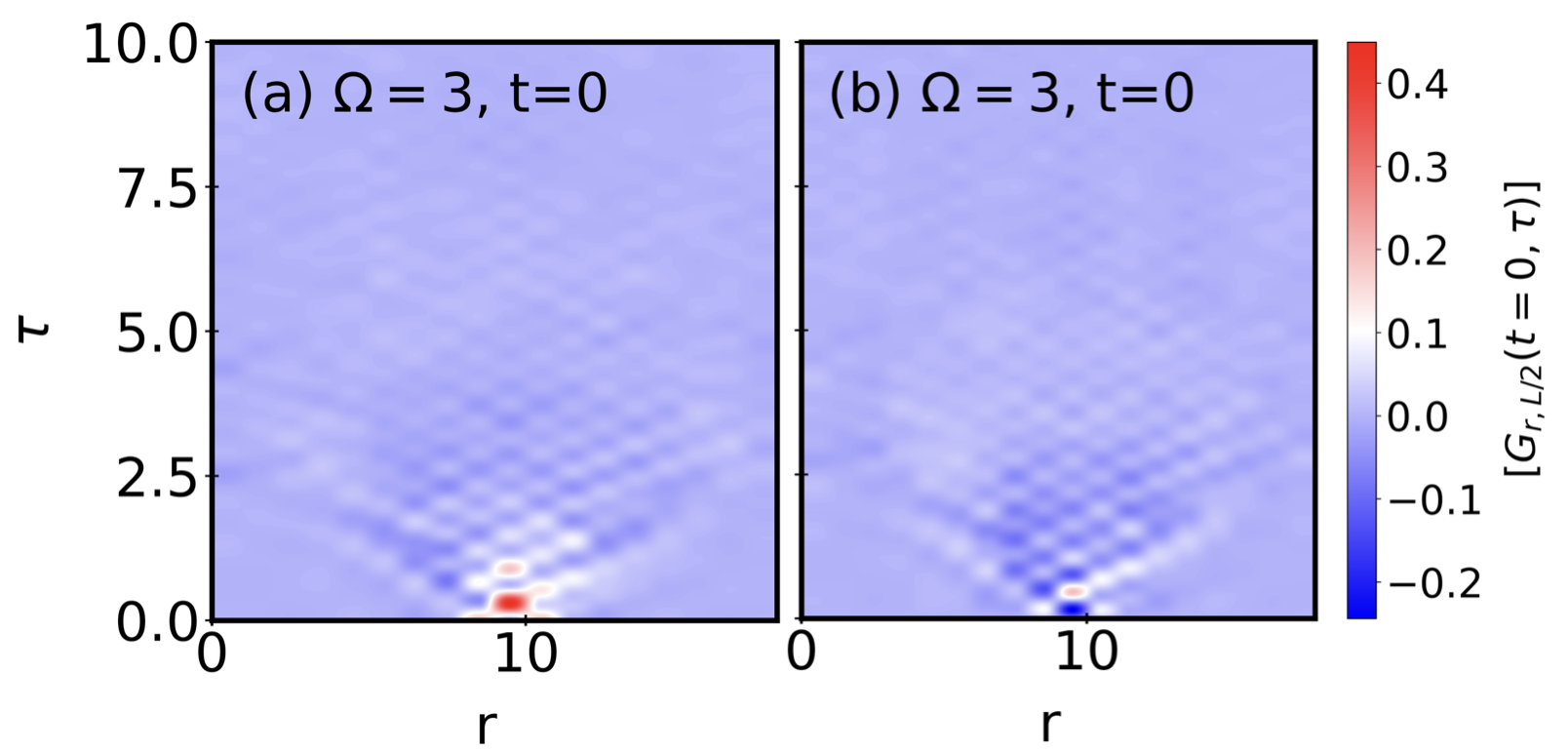}
    \caption{Lanczos results for $L=18$ (PBC) for the real-space correlation function $G_{r,L/2}(t,\tau) = \bra{\psi(t)} \left[ c^\dagger_r (\tau), c_{L/2} \right] \ket{\psi(t)}$ at waiting time $t=0$, with $A_0=1.0$ and $\Omega = 3.0$.
    (a) real part and (b) imaginary part. 
    }
    \label{fig:light_cone_Om3}
\end{figure}

In order to estimate the errors in the MPS results for the real time correlation function, we compared results for the coherent drive with $dt=0.04$ and $dt=0.005$.
For $\Omega=20$, we find in all cases that the difference between both calculations is substantially smaller than the magnitude of $G_{r,L/2}(t,\tau)$; the difference is (in the case of small values) of the order of a few percent.
For $\Omega=3$, the situation is more complicated, because at later times the value of $G_{r,L/2}(t,\tau)$ becomes very small, so that even small numerical errors lead to a larger relative deviation of the results. 
However, we find for waiting time $t=0$ that at short times the difference is $\lesssim 0.015$, while the maximal value of $G_{r,L/2}(t,\tau) \sim 0.16$.
At later times, the maximal value of $G_{r,L/2}(t,\tau) \sim 10^{-3}$, and the difference between both runs is of comparable magnitude.
Therefore, it is difficult to make quantitative statements at the longer times; however, it is clear that $G_{r,L/2}(t,\tau)$ decayed to a value much smaller than the one at the beginning of the time evolution, clearly indicating the loss of coherence in this case. 
This is further confirmed by Fig.~\ref{fig:light_cone_Om3}, in which Lanczos results for $\Omega=3$ for $L=18$ (PBC) clearly show the loss of coherence at waiting time $t=0$.
At later waiting times, the errors in the MPS calculations for $\Omega=20$ are similar to the ones at $t=0$, but larger for $\Omega=3$ due to the higher entanglement in the system; see Fig.~\ref{fig:light_cone_t0} for results at equilibrium and at waiting times $t=5$ and $t=20$.
However, the results are qualitatively very similar to the ones at waiting time $t=0$ shown in the main text (see also the discussion above).

\section{Energy $E(t)$ and momentum distribution function $\langle n_k \rangle (t)$}\label{app_E_nk_errorestimate} 

\begin{figure}[t]
    \centering
    \includegraphics[width=0.49\textwidth]{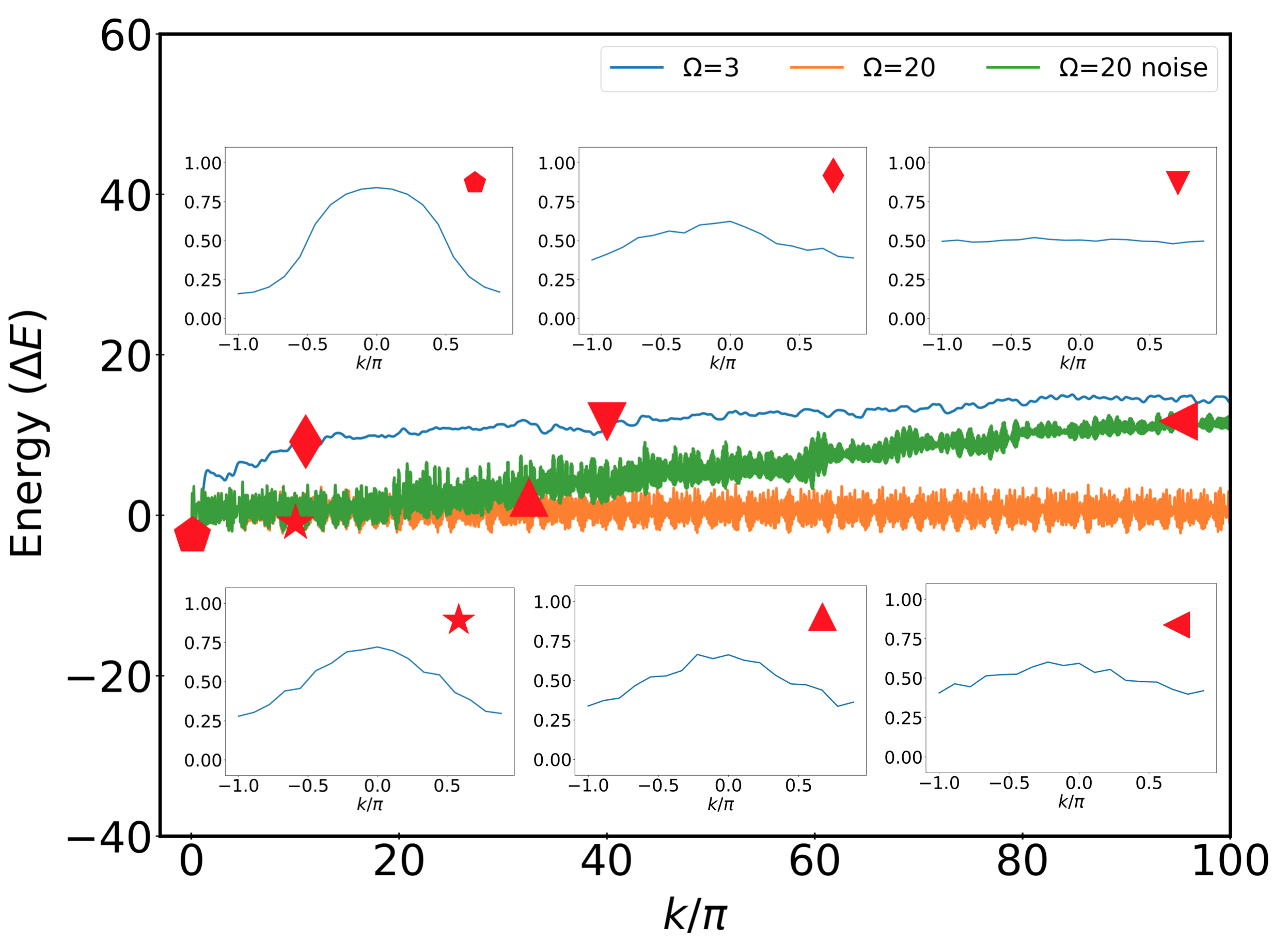}
    \caption{Time-dependent Lanczos results for systems with $L=18$ and PBC for the time evolution of the energy $E(t) = \langle H(t) \rangle$ (main figure) for the indicated different driven cases for $dt=0.005$, and $\langle n_k\rangle(t)$ (insets) for the indicated markers on the energy curves.
    }
    \label{fig:nk_energy_gain}
\end{figure}
\begin{figure}[t]
    \centering
    \includegraphics[width=0.97\linewidth]{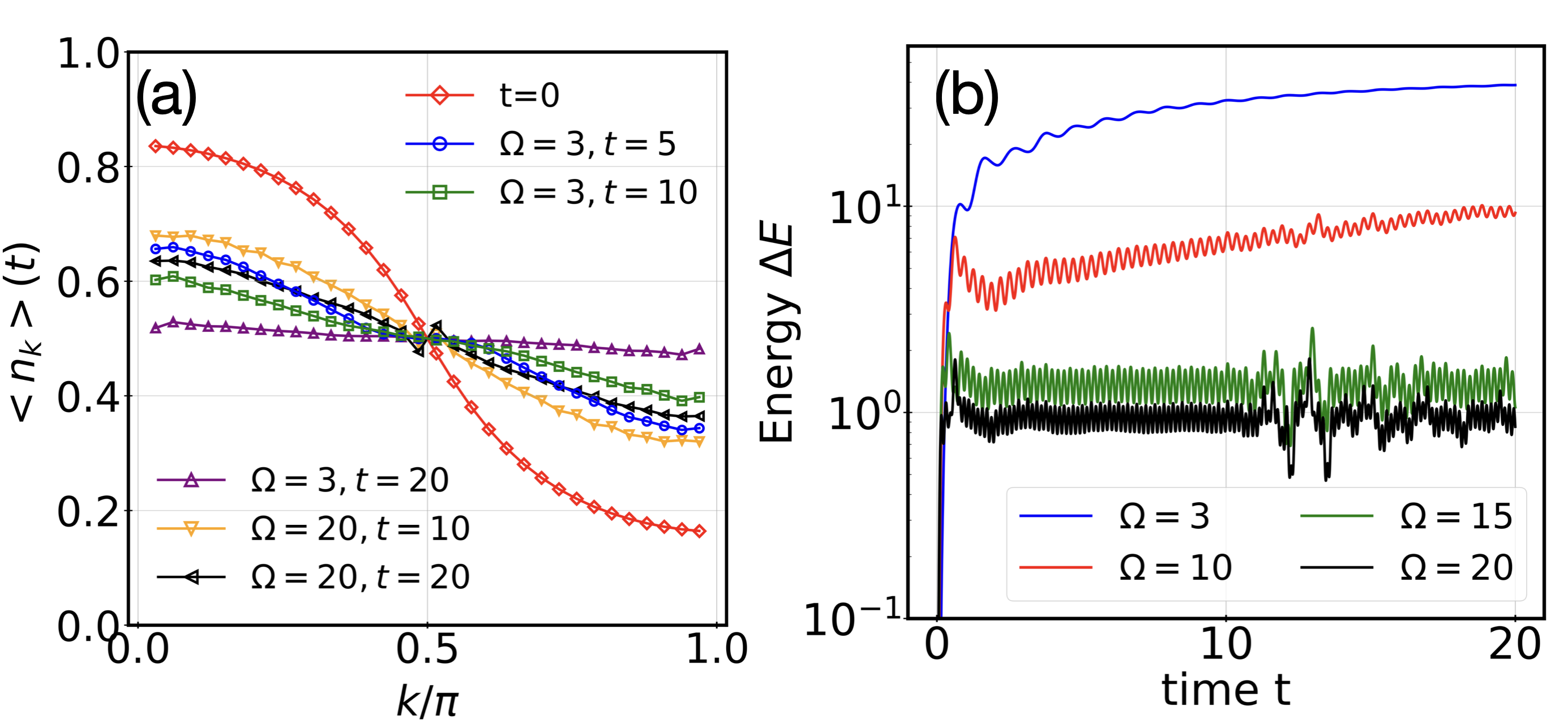}
    \caption{MPS results for $L=32$ and $V/t_h=5.0$ at half filling and with OBC: (a) momentum distribution function $\langle n_k \rangle(t)$ at different waiting times for low $\Omega/t_h=3.0$ and high frequency $\Omega/t_h=20.0$.
    (b) energy $\Delta E(t) = \langle H(t) \rangle - E_{\rm GS}$ as a function of time for the driven case with the different frequencies as indicated. 
    }
    \label{fig:QFI_Et_MPS}
\end{figure}
\begin{figure*}[t]
    \centering
    \includegraphics[width=0.7\textwidth]{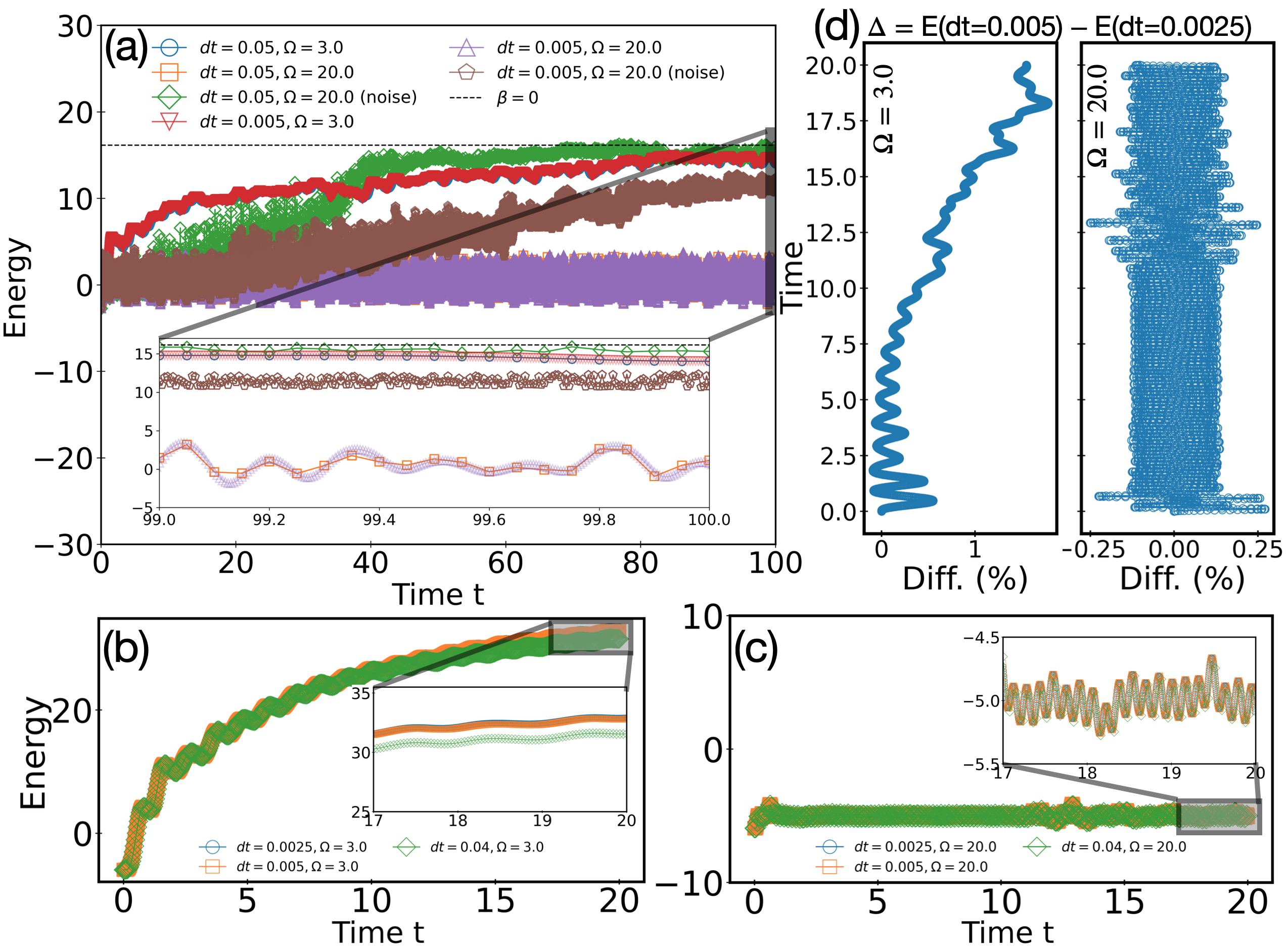}
    \caption{Energy $E(t) = \langle H(t) \rangle$ as a function of time for $V/t_h=5.0$. (a)
    Lanczos results for the time evolution of $E(t)$ 
    for the driven case for different values of $\Omega$ for $L=14$ (PBC) when using $dt=0.05$, or $dt=0.005$, respectively.
    The inset is a zoom into the interval $t = [99,100.0]$.
    MPS results for $E(t)$ in (b) for the driven case of $\Omega=3.0$ and in (c) for the driven case of $\Omega=20.0$ using $dt=0.04$, $dt=0.005$, and $dt=0.0025$.
    The insets in (b) and (c) zoom into the interval $t = [17,20.0]$.
    (d) relative difference for the energy for the choice of $dt=0.005$ and $dt=0.0025$, respectively.  
    }
    \label{fig:dt_vs_Et}
\end{figure*}

In Fig.~\ref{fig:nk_energy_gain}, we present the expectation value $E(t) = \langle H(t) \rangle$ for the driven system with and without noise, comparing the energy gain by plotting
the energy for $L=18$, periodic boundary conditions (PBC), and $V/t_h=5$ at half filling.
We consider driving frequencies $\Omega=3$, and $20$
(both with and without noise) and illustrate the resulting energy gain over time.
The insets of Fig.~\ref{fig:nk_energy_gain} display the momentum distribution function $\langle n_k \rangle(t) = \langle c_k^\dagger c_k^{\phantom{\dagger}} \rangle(t)$, with $c_k^{(\dagger)} = 1/\sqrt{L}\sum_{r=1}^L e^{(-)ik\cdot r} c_r^{(\dagger)}$
at the times indicated by the markers.
Note that, as we also mention in the main text, $\langle n_k \rangle(t)$ can be related to the Quantum Fisher Information (QFI) $F_Q(k,t)$ for the chosen witness operator $\hat{O}$ $= \sum_i (a_i c_i^{\dagger} + a_i^* c_i)$.
We use $F_Q(k,t) = \int d \omega A(\omega,k,t)$, at temperature $T=0$ where $A(\omega,k,t) = \text{Im} \int dt' e^{i\omega t'} i\langle[\hat{O}(k,t'),\hat{O}(k,0)]\rangle$, and 
integrating over $\omega$ results in the momentum distribution function $\langle n_k\rangle(t)$ \cite{Hauke2016,Baykusheva2023,Assaad2024}.
In Fig.~\ref{fig:nk_energy_gain}, the insets show 
loss of coherence as the system gains energy and the effective temperature increases over time. 
In the following, we use matrix product states (MPS)~\cite{Schollwck2011,Paeckel2019} to compute 
systems with open boundary conditions (OBC) and pinning field $\mu = V/t_h$ (see main text) for $L=32$.
We employ Lanczos time evolution~\cite{Manmana2005} for systems with $L=18$ and PBC to achieve longer simulation times than accessible with MPS.
In Fig.~\ref{fig:QFI_Et_MPS} (a) we show MPS results for $\langle n_k\rangle(t)$ for $\Omega=3.0$ and $20.0$ at various waiting times obtained with OBC (note that in the main text and in Fig.~\ref{fig:nk_energy_gain} we show results obtained with PBC; in Fig.~\ref{fig:QFI_Et_MPS}(a), we have only positive $k$-values due to the sine-transform).
For $\Omega=3.0$, we observe a strong flattening of the momentum distribution function, consistent with the insets of Fig.~\ref{fig:nk_energy_gain}.
Further, in Fig.~\ref{fig:QFI_Et_MPS} (b) we check $E(t)$ for $L=32$ using MPS for different values of $\Omega$, which shows that the energy gain decreases with increase of the driving frequency.
For the case of incoherent noise in the driving frequency, we consider a driving field with vector potential $A_V(t) = A_0 \sin((\Omega \pm \Omega_{\text{noise}})t),$ 
where the frequency is given by $\Omega = 20 \pm \Omega_{\text{noise}}, \quad \Omega_{\text{noise}} \in [0,0.1].$ 
We discretize the time as $ t = n \, dt $ with $ n \in \mathbb{N}_0 $ and apply  $A_V(n \, dt) = A_0 \sin((\Omega \pm \Omega_{\text{noise}}) n \, dt)$.
Therefore, at each time step of the simulation, we have a different value of $\Omega_{\rm noise}$. Hence, the frequency, with which the value of $\Omega_{\rm noise}$ is changed, depends on the value of $dt$, so that we can expect different time-dependent behavior of the observables for different values of $dt$ in the presence of noise. 
In Fig.~\ref{fig:dt_vs_Et}(a), we show the energy gain $ E(t)$  for discrete time steps $dt = 0.05$ and $dt = 0.00$5 for $L=14$ PBC.  
We observe that the choice of $dt$ affects the energy gain over time in the presence of noise, as mentioned above.

Furthermore, we check the dependence of our results on the size of $dt$ also without noise in the driving.
This is important, since we are treating a time-dependent Hamiltonian, and in addition to the usual MPS errors, we have an additional source of errors due to the discretization of $t$ in $H(t)$~\cite{ALVERMANN2011}. 
In Fig.~\ref{fig:dt_vs_Et}(a) we show Lanczos results for the different cases treated in the main text, but with values of $dt=0.05$ or $dt=0.005$, respectively, in order to estimate the error due to the discretization of $H(t)$.
As can be seen, in all cases the values of the energy $E(t)$ lie on top of each other, so that this discretization error seems to play a minor role.
In Fig.~\ref{fig:dt_vs_Et}(b) we turn to  MPS results for $L=32$ (OBC) for $\Omega=3.0$ and in Fig.~\ref{fig:dt_vs_Et}(c) for  $\Omega=20.0$, with discrete time steps $dt = 0.04$, $dt = 0.005$, and $dt = 0.0025$.
In these plots, the MPS results all carry the accumulated error due to (i) discretization of time in $H(t)$; (ii) error of the TDVP-ansatz due to the finite value of $dt$~\cite{Paeckel2019}; (iii) accumulated MPS-errors due to the finite bond dimension~\cite{Schollwck2011}. 
Hence, by comparing the time evolution of observables with different values of $dt$, we get an estimate for the accumulated errors of our numerical calculations, which can be used to estimate the overall accuracy of our results. 
As can be seen, only the values with the largest chosen value of $dt=0.04$ for $\Omega = 3$ show a clear discrepancy at later times, indicating that the results obtained with $dt=0.005$ shown in the main text have a high numerical accuracy.
In Fig.~\ref{fig:dt_vs_Et}(d) this is further quantified by showing the relative difference (in percentage) for the energy $E(t)$ with the choice of $dt = 0.005$, and $dt = 0.0025$.
For the coherent drive case, 
the maximal difference of these results at the end of the time evolutions shown is $\approx 2\%$ for $\Omega=3$ and $\approx 0.25\%$ for $\Omega=20$.
The smaller error in the high frequency case can be understood by having less heating and less growth of entanglement in this case (this can be seen in the growth of the discarded weight with time, which is much faster in the low frequency case, see further below). 
We, therefore, stick to the discrete time steps of $dt=0.005$ for all our calculations presented in the main text. 
\section{Further results at fixed $k$}\label{app_crosssectional_more_k}
\begin{figure}[t]
    \centering
    \includegraphics[width=0.45\textwidth]{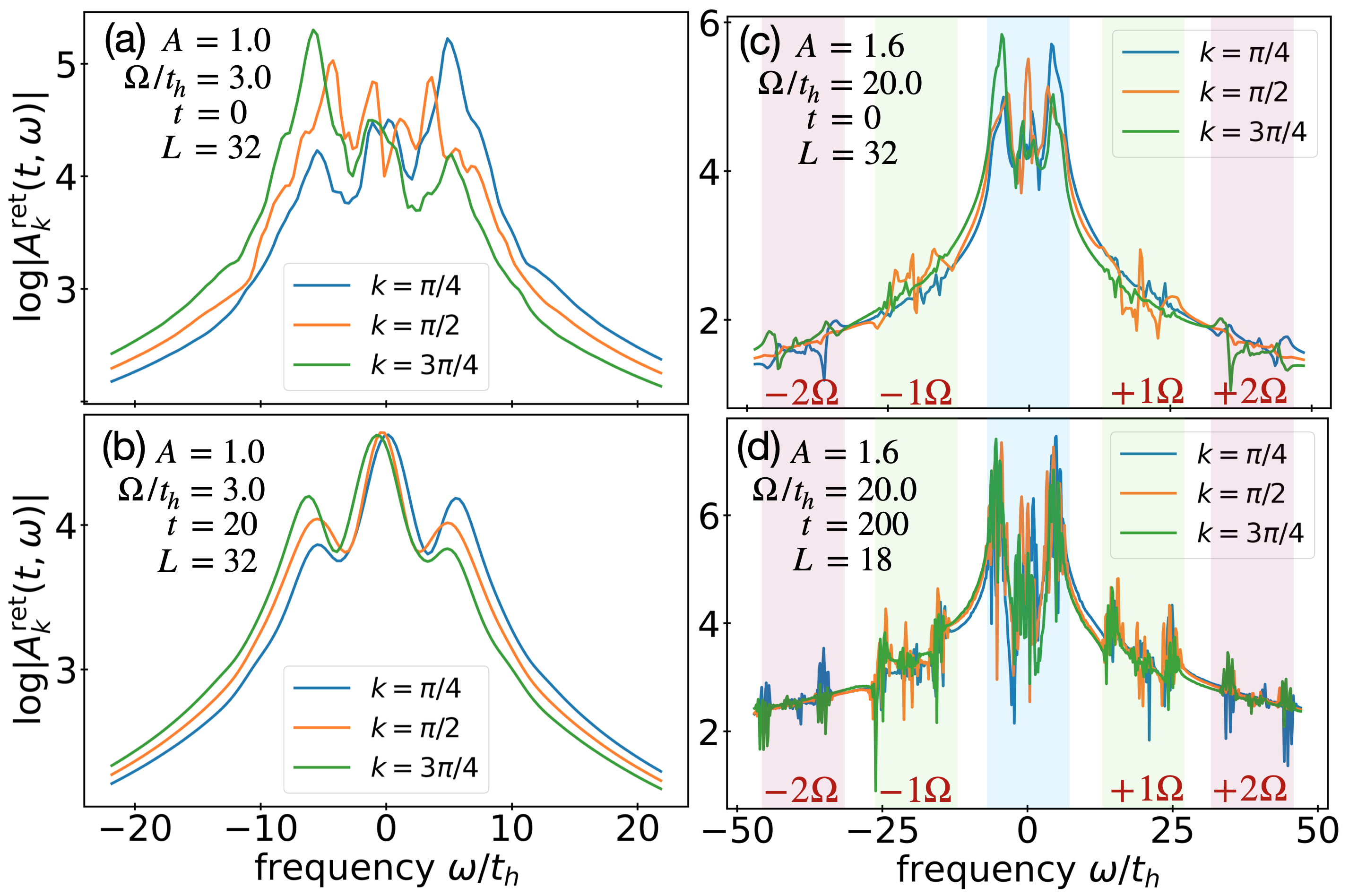}
    \caption{Results for $A_k^{\rm ret}(t,\omega)$ for $V/t_h=5.0$ at half filling at different values of $k$. (a) and (b) show the results for the driven case with $L=32$, $A=1.0$, and $\Omega=3.0$ at waiting time $t=0$ and $t=20$, respectively. (c) shows the plots for the driven case with $L=32$, $A=1.6$, and $\Omega=20.0$ for waiting time $t=0$. (d) shows the plots for the driven case with $L=18$, $A=1.6$, and $\Omega=20.0$ for waiting time $t=200$. The shaded regions show the location of the FBs as a guide to the eye at $\pm n\Omega$ in the pannels (c) and (d).    }
    \label{fig:cross_sectional_plots_low_high_freq_waightingtime}
\end{figure}

In Fig.~\ref{fig:cross_sectional_plots_low_high_freq_waightingtime} we show further results of $A^{\rm ret}_k(t,\omega)$ when keeping $k$ fixed and at different values of $k$ for $V/t_h=5.0$, see Figs.~\ref{fig:Low_freq_Eq_and_driven} and~\ref{fig:High_freq_drive} in the main text. 
In Fig.~\ref{fig:cross_sectional_plots_low_high_freq_waightingtime} (a) and (b) we show the results for $L=32$ (MPS), $A=1.0$, and $\Omega=3.0$ for waiting time $t=0$ and $t=20$. 
As we have mentioned in the main text, we observe high intensity in the gap region at finite waiting time $t=20$ across different momenta as shown in the plots.
Fig.~\ref{fig:cross_sectional_plots_low_high_freq_waightingtime} (c) shows the results for $L=32$ (MPS), $A=1.6$, and $\Omega=20.0$ for waiting time $t=0$.
We observe the intensities of the sidebands at $\pm n\Omega$ across different momenta shown in the plot.
Furthermore, we also show results at longer waiting time, $t=200$, in (d) for $L=18$, $A=1.6$, and $\Omega=20.0$. We again observe stable FBs at the different values of $k$ shown here. We highlight by  shaded regions as guides to the eye the location of the FBs in the (c) and (d) subplots. 

\newpage
\bibliography{ms_supp_bib.bib}

\begin{thebibliography}{101}%
\makeatletter
\providecommand \@ifxundefined [1]{%
 \@ifx{#1\undefined}
}%
\providecommand \@ifnum [1]{%
 \ifnum #1\expandafter \@firstoftwo
 \else \expandafter \@secondoftwo
 \fi
}%
\providecommand \@ifx [1]{%
 \ifx #1\expandafter \@firstoftwo
 \else \expandafter \@secondoftwo
 \fi
}%
\providecommand \natexlab [1]{#1}%
\providecommand \enquote  [1]{``#1''}%
\providecommand \bibnamefont  [1]{#1}%
\providecommand \bibfnamefont [1]{#1}%
\providecommand \citenamefont [1]{#1}%
\providecommand \href@noop [0]{\@secondoftwo}%
\providecommand \href [0]{\begingroup \@sanitize@url \@href}%
\providecommand \@href[1]{\@@startlink{#1}\@@href}%
\providecommand \@@href[1]{\endgroup#1\@@endlink}%
\providecommand \@sanitize@url [0]{\catcode `\\12\catcode `\$12\catcode
  `\&12\catcode `\#12\catcode `\^12\catcode `\_12\catcode `\%12\relax}%
\providecommand \@@startlink[1]{}%
\providecommand \@@endlink[0]{}%
\providecommand \url  [0]{\begingroup\@sanitize@url \@url }%
\providecommand \@url [1]{\endgroup\@href {#1}{\urlprefix }}%
\providecommand \urlprefix  [0]{URL }%
\providecommand \Eprint [0]{\href }%
\providecommand \doibase [0]{https://doi.org/}%
\providecommand \selectlanguage [0]{\@gobble}%
\providecommand \bibinfo  [0]{\@secondoftwo}%
\providecommand \bibfield  [0]{\@secondoftwo}%
\providecommand \translation [1]{[#1]}%
\providecommand \BibitemOpen [0]{}%
\providecommand \bibitemStop [0]{}%
\providecommand \bibitemNoStop [0]{.\EOS\space}%
\providecommand \EOS [0]{\spacefactor3000\relax}%
\providecommand \BibitemShut  [1]{\csname bibitem#1\endcsname}%
\let\auto@bib@innerbib\@empty
\bibitem [{\citenamefont {Basov}\ \emph {et~al.}(2017)\citenamefont {Basov},
  \citenamefont {Averitt},\ and\ \citenamefont {Hsieh}}]{Basov2017}%
  \BibitemOpen
  \bibfield  {author} {\bibinfo {author} {\bibfnamefont {D.~N.}\ \bibnamefont
  {Basov}}, \bibinfo {author} {\bibfnamefont {R.~D.}\ \bibnamefont {Averitt}},\
  and\ \bibinfo {author} {\bibfnamefont {D.}~\bibnamefont {Hsieh}},\ }\bibfield
   {title} {\bibinfo {title} {Towards properties on demand in quantum
  materials},\ }\href {https://doi.org/10.1038/nmat5017} {\bibfield  {journal}
  {\bibinfo  {journal} {Nature Materials}\ }\textbf {\bibinfo {volume} {16}},\
  \bibinfo {pages} {1077–1088} (\bibinfo {year} {2017})}\BibitemShut
  {NoStop}%
\bibitem [{\citenamefont {Oka}\ and\ \citenamefont
  {Kitamura}(2019)}]{OkaRev2019}%
  \BibitemOpen
  \bibfield  {author} {\bibinfo {author} {\bibfnamefont {T.}~\bibnamefont
  {Oka}}\ and\ \bibinfo {author} {\bibfnamefont {S.}~\bibnamefont {Kitamura}},\
  }\bibfield  {title} {\bibinfo {title} {Floquet engineering of quantum
  materials},\ }\href
  {https://doi.org/10.1146/annurev-conmatphys-031218-013423} {\bibfield
  {journal} {\bibinfo  {journal} {Annual Review of Condensed Matter Physics}\
  }\textbf {\bibinfo {volume} {10}},\ \bibinfo {pages} {387–408} (\bibinfo
  {year} {2019})}\BibitemShut {NoStop}%
\bibitem [{\citenamefont {de~la Torre}\ \emph {et~al.}(2021)\citenamefont
  {de~la Torre}, \citenamefont {Kennes}, \citenamefont {Claassen},
  \citenamefont {Gerber}, \citenamefont {McIver},\ and\ \citenamefont
  {Sentef}}]{delaTorre2021}%
  \BibitemOpen
  \bibfield  {author} {\bibinfo {author} {\bibfnamefont {A.}~\bibnamefont
  {de~la Torre}}, \bibinfo {author} {\bibfnamefont {D.~M.}\ \bibnamefont
  {Kennes}}, \bibinfo {author} {\bibfnamefont {M.}~\bibnamefont {Claassen}},
  \bibinfo {author} {\bibfnamefont {S.}~\bibnamefont {Gerber}}, \bibinfo
  {author} {\bibfnamefont {J.~W.}\ \bibnamefont {McIver}},\ and\ \bibinfo
  {author} {\bibfnamefont {M.~A.}\ \bibnamefont {Sentef}},\ }\bibfield  {title}
  {\bibinfo {title} {Colloquium: Nonthermal pathways to ultrafast control in
  quantum materials},\ }\bibfield  {journal} {\bibinfo  {journal} {Reviews of
  Modern Physics}\ }\textbf {\bibinfo {volume} {93}},\ \href
  {https://doi.org/10.1103/revmodphys.93.041002} {10.1103/revmodphys.93.041002}
  (\bibinfo {year} {2021})\BibitemShut {NoStop}%
\bibitem [{\citenamefont {Bao}\ \emph {et~al.}(2021)\citenamefont {Bao},
  \citenamefont {Tang}, \citenamefont {Sun},\ and\ \citenamefont
  {Zhou}}]{Bao2021}%
  \BibitemOpen
  \bibfield  {author} {\bibinfo {author} {\bibfnamefont {C.}~\bibnamefont
  {Bao}}, \bibinfo {author} {\bibfnamefont {P.}~\bibnamefont {Tang}}, \bibinfo
  {author} {\bibfnamefont {D.}~\bibnamefont {Sun}},\ and\ \bibinfo {author}
  {\bibfnamefont {S.}~\bibnamefont {Zhou}},\ }\bibfield  {title} {\bibinfo
  {title} {Light-induced emergent phenomena in 2d materials and topological
  materials},\ }\href {https://doi.org/10.1038/s42254-021-00388-1} {\bibfield
  {journal} {\bibinfo  {journal} {Nature Reviews Physics}\ }\textbf {\bibinfo
  {volume} {4}},\ \bibinfo {pages} {33–48} (\bibinfo {year}
  {2021})}\BibitemShut {NoStop}%
\bibitem [{\citenamefont {Valmispild}\ \emph {et~al.}(2024)\citenamefont
  {Valmispild}, \citenamefont {Gorelov}, \citenamefont {Eckstein},
  \citenamefont {Lichtenstein}, \citenamefont {Aoki}, \citenamefont
  {Katsnelson}, \citenamefont {Ivanov},\ and\ \citenamefont
  {Smirnova}}]{ValmispildNatPhotonics2024}%
  \BibitemOpen
  \bibfield  {author} {\bibinfo {author} {\bibfnamefont {V.~N.}\ \bibnamefont
  {Valmispild}}, \bibinfo {author} {\bibfnamefont {E.}~\bibnamefont {Gorelov}},
  \bibinfo {author} {\bibfnamefont {M.}~\bibnamefont {Eckstein}}, \bibinfo
  {author} {\bibfnamefont {A.~I.}\ \bibnamefont {Lichtenstein}}, \bibinfo
  {author} {\bibfnamefont {H.}~\bibnamefont {Aoki}}, \bibinfo {author}
  {\bibfnamefont {M.~I.}\ \bibnamefont {Katsnelson}}, \bibinfo {author}
  {\bibfnamefont {M.~Y.}\ \bibnamefont {Ivanov}},\ and\ \bibinfo {author}
  {\bibfnamefont {O.}~\bibnamefont {Smirnova}},\ }\bibfield  {title} {\bibinfo
  {title} {Sub-cycle multidimensional spectroscopy of strongly correlated
  materials},\ }\href {https://doi.org/10.1038/s41566-023-01371-1} {\bibfield
  {journal} {\bibinfo  {journal} {Nature Photonics}\ }\textbf {\bibinfo
  {volume} {18}},\ \bibinfo {pages} {432} (\bibinfo {year} {2024})}\BibitemShut
  {NoStop}%
\bibitem [{\citenamefont {Li}\ and\ \citenamefont
  {Eckstein}(2020)}]{PRL_Li2020}%
  \BibitemOpen
  \bibfield  {author} {\bibinfo {author} {\bibfnamefont {J.}~\bibnamefont
  {Li}}\ and\ \bibinfo {author} {\bibfnamefont {M.}~\bibnamefont {Eckstein}},\
  }\bibfield  {title} {\bibinfo {title} {Manipulating intertwined orders in
  solids with quantum light},\ }\href
  {https://doi.org/10.1103/PhysRevLett.125.217402} {\bibfield  {journal}
  {\bibinfo  {journal} {Phys. Rev. Lett.}\ }\textbf {\bibinfo {volume} {125}},\
  \bibinfo {pages} {217402} (\bibinfo {year} {2020})}\BibitemShut {NoStop}%
\bibitem [{\citenamefont {Murakami}\ \emph {et~al.}(2023)\citenamefont
  {Murakami}, \citenamefont {Golež}, \citenamefont {Eckstein},\ and\
  \citenamefont {Werner}}]{Reviewmurakami2023}%
  \BibitemOpen
  \bibfield  {author} {\bibinfo {author} {\bibfnamefont {Y.}~\bibnamefont
  {Murakami}}, \bibinfo {author} {\bibfnamefont {D.}~\bibnamefont {Golež}},
  \bibinfo {author} {\bibfnamefont {M.}~\bibnamefont {Eckstein}},\ and\
  \bibinfo {author} {\bibfnamefont {P.}~\bibnamefont {Werner}},\ }\href
  {https://arxiv.org/abs/2310.05201} {\bibinfo {title} {Photo-induced
  nonequilibrium states in mott insulators}} (\bibinfo {year} {2023}),\ \Eprint
  {https://arxiv.org/abs/2310.05201} {arXiv:2310.05201 [cond-mat.str-el]}
  \BibitemShut {NoStop}%
\bibitem [{\citenamefont {Grushin}\ \emph {et~al.}(2014)\citenamefont
  {Grushin}, \citenamefont {G\'omez-Le\'on},\ and\ \citenamefont
  {Neupert}}]{Grushin2014}%
  \BibitemOpen
  \bibfield  {author} {\bibinfo {author} {\bibfnamefont {A.~G.}\ \bibnamefont
  {Grushin}}, \bibinfo {author} {\bibfnamefont {A.}~\bibnamefont
  {G\'omez-Le\'on}},\ and\ \bibinfo {author} {\bibfnamefont {T.}~\bibnamefont
  {Neupert}},\ }\bibfield  {title} {\bibinfo {title} {Floquet fractional chern
  insulators},\ }\bibfield  {journal} {\bibinfo  {journal} {Physical Review
  Letters}\ }\textbf {\bibinfo {volume} {112}},\ \href
  {https://doi.org/10.1103/physrevlett.112.156801}
  {10.1103/physrevlett.112.156801} (\bibinfo {year} {2014})\BibitemShut
  {NoStop}%
\bibitem [{\citenamefont {Jotzu}\ \emph {et~al.}(2014)\citenamefont {Jotzu},
  \citenamefont {Messer}, \citenamefont {Desbuquois}, \citenamefont {Lebrat},
  \citenamefont {Uehlinger}, \citenamefont {Greif},\ and\ \citenamefont
  {Esslinger}}]{Jotzu2014}%
  \BibitemOpen
  \bibfield  {author} {\bibinfo {author} {\bibfnamefont {G.}~\bibnamefont
  {Jotzu}}, \bibinfo {author} {\bibfnamefont {M.}~\bibnamefont {Messer}},
  \bibinfo {author} {\bibfnamefont {R.}~\bibnamefont {Desbuquois}}, \bibinfo
  {author} {\bibfnamefont {M.}~\bibnamefont {Lebrat}}, \bibinfo {author}
  {\bibfnamefont {T.}~\bibnamefont {Uehlinger}}, \bibinfo {author}
  {\bibfnamefont {D.}~\bibnamefont {Greif}},\ and\ \bibinfo {author}
  {\bibfnamefont {T.}~\bibnamefont {Esslinger}},\ }\bibfield  {title} {\bibinfo
  {title} {Experimental realization of the topological haldane model with
  ultracold fermions},\ }\href {https://doi.org/10.1038/nature13915} {\bibfield
   {journal} {\bibinfo  {journal} {Nature}\ }\textbf {\bibinfo {volume}
  {515}},\ \bibinfo {pages} {237–240} (\bibinfo {year} {2014})}\BibitemShut
  {NoStop}%
\bibitem [{\citenamefont {Mikami}\ \emph {et~al.}(2016)\citenamefont {Mikami},
  \citenamefont {Kitamura}, \citenamefont {Yasuda}, \citenamefont {Tsuji},
  \citenamefont {Oka},\ and\ \citenamefont {Aoki}}]{Mikami2016}%
  \BibitemOpen
  \bibfield  {author} {\bibinfo {author} {\bibfnamefont {T.}~\bibnamefont
  {Mikami}}, \bibinfo {author} {\bibfnamefont {S.}~\bibnamefont {Kitamura}},
  \bibinfo {author} {\bibfnamefont {K.}~\bibnamefont {Yasuda}}, \bibinfo
  {author} {\bibfnamefont {N.}~\bibnamefont {Tsuji}}, \bibinfo {author}
  {\bibfnamefont {T.}~\bibnamefont {Oka}},\ and\ \bibinfo {author}
  {\bibfnamefont {H.}~\bibnamefont {Aoki}},\ }\bibfield  {title} {\bibinfo
  {title} {Brillouin-wigner theory for high-frequency expansion in periodically
  driven systems: Application to floquet topological insulators},\ }\bibfield
  {journal} {\bibinfo  {journal} {Physical Review B}\ }\textbf {\bibinfo
  {volume} {93}},\ \href {https://doi.org/10.1103/physrevb.93.144307}
  {10.1103/physrevb.93.144307} (\bibinfo {year} {2016})\BibitemShut {NoStop}%
\bibitem [{\citenamefont {Lindner}\ \emph {et~al.}(2011)\citenamefont
  {Lindner}, \citenamefont {Refael},\ and\ \citenamefont
  {Galitski}}]{Lindner2011}%
  \BibitemOpen
  \bibfield  {author} {\bibinfo {author} {\bibfnamefont {N.~H.}\ \bibnamefont
  {Lindner}}, \bibinfo {author} {\bibfnamefont {G.}~\bibnamefont {Refael}},\
  and\ \bibinfo {author} {\bibfnamefont {V.}~\bibnamefont {Galitski}},\
  }\bibfield  {title} {\bibinfo {title} {Floquet topological insulator in
  semiconductor quantum wells},\ }\href {https://doi.org/10.1038/nphys1926}
  {\bibfield  {journal} {\bibinfo  {journal} {Nature Physics}\ }\textbf
  {\bibinfo {volume} {7}},\ \bibinfo {pages} {490–495} (\bibinfo {year}
  {2011})}\BibitemShut {NoStop}%
\bibitem [{\citenamefont {McIver}\ \emph {et~al.}(2019)\citenamefont {McIver},
  \citenamefont {Schulte}, \citenamefont {Stein}, \citenamefont {Matsuyama},
  \citenamefont {Jotzu}, \citenamefont {Meier},\ and\ \citenamefont
  {Cavalleri}}]{McIver2019}%
  \BibitemOpen
  \bibfield  {author} {\bibinfo {author} {\bibfnamefont {J.~W.}\ \bibnamefont
  {McIver}}, \bibinfo {author} {\bibfnamefont {B.}~\bibnamefont {Schulte}},
  \bibinfo {author} {\bibfnamefont {F.-U.}\ \bibnamefont {Stein}}, \bibinfo
  {author} {\bibfnamefont {T.}~\bibnamefont {Matsuyama}}, \bibinfo {author}
  {\bibfnamefont {G.}~\bibnamefont {Jotzu}}, \bibinfo {author} {\bibfnamefont
  {G.}~\bibnamefont {Meier}},\ and\ \bibinfo {author} {\bibfnamefont
  {A.}~\bibnamefont {Cavalleri}},\ }\bibfield  {title} {\bibinfo {title}
  {Light-induced anomalous hall effect in graphene},\ }\href
  {https://doi.org/10.1038/s41567-019-0698-y} {\bibfield  {journal} {\bibinfo
  {journal} {Nature Physics}\ }\textbf {\bibinfo {volume} {16}},\ \bibinfo
  {pages} {38–41} (\bibinfo {year} {2019})}\BibitemShut {NoStop}%
\bibitem [{\citenamefont {Dehghani}\ \emph {et~al.}(2015)\citenamefont
  {Dehghani}, \citenamefont {Oka},\ and\ \citenamefont
  {Mitra}}]{Dehghani2015HC}%
  \BibitemOpen
  \bibfield  {author} {\bibinfo {author} {\bibfnamefont {H.}~\bibnamefont
  {Dehghani}}, \bibinfo {author} {\bibfnamefont {T.}~\bibnamefont {Oka}},\ and\
  \bibinfo {author} {\bibfnamefont {A.}~\bibnamefont {Mitra}},\ }\bibfield
  {title} {\bibinfo {title} {Out-of-equilibrium electrons and the hall
  conductance of a floquet topological insulator},\ }\bibfield  {journal}
  {\bibinfo  {journal} {Physical Review B}\ }\textbf {\bibinfo {volume} {91}},\
  \href {https://doi.org/10.1103/physrevb.91.155422}
  {10.1103/physrevb.91.155422} (\bibinfo {year} {2015})\BibitemShut {NoStop}%
\bibitem [{\citenamefont {Dehghani}\ and\ \citenamefont
  {Mitra}(2015)}]{Dehghani2015TI}%
  \BibitemOpen
  \bibfield  {author} {\bibinfo {author} {\bibfnamefont {H.}~\bibnamefont
  {Dehghani}}\ and\ \bibinfo {author} {\bibfnamefont {A.}~\bibnamefont
  {Mitra}},\ }\bibfield  {title} {\bibinfo {title} {Optical hall conductivity
  of a floquet topological insulator},\ }\bibfield  {journal} {\bibinfo
  {journal} {Physical Review B}\ }\textbf {\bibinfo {volume} {92}},\ \href
  {https://doi.org/10.1103/physrevb.92.165111} {10.1103/physrevb.92.165111}
  (\bibinfo {year} {2015})\BibitemShut {NoStop}%
\bibitem [{\citenamefont {Oka}\ and\ \citenamefont {Aoki}(2009)}]{Oka2009}%
  \BibitemOpen
  \bibfield  {author} {\bibinfo {author} {\bibfnamefont {T.}~\bibnamefont
  {Oka}}\ and\ \bibinfo {author} {\bibfnamefont {H.}~\bibnamefont {Aoki}},\
  }\bibfield  {title} {\bibinfo {title} {Photovoltaic hall effect in
  graphene},\ }\bibfield  {journal} {\bibinfo  {journal} {Physical Review B}\
  }\textbf {\bibinfo {volume} {79}},\ \href
  {https://doi.org/10.1103/physrevb.79.081406} {10.1103/physrevb.79.081406}
  (\bibinfo {year} {2009})\BibitemShut {NoStop}%
\bibitem [{\citenamefont {Bloch}\ \emph {et~al.}(2008)\citenamefont {Bloch},
  \citenamefont {Dalibard},\ and\ \citenamefont {Zwerger}}]{ReviewBloch2008}%
  \BibitemOpen
  \bibfield  {author} {\bibinfo {author} {\bibfnamefont {I.}~\bibnamefont
  {Bloch}}, \bibinfo {author} {\bibfnamefont {J.}~\bibnamefont {Dalibard}},\
  and\ \bibinfo {author} {\bibfnamefont {W.}~\bibnamefont {Zwerger}},\
  }\bibfield  {title} {\bibinfo {title} {Many-body physics with ultracold
  gases},\ }\href {https://doi.org/10.1103/RevModPhys.80.885} {\bibfield
  {journal} {\bibinfo  {journal} {Rev. Mod. Phys.}\ }\textbf {\bibinfo {volume}
  {80}},\ \bibinfo {pages} {885} (\bibinfo {year} {2008})}\BibitemShut
  {NoStop}%
\bibitem [{\citenamefont {Wintersperger}\ \emph {et~al.}(2020)\citenamefont
  {Wintersperger}, \citenamefont {Bukov}, \citenamefont {N\"{a}ger},
  \citenamefont {Lellouch}, \citenamefont {Demler}, \citenamefont {Schneider},
  \citenamefont {Bloch}, \citenamefont {Goldman},\ and\ \citenamefont
  {Aidelsburger}}]{Wintersperger2020}%
  \BibitemOpen
  \bibfield  {author} {\bibinfo {author} {\bibfnamefont {K.}~\bibnamefont
  {Wintersperger}}, \bibinfo {author} {\bibfnamefont {M.}~\bibnamefont
  {Bukov}}, \bibinfo {author} {\bibfnamefont {J.}~\bibnamefont {N\"{a}ger}},
  \bibinfo {author} {\bibfnamefont {S.}~\bibnamefont {Lellouch}}, \bibinfo
  {author} {\bibfnamefont {E.}~\bibnamefont {Demler}}, \bibinfo {author}
  {\bibfnamefont {U.}~\bibnamefont {Schneider}}, \bibinfo {author}
  {\bibfnamefont {I.}~\bibnamefont {Bloch}}, \bibinfo {author} {\bibfnamefont
  {N.}~\bibnamefont {Goldman}},\ and\ \bibinfo {author} {\bibfnamefont
  {M.}~\bibnamefont {Aidelsburger}},\ }\bibfield  {title} {\bibinfo {title}
  {Parametric instabilities of interacting bosons in periodically driven 1d
  optical lattices},\ }\bibfield  {journal} {\bibinfo  {journal} {Physical
  Review X}\ }\textbf {\bibinfo {volume} {10}},\ \href
  {https://doi.org/10.1103/physrevx.10.011030} {10.1103/physrevx.10.011030}
  (\bibinfo {year} {2020})\BibitemShut {NoStop}%
\bibitem [{\citenamefont {Weitenberg}\ and\ \citenamefont
  {Simonet}(2021)}]{Weitenberg2021}%
  \BibitemOpen
  \bibfield  {author} {\bibinfo {author} {\bibfnamefont {C.}~\bibnamefont
  {Weitenberg}}\ and\ \bibinfo {author} {\bibfnamefont {J.}~\bibnamefont
  {Simonet}},\ }\bibfield  {title} {\bibinfo {title} {Tailoring quantum gases
  by floquet engineering},\ }\href {https://doi.org/10.1038/s41567-021-01316-x}
  {\bibfield  {journal} {\bibinfo  {journal} {Nature Physics}\ }\textbf
  {\bibinfo {volume} {17}},\ \bibinfo {pages} {1342–1348} (\bibinfo {year}
  {2021})}\BibitemShut {NoStop}%
\bibitem [{\citenamefont {Lei}\ \emph {et~al.}(2024)\citenamefont {Lei},
  \citenamefont {Fukumori}, \citenamefont {Wu}, \citenamefont {Barnes},
  \citenamefont {Economou}, \citenamefont {Choi},\ and\ \citenamefont
  {Faraon}}]{Lei2024}%
  \BibitemOpen
  \bibfield  {author} {\bibinfo {author} {\bibfnamefont {M.}~\bibnamefont
  {Lei}}, \bibinfo {author} {\bibfnamefont {R.}~\bibnamefont {Fukumori}},
  \bibinfo {author} {\bibfnamefont {C.-J.}\ \bibnamefont {Wu}}, \bibinfo
  {author} {\bibfnamefont {E.}~\bibnamefont {Barnes}}, \bibinfo {author}
  {\bibfnamefont {S.}~\bibnamefont {Economou}}, \bibinfo {author}
  {\bibfnamefont {J.}~\bibnamefont {Choi}},\ and\ \bibinfo {author}
  {\bibfnamefont {A.}~\bibnamefont {Faraon}},\ }\href
  {https://arxiv.org/abs/2408.00252} {\bibinfo {title} {Quantum thermalization
  and floquet engineering in a spin ensemble with a clock transition}}
  (\bibinfo {year} {2024}),\ \Eprint {https://arxiv.org/abs/2408.00252}
  {arXiv:2408.00252 [quant-ph]} \BibitemShut {NoStop}%
\bibitem [{\citenamefont {Di~Carli}\ \emph {et~al.}(2023)\citenamefont
  {Di~Carli}, \citenamefont {Cruickshank}, \citenamefont {Mitchell},
  \citenamefont {La~Rooij}, \citenamefont {Kuhr}, \citenamefont {Creffield},\
  and\ \citenamefont {Haller}}]{DiCarli2023}%
  \BibitemOpen
  \bibfield  {author} {\bibinfo {author} {\bibfnamefont {A.}~\bibnamefont
  {Di~Carli}}, \bibinfo {author} {\bibfnamefont {R.}~\bibnamefont
  {Cruickshank}}, \bibinfo {author} {\bibfnamefont {M.}~\bibnamefont
  {Mitchell}}, \bibinfo {author} {\bibfnamefont {A.}~\bibnamefont {La~Rooij}},
  \bibinfo {author} {\bibfnamefont {S.}~\bibnamefont {Kuhr}}, \bibinfo {author}
  {\bibfnamefont {C.~E.}\ \bibnamefont {Creffield}},\ and\ \bibinfo {author}
  {\bibfnamefont {E.}~\bibnamefont {Haller}},\ }\bibfield  {title} {\bibinfo
  {title} {Instabilities of interacting matter waves in optical lattices with
  floquet driving},\ }\bibfield  {journal} {\bibinfo  {journal} {Physical
  Review Research}\ }\textbf {\bibinfo {volume} {5}},\ \href
  {https://doi.org/10.1103/physrevresearch.5.033024}
  {10.1103/physrevresearch.5.033024} (\bibinfo {year} {2023})\BibitemShut
  {NoStop}%
\bibitem [{\citenamefont {Eckardt}(2017)}]{Eckardt2017}%
  \BibitemOpen
  \bibfield  {author} {\bibinfo {author} {\bibfnamefont {A.}~\bibnamefont
  {Eckardt}},\ }\bibfield  {title} {\bibinfo {title} {Colloquium: Atomic
  quantum gases in periodically driven optical lattices},\ }\bibfield
  {journal} {\bibinfo  {journal} {Reviews of Modern Physics}\ }\textbf
  {\bibinfo {volume} {89}},\ \href
  {https://doi.org/10.1103/revmodphys.89.011004} {10.1103/revmodphys.89.011004}
  (\bibinfo {year} {2017})\BibitemShut {NoStop}%
\bibitem [{\citenamefont {Sentef}\ \emph {et~al.}(2015)\citenamefont {Sentef},
  \citenamefont {Claassen}, \citenamefont {Kemper}, \citenamefont {Moritz},
  \citenamefont {Oka}, \citenamefont {Freericks},\ and\ \citenamefont
  {Devereaux}}]{Sentef2015}%
  \BibitemOpen
  \bibfield  {author} {\bibinfo {author} {\bibfnamefont {M.}~\bibnamefont
  {Sentef}}, \bibinfo {author} {\bibfnamefont {M.}~\bibnamefont {Claassen}},
  \bibinfo {author} {\bibfnamefont {A.}~\bibnamefont {Kemper}}, \bibinfo
  {author} {\bibfnamefont {B.}~\bibnamefont {Moritz}}, \bibinfo {author}
  {\bibfnamefont {T.}~\bibnamefont {Oka}}, \bibinfo {author} {\bibfnamefont
  {J.}~\bibnamefont {Freericks}},\ and\ \bibinfo {author} {\bibfnamefont
  {T.}~\bibnamefont {Devereaux}},\ }\bibfield  {title} {\bibinfo {title}
  {Theory of floquet band formation and local pseudospin textures in pump-probe
  photoemission of graphene},\ }\bibfield  {journal} {\bibinfo  {journal}
  {Nature Communications}\ }\textbf {\bibinfo {volume} {6}},\ \href
  {https://doi.org/10.1038/ncomms8047} {10.1038/ncomms8047} (\bibinfo {year}
  {2015})\BibitemShut {NoStop}%
\bibitem [{\citenamefont {Wang}\ \emph {et~al.}(2013)\citenamefont {Wang},
  \citenamefont {Steinberg}, \citenamefont {Jarillo-Herrero},\ and\
  \citenamefont {Gedik}}]{Wang2013}%
  \BibitemOpen
  \bibfield  {author} {\bibinfo {author} {\bibfnamefont {Y.~H.}\ \bibnamefont
  {Wang}}, \bibinfo {author} {\bibfnamefont {H.}~\bibnamefont {Steinberg}},
  \bibinfo {author} {\bibfnamefont {P.}~\bibnamefont {Jarillo-Herrero}},\ and\
  \bibinfo {author} {\bibfnamefont {N.}~\bibnamefont {Gedik}},\ }\bibfield
  {title} {\bibinfo {title} {Observation of floquet-bloch states on the surface
  of a topological insulator},\ }\href
  {https://doi.org/10.1126/science.1239834} {\bibfield  {journal} {\bibinfo
  {journal} {Science}\ }\textbf {\bibinfo {volume} {342}},\ \bibinfo {pages}
  {453–457} (\bibinfo {year} {2013})}\BibitemShut {NoStop}%
\bibitem [{\citenamefont {Mahmood}\ \emph {et~al.}(2016)\citenamefont
  {Mahmood}, \citenamefont {Chan}, \citenamefont {Alpichshev}, \citenamefont
  {Gardner}, \citenamefont {Lee}, \citenamefont {Lee},\ and\ \citenamefont
  {Gedik}}]{Mahmood2016}%
  \BibitemOpen
  \bibfield  {author} {\bibinfo {author} {\bibfnamefont {F.}~\bibnamefont
  {Mahmood}}, \bibinfo {author} {\bibfnamefont {C.-K.}\ \bibnamefont {Chan}},
  \bibinfo {author} {\bibfnamefont {Z.}~\bibnamefont {Alpichshev}}, \bibinfo
  {author} {\bibfnamefont {D.}~\bibnamefont {Gardner}}, \bibinfo {author}
  {\bibfnamefont {Y.}~\bibnamefont {Lee}}, \bibinfo {author} {\bibfnamefont
  {P.~A.}\ \bibnamefont {Lee}},\ and\ \bibinfo {author} {\bibfnamefont
  {N.}~\bibnamefont {Gedik}},\ }\bibfield  {title} {\bibinfo {title} {Selective
  scattering between floquet–bloch and volkov states in a topological
  insulator},\ }\href {https://doi.org/10.1038/nphys3609} {\bibfield  {journal}
  {\bibinfo  {journal} {Nature Physics}\ }\textbf {\bibinfo {volume} {12}},\
  \bibinfo {pages} {306–310} (\bibinfo {year} {2016})}\BibitemShut {NoStop}%
\bibitem [{\citenamefont {De~Giovannini}\ \emph {et~al.}(2016)\citenamefont
  {De~Giovannini}, \citenamefont {H\"{u}bener},\ and\ \citenamefont
  {Rubio}}]{DeGiovannini2016}%
  \BibitemOpen
  \bibfield  {author} {\bibinfo {author} {\bibfnamefont {U.}~\bibnamefont
  {De~Giovannini}}, \bibinfo {author} {\bibfnamefont {H.}~\bibnamefont
  {H\"{u}bener}},\ and\ \bibinfo {author} {\bibfnamefont {A.}~\bibnamefont
  {Rubio}},\ }\bibfield  {title} {\bibinfo {title} {Monitoring electron-photon
  dressing in wse2},\ }\href {https://doi.org/10.1021/acs.nanolett.6b04419}
  {\bibfield  {journal} {\bibinfo  {journal} {Nano Letters}\ }\textbf {\bibinfo
  {volume} {16}},\ \bibinfo {pages} {7993–7998} (\bibinfo {year}
  {2016})}\BibitemShut {NoStop}%
\bibitem [{\citenamefont {Aeschlimann}\ \emph {et~al.}(2021)\citenamefont
  {Aeschlimann}, \citenamefont {Sato}, \citenamefont {Krause}, \citenamefont
  {Ch\'avez-Cervantes}, \citenamefont {De~Giovannini}, \citenamefont
  {H\"ubener}, \citenamefont {Forti}, \citenamefont {Coletti}, \citenamefont
  {Hanff}, \citenamefont {Rossnagel}, \citenamefont {Rubio},\ and\
  \citenamefont {Gierz}}]{Gierz2021}%
  \BibitemOpen
  \bibfield  {author} {\bibinfo {author} {\bibfnamefont {S.}~\bibnamefont
  {Aeschlimann}}, \bibinfo {author} {\bibfnamefont {S.~A.}\ \bibnamefont
  {Sato}}, \bibinfo {author} {\bibfnamefont {R.}~\bibnamefont {Krause}},
  \bibinfo {author} {\bibfnamefont {M.}~\bibnamefont {Ch\'avez-Cervantes}},
  \bibinfo {author} {\bibfnamefont {U.}~\bibnamefont {De~Giovannini}}, \bibinfo
  {author} {\bibfnamefont {H.}~\bibnamefont {H\"ubener}}, \bibinfo {author}
  {\bibfnamefont {S.}~\bibnamefont {Forti}}, \bibinfo {author} {\bibfnamefont
  {C.}~\bibnamefont {Coletti}}, \bibinfo {author} {\bibfnamefont
  {K.}~\bibnamefont {Hanff}}, \bibinfo {author} {\bibfnamefont
  {K.}~\bibnamefont {Rossnagel}}, \bibinfo {author} {\bibfnamefont
  {A.}~\bibnamefont {Rubio}},\ and\ \bibinfo {author} {\bibfnamefont
  {I.}~\bibnamefont {Gierz}},\ }\bibfield  {title} {\bibinfo {title} {Survival
  of floquet–bloch states in the presence of scattering},\ }\href
  {https://doi.org/10.1021/acs.nanolett.1c00801} {\bibfield  {journal}
  {\bibinfo  {journal} {Nano Letters}\ }\textbf {\bibinfo {volume} {21}},\
  \bibinfo {pages} {5028–5035} (\bibinfo {year} {2021})}\BibitemShut
  {NoStop}%
\bibitem [{\citenamefont {Merboldt}\ \emph {et~al.}(2025)\citenamefont
  {Merboldt}, \citenamefont {Sch\"{u}ler}, \citenamefont {Schmitt},
  \citenamefont {Bange}, \citenamefont {Bennecke}, \citenamefont {Gadge},
  \citenamefont {Pierz}, \citenamefont {Schumacher}, \citenamefont {Momeni},
  \citenamefont {Steil}, \citenamefont {Manmana}, \citenamefont {Sentef},
  \citenamefont {Reutzel},\ and\ \citenamefont {Mathias}}]{Marco2024}%
  \BibitemOpen
  \bibfield  {author} {\bibinfo {author} {\bibfnamefont {M.}~\bibnamefont
  {Merboldt}}, \bibinfo {author} {\bibfnamefont {M.}~\bibnamefont
  {Sch\"{u}ler}}, \bibinfo {author} {\bibfnamefont {D.}~\bibnamefont
  {Schmitt}}, \bibinfo {author} {\bibfnamefont {J.~P.}\ \bibnamefont {Bange}},
  \bibinfo {author} {\bibfnamefont {W.}~\bibnamefont {Bennecke}}, \bibinfo
  {author} {\bibfnamefont {K.}~\bibnamefont {Gadge}}, \bibinfo {author}
  {\bibfnamefont {K.}~\bibnamefont {Pierz}}, \bibinfo {author} {\bibfnamefont
  {H.~W.}\ \bibnamefont {Schumacher}}, \bibinfo {author} {\bibfnamefont
  {D.}~\bibnamefont {Momeni}}, \bibinfo {author} {\bibfnamefont
  {D.}~\bibnamefont {Steil}}, \bibinfo {author} {\bibfnamefont {S.~R.}\
  \bibnamefont {Manmana}}, \bibinfo {author} {\bibfnamefont {M.~A.}\
  \bibnamefont {Sentef}}, \bibinfo {author} {\bibfnamefont {M.}~\bibnamefont
  {Reutzel}},\ and\ \bibinfo {author} {\bibfnamefont {S.}~\bibnamefont
  {Mathias}},\ }\bibfield  {title} {\bibinfo {title} {Observation of floquet
  states in graphene},\ }\bibfield  {journal} {\bibinfo  {journal} {Nature
  Physics}\ }\href {https://doi.org/10.1038/s41567-025-02889-7}
  {10.1038/s41567-025-02889-7} (\bibinfo {year} {2025})\BibitemShut {NoStop}%
\bibitem [{\citenamefont {Choi}\ \emph {et~al.}(2025)\citenamefont {Choi},
  \citenamefont {Mogi}, \citenamefont {De~Giovannini}, \citenamefont {Azoury},
  \citenamefont {Lv}, \citenamefont {Su}, \citenamefont {H\"{u}bener},
  \citenamefont {Rubio},\ and\ \citenamefont {Gedik}}]{Choi2025}%
  \BibitemOpen
  \bibfield  {author} {\bibinfo {author} {\bibfnamefont {D.}~\bibnamefont
  {Choi}}, \bibinfo {author} {\bibfnamefont {M.}~\bibnamefont {Mogi}}, \bibinfo
  {author} {\bibfnamefont {U.}~\bibnamefont {De~Giovannini}}, \bibinfo {author}
  {\bibfnamefont {D.}~\bibnamefont {Azoury}}, \bibinfo {author} {\bibfnamefont
  {B.}~\bibnamefont {Lv}}, \bibinfo {author} {\bibfnamefont {Y.}~\bibnamefont
  {Su}}, \bibinfo {author} {\bibfnamefont {H.}~\bibnamefont {H\"{u}bener}},
  \bibinfo {author} {\bibfnamefont {A.}~\bibnamefont {Rubio}},\ and\ \bibinfo
  {author} {\bibfnamefont {N.}~\bibnamefont {Gedik}},\ }\bibfield  {title}
  {\bibinfo {title} {Observation of floquet–bloch states in monolayer
  graphene},\ }\bibfield  {journal} {\bibinfo  {journal} {Nature Physics}\
  }\href {https://doi.org/10.1038/s41567-025-02888-8}
  {10.1038/s41567-025-02888-8} (\bibinfo {year} {2025})\BibitemShut {NoStop}%
\bibitem [{\citenamefont {Bielinski}\ \emph {et~al.}(2025)\citenamefont
  {Bielinski}, \citenamefont {Chari}, \citenamefont {May-Mann}, \citenamefont
  {Kim}, \citenamefont {Zwettler}, \citenamefont {Deng}, \citenamefont
  {Aishwarya}, \citenamefont {Roychowdhury}, \citenamefont {Shekhar},
  \citenamefont {Hashimoto}, \citenamefont {Lu}, \citenamefont {Yan},
  \citenamefont {Felser}, \citenamefont {Madhavan}, \citenamefont {Shen},
  \citenamefont {Hughes},\ and\ \citenamefont {Mahmood}}]{Bielinski2025}%
  \BibitemOpen
  \bibfield  {author} {\bibinfo {author} {\bibfnamefont {N.}~\bibnamefont
  {Bielinski}}, \bibinfo {author} {\bibfnamefont {R.}~\bibnamefont {Chari}},
  \bibinfo {author} {\bibfnamefont {J.}~\bibnamefont {May-Mann}}, \bibinfo
  {author} {\bibfnamefont {S.}~\bibnamefont {Kim}}, \bibinfo {author}
  {\bibfnamefont {J.}~\bibnamefont {Zwettler}}, \bibinfo {author}
  {\bibfnamefont {Y.}~\bibnamefont {Deng}}, \bibinfo {author} {\bibfnamefont
  {A.}~\bibnamefont {Aishwarya}}, \bibinfo {author} {\bibfnamefont
  {S.}~\bibnamefont {Roychowdhury}}, \bibinfo {author} {\bibfnamefont
  {C.}~\bibnamefont {Shekhar}}, \bibinfo {author} {\bibfnamefont
  {M.}~\bibnamefont {Hashimoto}}, \bibinfo {author} {\bibfnamefont
  {D.}~\bibnamefont {Lu}}, \bibinfo {author} {\bibfnamefont {J.}~\bibnamefont
  {Yan}}, \bibinfo {author} {\bibfnamefont {C.}~\bibnamefont {Felser}},
  \bibinfo {author} {\bibfnamefont {V.}~\bibnamefont {Madhavan}}, \bibinfo
  {author} {\bibfnamefont {Z.-X.}\ \bibnamefont {Shen}}, \bibinfo {author}
  {\bibfnamefont {T.~L.}\ \bibnamefont {Hughes}},\ and\ \bibinfo {author}
  {\bibfnamefont {F.}~\bibnamefont {Mahmood}},\ }\bibfield  {title} {\bibinfo
  {title} {Floquet–bloch manipulation of the dirac gap in a topological
  antiferromagnet},\ }\bibfield  {journal} {\bibinfo  {journal} {Nature
  Physics}\ }\href {https://doi.org/10.1038/s41567-024-02769-6}
  {10.1038/s41567-024-02769-6} (\bibinfo {year} {2025})\BibitemShut {NoStop}%
\bibitem [{\citenamefont {Fragkos}\ \emph {et~al.}(2025)\citenamefont
  {Fragkos}, \citenamefont {Fabre}, \citenamefont {Tkach}, \citenamefont
  {Petit}, \citenamefont {Descamps}, \citenamefont {Sch\"{o}nhense},
  \citenamefont {Mairesse}, \citenamefont {Sch\"{u}ler},\ and\ \citenamefont
  {Beaulieu}}]{fragkos2024}%
  \BibitemOpen
  \bibfield  {author} {\bibinfo {author} {\bibfnamefont {S.}~\bibnamefont
  {Fragkos}}, \bibinfo {author} {\bibfnamefont {B.}~\bibnamefont {Fabre}},
  \bibinfo {author} {\bibfnamefont {O.}~\bibnamefont {Tkach}}, \bibinfo
  {author} {\bibfnamefont {S.}~\bibnamefont {Petit}}, \bibinfo {author}
  {\bibfnamefont {D.}~\bibnamefont {Descamps}}, \bibinfo {author}
  {\bibfnamefont {G.}~\bibnamefont {Sch\"{o}nhense}}, \bibinfo {author}
  {\bibfnamefont {Y.}~\bibnamefont {Mairesse}}, \bibinfo {author}
  {\bibfnamefont {M.}~\bibnamefont {Sch\"{u}ler}},\ and\ \bibinfo {author}
  {\bibfnamefont {S.}~\bibnamefont {Beaulieu}},\ }\bibfield  {title} {\bibinfo
  {title} {Floquet-bloch valleytronics},\ }\bibfield  {journal} {\bibinfo
  {journal} {Nature Communications}\ }\textbf {\bibinfo {volume} {16}},\ \href
  {https://doi.org/10.1038/s41467-025-61076-7} {10.1038/s41467-025-61076-7}
  (\bibinfo {year} {2025})\BibitemShut {NoStop}%
\bibitem [{\citenamefont {Sch\"{u}ler}\ \emph {et~al.}(2020)\citenamefont
  {Sch\"{u}ler}, \citenamefont {De~Giovannini}, \citenamefont {H\"{u}bener},
  \citenamefont {Rubio}, \citenamefont {Sentef}, \citenamefont {Devereaux},\
  and\ \citenamefont {Werner}}]{Schler2020}%
  \BibitemOpen
  \bibfield  {author} {\bibinfo {author} {\bibfnamefont {M.}~\bibnamefont
  {Sch\"{u}ler}}, \bibinfo {author} {\bibfnamefont {U.}~\bibnamefont
  {De~Giovannini}}, \bibinfo {author} {\bibfnamefont {H.}~\bibnamefont
  {H\"{u}bener}}, \bibinfo {author} {\bibfnamefont {A.}~\bibnamefont {Rubio}},
  \bibinfo {author} {\bibfnamefont {M.~A.}\ \bibnamefont {Sentef}}, \bibinfo
  {author} {\bibfnamefont {T.~P.}\ \bibnamefont {Devereaux}},\ and\ \bibinfo
  {author} {\bibfnamefont {P.}~\bibnamefont {Werner}},\ }\bibfield  {title}
  {\bibinfo {title} {How circular dichroism in time- and angle-resolved
  photoemission can be used to spectroscopically detect transient topological
  states in graphene},\ }\bibfield  {journal} {\bibinfo  {journal} {Physical
  Review X}\ }\textbf {\bibinfo {volume} {10}},\ \href
  {https://doi.org/10.1103/physrevx.10.041013} {10.1103/physrevx.10.041013}
  (\bibinfo {year} {2020})\BibitemShut {NoStop}%
\bibitem [{\citenamefont {Sch\"{u}ler}\ and\ \citenamefont
  {Beaulieu}(2022)}]{Schler2022}%
  \BibitemOpen
  \bibfield  {author} {\bibinfo {author} {\bibfnamefont {M.}~\bibnamefont
  {Sch\"{u}ler}}\ and\ \bibinfo {author} {\bibfnamefont {S.}~\bibnamefont
  {Beaulieu}},\ }\bibfield  {title} {\bibinfo {title} {Probing topological
  floquet states in wse2 using circular dichroism in time- and angle-resolved
  photoemission spectroscopy},\ }\bibfield  {journal} {\bibinfo  {journal}
  {Communications Physics}\ }\textbf {\bibinfo {volume} {5}},\ \href
  {https://doi.org/10.1038/s42005-022-00944-w} {10.1038/s42005-022-00944-w}
  (\bibinfo {year} {2022})\BibitemShut {NoStop}%
\bibitem [{\citenamefont {Murakami}\ and\ \citenamefont
  {Werner}(2018)}]{Murakami2018}%
  \BibitemOpen
  \bibfield  {author} {\bibinfo {author} {\bibfnamefont {Y.}~\bibnamefont
  {Murakami}}\ and\ \bibinfo {author} {\bibfnamefont {P.}~\bibnamefont
  {Werner}},\ }\bibfield  {title} {\bibinfo {title} {Nonequilibrium steady
  states of electric field driven mott insulators},\ }\bibfield  {journal}
  {\bibinfo  {journal} {Physical Review B}\ }\textbf {\bibinfo {volume} {98}},\
  \href {https://doi.org/10.1103/physrevb.98.075102}
  {10.1103/physrevb.98.075102} (\bibinfo {year} {2018})\BibitemShut {NoStop}%
\bibitem [{\citenamefont {Tsuji}\ \emph {et~al.}(2008)\citenamefont {Tsuji},
  \citenamefont {Oka},\ and\ \citenamefont {Aoki}}]{Tsuji2008}%
  \BibitemOpen
  \bibfield  {author} {\bibinfo {author} {\bibfnamefont {N.}~\bibnamefont
  {Tsuji}}, \bibinfo {author} {\bibfnamefont {T.}~\bibnamefont {Oka}},\ and\
  \bibinfo {author} {\bibfnamefont {H.}~\bibnamefont {Aoki}},\ }\bibfield
  {title} {\bibinfo {title} {Correlated electron systems periodically driven
  out of equilibrium:floquet+dmft formalism},\ }\bibfield  {journal} {\bibinfo
  {journal} {Physical Review B}\ }\textbf {\bibinfo {volume} {78}},\ \href
  {https://doi.org/10.1103/physrevb.78.235124} {10.1103/physrevb.78.235124}
  (\bibinfo {year} {2008})\BibitemShut {NoStop}%
\bibitem [{\citenamefont {Aoki}\ \emph {et~al.}(2014)\citenamefont {Aoki},
  \citenamefont {Tsuji}, \citenamefont {Eckstein}, \citenamefont {Kollar},
  \citenamefont {Oka},\ and\ \citenamefont {Werner}}]{Aoki2014}%
  \BibitemOpen
  \bibfield  {author} {\bibinfo {author} {\bibfnamefont {H.}~\bibnamefont
  {Aoki}}, \bibinfo {author} {\bibfnamefont {N.}~\bibnamefont {Tsuji}},
  \bibinfo {author} {\bibfnamefont {M.}~\bibnamefont {Eckstein}}, \bibinfo
  {author} {\bibfnamefont {M.}~\bibnamefont {Kollar}}, \bibinfo {author}
  {\bibfnamefont {T.}~\bibnamefont {Oka}},\ and\ \bibinfo {author}
  {\bibfnamefont {P.}~\bibnamefont {Werner}},\ }\bibfield  {title} {\bibinfo
  {title} {Nonequilibrium dynamical mean-field theory and its applications},\
  }\href {https://doi.org/10.1103/revmodphys.86.779} {\bibfield  {journal}
  {\bibinfo  {journal} {Reviews of Modern Physics}\ }\textbf {\bibinfo {volume}
  {86}},\ \bibinfo {pages} {779–837} (\bibinfo {year} {2014})}\BibitemShut
  {NoStop}%
\bibitem [{\citenamefont {Puviani}\ and\ \citenamefont
  {Manghi}(2016)}]{Puviani2016}%
  \BibitemOpen
  \bibfield  {author} {\bibinfo {author} {\bibfnamefont {M.}~\bibnamefont
  {Puviani}}\ and\ \bibinfo {author} {\bibfnamefont {F.}~\bibnamefont
  {Manghi}},\ }\bibfield  {title} {\bibinfo {title} {Periodically driven
  interacting electrons in one dimension: Many-body floquet approach},\
  }\bibfield  {journal} {\bibinfo  {journal} {Physical Review B}\ }\textbf
  {\bibinfo {volume} {94}},\ \href {https://doi.org/10.1103/physrevb.94.161111}
  {10.1103/physrevb.94.161111} (\bibinfo {year} {2016})\BibitemShut {NoStop}%
\bibitem [{\citenamefont {Dunlap}\ and\ \citenamefont
  {Kenkre}(1986)}]{Dunlap1986}%
  \BibitemOpen
  \bibfield  {author} {\bibinfo {author} {\bibfnamefont {D.~H.}\ \bibnamefont
  {Dunlap}}\ and\ \bibinfo {author} {\bibfnamefont {V.~M.}\ \bibnamefont
  {Kenkre}},\ }\bibfield  {title} {\bibinfo {title} {Dynamic localization of a
  charged particle moving under the influence of an electric field},\ }\href
  {https://doi.org/10.1103/physrevb.34.3625} {\bibfield  {journal} {\bibinfo
  {journal} {Physical Review B}\ }\textbf {\bibinfo {volume} {34}},\ \bibinfo
  {pages} {3625–3633} (\bibinfo {year} {1986})}\BibitemShut {NoStop}%
\bibitem [{\citenamefont {Osterkorn}\ \emph {et~al.}(2023)\citenamefont
  {Osterkorn}, \citenamefont {Meyer},\ and\ \citenamefont
  {Manmana}}]{Osterkorn2023}%
  \BibitemOpen
  \bibfield  {author} {\bibinfo {author} {\bibfnamefont {A.}~\bibnamefont
  {Osterkorn}}, \bibinfo {author} {\bibfnamefont {C.}~\bibnamefont {Meyer}},\
  and\ \bibinfo {author} {\bibfnamefont {S.~R.}\ \bibnamefont {Manmana}},\
  }\bibfield  {title} {\bibinfo {title} {In-gap band formation in a
  periodically driven charge density wave insulator},\ }\bibfield  {journal}
  {\bibinfo  {journal} {Communications Physics}\ }\textbf {\bibinfo {volume}
  {6}},\ \href {https://doi.org/10.1038/s42005-023-01346-2}
  {10.1038/s42005-023-01346-2} (\bibinfo {year} {2023})\BibitemShut {NoStop}%
\bibitem [{\citenamefont {Ponte}\ \emph
  {et~al.}(2015{\natexlab{a}})\citenamefont {Ponte}, \citenamefont {Papi\'c},
  \citenamefont {Huveneers},\ and\ \citenamefont {Abanin}}]{Ponte2015}%
  \BibitemOpen
  \bibfield  {author} {\bibinfo {author} {\bibfnamefont {P.}~\bibnamefont
  {Ponte}}, \bibinfo {author} {\bibfnamefont {Z.}~\bibnamefont {Papi\'c}},
  \bibinfo {author} {\bibfnamefont {F.}~\bibnamefont {Huveneers}},\ and\
  \bibinfo {author} {\bibfnamefont {D.~A.}\ \bibnamefont {Abanin}},\ }\bibfield
   {title} {\bibinfo {title} {Many-body localization in periodically driven
  systems},\ }\bibfield  {journal} {\bibinfo  {journal} {Physical Review
  Letters}\ }\textbf {\bibinfo {volume} {114}},\ \href
  {https://doi.org/10.1103/physrevlett.114.140401}
  {10.1103/physrevlett.114.140401} (\bibinfo {year}
  {2015}{\natexlab{a}})\BibitemShut {NoStop}%
\bibitem [{\citenamefont {Lazarides}\ \emph {et~al.}(2015)\citenamefont
  {Lazarides}, \citenamefont {Das},\ and\ \citenamefont
  {Moessner}}]{Lazarides2015}%
  \BibitemOpen
  \bibfield  {author} {\bibinfo {author} {\bibfnamefont {A.}~\bibnamefont
  {Lazarides}}, \bibinfo {author} {\bibfnamefont {A.}~\bibnamefont {Das}},\
  and\ \bibinfo {author} {\bibfnamefont {R.}~\bibnamefont {Moessner}},\
  }\bibfield  {title} {\bibinfo {title} {Fate of many-body localization under
  periodic driving},\ }\bibfield  {journal} {\bibinfo  {journal} {Physical
  Review Letters}\ }\textbf {\bibinfo {volume} {115}},\ \href
  {https://doi.org/10.1103/physrevlett.115.030402}
  {10.1103/physrevlett.115.030402} (\bibinfo {year} {2015})\BibitemShut
  {NoStop}%
\bibitem [{\citenamefont {Sierant}\ \emph {et~al.}(2023)\citenamefont
  {Sierant}, \citenamefont {Lewenstein}, \citenamefont {Scardicchio},\ and\
  \citenamefont {Zakrzewski}}]{Sierant2023}%
  \BibitemOpen
  \bibfield  {author} {\bibinfo {author} {\bibfnamefont {P.}~\bibnamefont
  {Sierant}}, \bibinfo {author} {\bibfnamefont {M.}~\bibnamefont {Lewenstein}},
  \bibinfo {author} {\bibfnamefont {A.}~\bibnamefont {Scardicchio}},\ and\
  \bibinfo {author} {\bibfnamefont {J.}~\bibnamefont {Zakrzewski}},\ }\bibfield
   {title} {\bibinfo {title} {Stability of many-body localization in floquet
  systems},\ }\bibfield  {journal} {\bibinfo  {journal} {Physical Review B}\
  }\textbf {\bibinfo {volume} {107}},\ \href
  {https://doi.org/10.1103/physrevb.107.115132} {10.1103/physrevb.107.115132}
  (\bibinfo {year} {2023})\BibitemShut {NoStop}%
\bibitem [{\citenamefont {Zhang}\ \emph {et~al.}(2016)\citenamefont {Zhang},
  \citenamefont {Khemani},\ and\ \citenamefont {Huse}}]{Zhang2016}%
  \BibitemOpen
  \bibfield  {author} {\bibinfo {author} {\bibfnamefont {L.}~\bibnamefont
  {Zhang}}, \bibinfo {author} {\bibfnamefont {V.}~\bibnamefont {Khemani}},\
  and\ \bibinfo {author} {\bibfnamefont {D.~A.}\ \bibnamefont {Huse}},\
  }\bibfield  {title} {\bibinfo {title} {A floquet model for the many-body
  localization transition},\ }\bibfield  {journal} {\bibinfo  {journal}
  {Physical Review B}\ }\textbf {\bibinfo {volume} {94}},\ \href
  {https://doi.org/10.1103/physrevb.94.224202} {10.1103/physrevb.94.224202}
  (\bibinfo {year} {2016})\BibitemShut {NoStop}%
\bibitem [{\citenamefont {Decker}\ \emph {et~al.}(2020)\citenamefont {Decker},
  \citenamefont {Karrasch}, \citenamefont {Eisert},\ and\ \citenamefont
  {Kennes}}]{Decker2020}%
  \BibitemOpen
  \bibfield  {author} {\bibinfo {author} {\bibfnamefont {K.~S.~C.}\
  \bibnamefont {Decker}}, \bibinfo {author} {\bibfnamefont {C.}~\bibnamefont
  {Karrasch}}, \bibinfo {author} {\bibfnamefont {J.}~\bibnamefont {Eisert}},\
  and\ \bibinfo {author} {\bibfnamefont {D.~M.}\ \bibnamefont {Kennes}},\
  }\bibfield  {title} {\bibinfo {title} {Floquet engineering topological
  many-body localized systems},\ }\bibfield  {journal} {\bibinfo  {journal}
  {Physical Review Letters}\ }\textbf {\bibinfo {volume} {124}},\ \href
  {https://doi.org/10.1103/physrevlett.124.190601}
  {10.1103/physrevlett.124.190601} (\bibinfo {year} {2020})\BibitemShut
  {NoStop}%
\bibitem [{\citenamefont {Bukov}\ \emph {et~al.}(2015)\citenamefont {Bukov},
  \citenamefont {D'Alessio},\ and\ \citenamefont {Polkovnikov}}]{Bukov2015}%
  \BibitemOpen
  \bibfield  {author} {\bibinfo {author} {\bibfnamefont {M.}~\bibnamefont
  {Bukov}}, \bibinfo {author} {\bibfnamefont {L.}~\bibnamefont {D'Alessio}},\
  and\ \bibinfo {author} {\bibfnamefont {A.}~\bibnamefont {Polkovnikov}},\
  }\bibfield  {title} {\bibinfo {title} {Universal high-frequency behavior of
  periodically driven systems: from dynamical stabilization to floquet
  engineering},\ }\href {https://doi.org/10.1080/00018732.2015.1055918}
  {\bibfield  {journal} {\bibinfo  {journal} {Advances in Physics}\ }\textbf
  {\bibinfo {volume} {64}},\ \bibinfo {pages} {139} (\bibinfo {year} {2015})},\
  \Eprint {https://arxiv.org/abs/https://doi.org/10.1080/00018732.2015.1055918}
  {https://doi.org/10.1080/00018732.2015.1055918} \BibitemShut {NoStop}%
\bibitem [{\citenamefont {Kennes}\ \emph {et~al.}(2018)\citenamefont {Kennes},
  \citenamefont {de~la Torre}, \citenamefont {Ron}, \citenamefont {Hsieh},\
  and\ \citenamefont {Millis}}]{Kennes2018}%
  \BibitemOpen
  \bibfield  {author} {\bibinfo {author} {\bibfnamefont {D.}~\bibnamefont
  {Kennes}}, \bibinfo {author} {\bibfnamefont {A.}~\bibnamefont {de~la Torre}},
  \bibinfo {author} {\bibfnamefont {A.}~\bibnamefont {Ron}}, \bibinfo {author}
  {\bibfnamefont {D.}~\bibnamefont {Hsieh}},\ and\ \bibinfo {author}
  {\bibfnamefont {A.}~\bibnamefont {Millis}},\ }\bibfield  {title} {\bibinfo
  {title} {Floquet engineering in quantum chains},\ }\bibfield  {journal}
  {\bibinfo  {journal} {Physical Review Letters}\ }\textbf {\bibinfo {volume}
  {120}},\ \href {https://doi.org/10.1103/physrevlett.120.127601}
  {10.1103/physrevlett.120.127601} (\bibinfo {year} {2018})\BibitemShut
  {NoStop}%
\bibitem [{\citenamefont {Broers}\ and\ \citenamefont
  {Mathey}(2021)}]{Broers2021}%
  \BibitemOpen
  \bibfield  {author} {\bibinfo {author} {\bibfnamefont {L.}~\bibnamefont
  {Broers}}\ and\ \bibinfo {author} {\bibfnamefont {L.}~\bibnamefont
  {Mathey}},\ }\bibfield  {title} {\bibinfo {title} {Observing light-induced
  floquet band gaps in the longitudinal conductivity of graphene},\ }\bibfield
  {journal} {\bibinfo  {journal} {Communications Physics}\ }\textbf {\bibinfo
  {volume} {4}},\ \href {https://doi.org/10.1038/s42005-021-00746-6}
  {10.1038/s42005-021-00746-6} (\bibinfo {year} {2021})\BibitemShut {NoStop}%
\bibitem [{\citenamefont {Broers}\ and\ \citenamefont
  {Mathey}(2022)}]{Broers2022}%
  \BibitemOpen
  \bibfield  {author} {\bibinfo {author} {\bibfnamefont {L.}~\bibnamefont
  {Broers}}\ and\ \bibinfo {author} {\bibfnamefont {L.}~\bibnamefont
  {Mathey}},\ }\bibfield  {title} {\bibinfo {title} {Detecting light-induced
  floquet band gaps of graphene via trarpes},\ }\bibfield  {journal} {\bibinfo
  {journal} {Physical Review Research}\ }\textbf {\bibinfo {volume} {4}},\
  \href {https://doi.org/10.1103/physrevresearch.4.013057}
  {10.1103/physrevresearch.4.013057} (\bibinfo {year} {2022})\BibitemShut
  {NoStop}%
\bibitem [{\citenamefont {Tsuji}(2024)}]{Tsuji2024}%
  \BibitemOpen
  \bibfield  {author} {\bibinfo {author} {\bibfnamefont {N.}~\bibnamefont
  {Tsuji}},\ }\href {https://doi.org/10.1016/b978-0-323-90800-9.00241-9} {\emph
  {\bibinfo {title} {Encyclopedia of Condensed Matter Physics}}}\ (\bibinfo
  {publisher} {Elsevier},\ \bibinfo {year} {2024})\ p.\ \bibinfo {pages}
  {967–980}\BibitemShut {NoStop}%
\bibitem [{\citenamefont {Bukov}\ \emph {et~al.}(2016)\citenamefont {Bukov},
  \citenamefont {Heyl}, \citenamefont {Huse},\ and\ \citenamefont
  {Polkovnikov}}]{Bukov2016}%
  \BibitemOpen
  \bibfield  {author} {\bibinfo {author} {\bibfnamefont {M.}~\bibnamefont
  {Bukov}}, \bibinfo {author} {\bibfnamefont {M.}~\bibnamefont {Heyl}},
  \bibinfo {author} {\bibfnamefont {D.~A.}\ \bibnamefont {Huse}},\ and\
  \bibinfo {author} {\bibfnamefont {A.}~\bibnamefont {Polkovnikov}},\
  }\bibfield  {title} {\bibinfo {title} {Heating and many-body resonances in a
  periodically driven two-band system},\ }\bibfield  {journal} {\bibinfo
  {journal} {Physical Review B}\ }\textbf {\bibinfo {volume} {93}},\ \href
  {https://doi.org/10.1103/physrevb.93.155132} {10.1103/physrevb.93.155132}
  (\bibinfo {year} {2016})\BibitemShut {NoStop}%
\bibitem [{\citenamefont {Zhao}\ \emph {et~al.}(2022)\citenamefont {Zhao},
  \citenamefont {Knolle}, \citenamefont {Moessner},\ and\ \citenamefont
  {Mintert}}]{Zhao2022}%
  \BibitemOpen
  \bibfield  {author} {\bibinfo {author} {\bibfnamefont {H.}~\bibnamefont
  {Zhao}}, \bibinfo {author} {\bibfnamefont {J.}~\bibnamefont {Knolle}},
  \bibinfo {author} {\bibfnamefont {R.}~\bibnamefont {Moessner}},\ and\
  \bibinfo {author} {\bibfnamefont {F.}~\bibnamefont {Mintert}},\ }\bibfield
  {title} {\bibinfo {title} {Suppression of interband heating for random
  driving},\ }\bibfield  {journal} {\bibinfo  {journal} {Physical Review
  Letters}\ }\textbf {\bibinfo {volume} {129}},\ \href
  {https://doi.org/10.1103/physrevlett.129.120605}
  {10.1103/physrevlett.129.120605} (\bibinfo {year} {2022})\BibitemShut
  {NoStop}%
\bibitem [{\citenamefont {Abanin}\ \emph {et~al.}(2015)\citenamefont {Abanin},
  \citenamefont {De~Roeck},\ and\ \citenamefont {Huveneers}}]{Abanin2015}%
  \BibitemOpen
  \bibfield  {author} {\bibinfo {author} {\bibfnamefont {D.~A.}\ \bibnamefont
  {Abanin}}, \bibinfo {author} {\bibfnamefont {W.}~\bibnamefont {De~Roeck}},\
  and\ \bibinfo {author} {\bibfnamefont {F.}~\bibnamefont {Huveneers}},\
  }\bibfield  {title} {\bibinfo {title} {Exponentially slow heating in
  periodically driven many-body systems},\ }\bibfield  {journal} {\bibinfo
  {journal} {Physical Review Letters}\ }\textbf {\bibinfo {volume} {115}},\
  \href {https://doi.org/10.1103/physrevlett.115.256803}
  {10.1103/physrevlett.115.256803} (\bibinfo {year} {2015})\BibitemShut
  {NoStop}%
\bibitem [{\citenamefont {Haldar}\ \emph {et~al.}(2018)\citenamefont {Haldar},
  \citenamefont {Moessner},\ and\ \citenamefont {Das}}]{Haldar2018}%
  \BibitemOpen
  \bibfield  {author} {\bibinfo {author} {\bibfnamefont {A.}~\bibnamefont
  {Haldar}}, \bibinfo {author} {\bibfnamefont {R.}~\bibnamefont {Moessner}},\
  and\ \bibinfo {author} {\bibfnamefont {A.}~\bibnamefont {Das}},\ }\bibfield
  {title} {\bibinfo {title} {Onset of floquet thermalization},\ }\bibfield
  {journal} {\bibinfo  {journal} {Physical Review B}\ }\textbf {\bibinfo
  {volume} {97}},\ \href {https://doi.org/10.1103/physrevb.97.245122}
  {10.1103/physrevb.97.245122} (\bibinfo {year} {2018})\BibitemShut {NoStop}%
\bibitem [{\citenamefont {Fleckenstein}\ and\ \citenamefont
  {Bukov}(2021{\natexlab{a}})}]{Fleckenstein2021}%
  \BibitemOpen
  \bibfield  {author} {\bibinfo {author} {\bibfnamefont {C.}~\bibnamefont
  {Fleckenstein}}\ and\ \bibinfo {author} {\bibfnamefont {M.}~\bibnamefont
  {Bukov}},\ }\bibfield  {title} {\bibinfo {title} {Prethermalization and
  thermalization in periodically driven many-body systems away from the
  high-frequency limit},\ }\bibfield  {journal} {\bibinfo  {journal} {Physical
  Review B}\ }\textbf {\bibinfo {volume} {103}},\ \href
  {https://doi.org/10.1103/physrevb.103.l140302} {10.1103/physrevb.103.l140302}
  (\bibinfo {year} {2021}{\natexlab{a}})\BibitemShut {NoStop}%
\bibitem [{\citenamefont {Peronaci}\ \emph {et~al.}(2018)\citenamefont
  {Peronaci}, \citenamefont {Schiró},\ and\ \citenamefont
  {Parcollet}}]{Peronaci2018}%
  \BibitemOpen
  \bibfield  {author} {\bibinfo {author} {\bibfnamefont {F.}~\bibnamefont
  {Peronaci}}, \bibinfo {author} {\bibfnamefont {M.}~\bibnamefont {Schiró}},\
  and\ \bibinfo {author} {\bibfnamefont {O.}~\bibnamefont {Parcollet}},\
  }\bibfield  {title} {\bibinfo {title} {Resonant thermalization of
  periodically driven strongly correlated electrons},\ }\bibfield  {journal}
  {\bibinfo  {journal} {Physical Review Letters}\ }\textbf {\bibinfo {volume}
  {120}},\ \href {https://doi.org/10.1103/physrevlett.120.197601}
  {10.1103/physrevlett.120.197601} (\bibinfo {year} {2018})\BibitemShut
  {NoStop}%
\bibitem [{\citenamefont {Machado}\ \emph {et~al.}(2019)\citenamefont
  {Machado}, \citenamefont {Kahanamoku-Meyer}, \citenamefont {Else},
  \citenamefont {Nayak},\ and\ \citenamefont {Yao}}]{Machado2019}%
  \BibitemOpen
  \bibfield  {author} {\bibinfo {author} {\bibfnamefont {F.}~\bibnamefont
  {Machado}}, \bibinfo {author} {\bibfnamefont {G.~D.}\ \bibnamefont
  {Kahanamoku-Meyer}}, \bibinfo {author} {\bibfnamefont {D.~V.}\ \bibnamefont
  {Else}}, \bibinfo {author} {\bibfnamefont {C.}~\bibnamefont {Nayak}},\ and\
  \bibinfo {author} {\bibfnamefont {N.~Y.}\ \bibnamefont {Yao}},\ }\bibfield
  {title} {\bibinfo {title} {Exponentially slow heating in short and long-range
  interacting floquet systems},\ }\href
  {https://doi.org/10.1103/PhysRevResearch.1.033202} {\bibfield  {journal}
  {\bibinfo  {journal} {Phys. Rev. Res.}\ }\textbf {\bibinfo {volume} {1}},\
  \bibinfo {pages} {033202} (\bibinfo {year} {2019})}\BibitemShut {NoStop}%
\bibitem [{\citenamefont {Hauke}\ \emph {et~al.}(2016)\citenamefont {Hauke},
  \citenamefont {Heyl}, \citenamefont {Tagliacozzo},\ and\ \citenamefont
  {Zoller}}]{Hauke2016}%
  \BibitemOpen
  \bibfield  {author} {\bibinfo {author} {\bibfnamefont {P.}~\bibnamefont
  {Hauke}}, \bibinfo {author} {\bibfnamefont {M.}~\bibnamefont {Heyl}},
  \bibinfo {author} {\bibfnamefont {L.}~\bibnamefont {Tagliacozzo}},\ and\
  \bibinfo {author} {\bibfnamefont {P.}~\bibnamefont {Zoller}},\ }\bibfield
  {title} {\bibinfo {title} {Measuring multipartite entanglement through
  dynamic susceptibilities},\ }\href {https://doi.org/10.1038/nphys3700}
  {\bibfield  {journal} {\bibinfo  {journal} {Nature Physics}\ }\textbf
  {\bibinfo {volume} {12}},\ \bibinfo {pages} {778–782} (\bibinfo {year}
  {2016})}\BibitemShut {NoStop}%
\bibitem [{\citenamefont {Schollw\"{o}ck}(2011)}]{Schollwck2011}%
  \BibitemOpen
  \bibfield  {author} {\bibinfo {author} {\bibfnamefont {U.}~\bibnamefont
  {Schollw\"{o}ck}},\ }\bibfield  {title} {\bibinfo {title} {The density-matrix
  renormalization group in the age of matrix product states},\ }\href
  {https://doi.org/10.1016/j.aop.2010.09.012} {\bibfield  {journal} {\bibinfo
  {journal} {Annals of Physics}\ }\textbf {\bibinfo {volume} {326}},\ \bibinfo
  {pages} {96–192} (\bibinfo {year} {2011})}\BibitemShut {NoStop}%
\bibitem [{\citenamefont {Paeckel}\ \emph {et~al.}(2019)\citenamefont
  {Paeckel}, \citenamefont {K\"{o}hler}, \citenamefont {Swoboda}, \citenamefont
  {Manmana}, \citenamefont {Schollw\"{o}ck},\ and\ \citenamefont
  {Hubig}}]{Paeckel2019}%
  \BibitemOpen
  \bibfield  {author} {\bibinfo {author} {\bibfnamefont {S.}~\bibnamefont
  {Paeckel}}, \bibinfo {author} {\bibfnamefont {T.}~\bibnamefont {K\"{o}hler}},
  \bibinfo {author} {\bibfnamefont {A.}~\bibnamefont {Swoboda}}, \bibinfo
  {author} {\bibfnamefont {S.~R.}\ \bibnamefont {Manmana}}, \bibinfo {author}
  {\bibfnamefont {U.}~\bibnamefont {Schollw\"{o}ck}},\ and\ \bibinfo {author}
  {\bibfnamefont {C.}~\bibnamefont {Hubig}},\ }\bibfield  {title} {\bibinfo
  {title} {Time-evolution methods for matrix-product states},\ }\href
  {https://doi.org/10.1016/j.aop.2019.167998} {\bibfield  {journal} {\bibinfo
  {journal} {Annals of Physics}\ }\textbf {\bibinfo {volume} {411}},\ \bibinfo
  {pages} {167998} (\bibinfo {year} {2019})}\BibitemShut {NoStop}%
\bibitem [{\citenamefont {Noack}\ and\ \citenamefont
  {Manmana}(2005)}]{Noack2005}%
  \BibitemOpen
  \bibfield  {author} {\bibinfo {author} {\bibfnamefont {R.~M.}\ \bibnamefont
  {Noack}}\ and\ \bibinfo {author} {\bibfnamefont {S.~R.}\ \bibnamefont
  {Manmana}},\ }\bibfield  {title} {\bibinfo {title} {Diagonalization‐ and
  numerical renormalization‐group‐based methods for interacting quantum
  systems},\ }\href {https://doi.org/10.1063/1.2080349} {\bibfield  {journal}
  {\bibinfo  {journal} {AIP Conference Proceedings}\ }\textbf {\bibinfo
  {volume} {789}},\ \bibinfo {pages} {93} (\bibinfo {year} {2005})},\ \Eprint
  {https://arxiv.org/abs/https://pubs.aip.org/aip/acp/article-pdf/789/1/93/11444794/93\_1\_online.pdf}
  {https://pubs.aip.org/aip/acp/article-pdf/789/1/93/11444794/93\_1\_online.pdf}
  \BibitemShut {NoStop}%
\bibitem [{\citenamefont {Sandvik}(2010)}]{Sandvik2010}%
  \BibitemOpen
  \bibfield  {author} {\bibinfo {author} {\bibfnamefont {A.~W.}\ \bibnamefont
  {Sandvik}},\ }\bibfield  {title} {\bibinfo {title} {Computational studies of
  quantum spin systems},\ }\href {https://doi.org/10.1063/1.3518900} {\bibfield
   {journal} {\bibinfo  {journal} {AIP Conference Proceedings}\ }\textbf
  {\bibinfo {volume} {1297}},\ \bibinfo {pages} {135} (\bibinfo {year}
  {2010})},\ \Eprint
  {https://arxiv.org/abs/https://pubs.aip.org/aip/acp/article-pdf/1297/1/135/11407753/135\_1\_online.pdf}
  {https://pubs.aip.org/aip/acp/article-pdf/1297/1/135/11407753/135\_1\_online.pdf}
  \BibitemShut {NoStop}%
\bibitem [{\citenamefont {Weinberg}\ and\ \citenamefont
  {Bukov}(2019)}]{Weinberg2019}%
  \BibitemOpen
  \bibfield  {author} {\bibinfo {author} {\bibfnamefont {P.}~\bibnamefont
  {Weinberg}}\ and\ \bibinfo {author} {\bibfnamefont {M.}~\bibnamefont
  {Bukov}},\ }\bibfield  {title} {\bibinfo {title} {Quspin: a python package
  for dynamics and exact diagonalisation of quantum many body systems. part ii:
  bosons, fermions and higher spins},\ }\bibfield  {journal} {\bibinfo
  {journal} {SciPost Physics}\ }\textbf {\bibinfo {volume} {7}},\ \href
  {https://doi.org/10.21468/scipostphys.7.2.020} {10.21468/scipostphys.7.2.020}
  (\bibinfo {year} {2019})\BibitemShut {NoStop}%
\bibitem [{\citenamefont {Scheie}\ \emph {et~al.}(2022)\citenamefont {Scheie},
  \citenamefont {Laurell}, \citenamefont {Lake}, \citenamefont {Nagler},
  \citenamefont {Stone}, \citenamefont {Caux},\ and\ \citenamefont
  {Tennant}}]{Tennant2022}%
  \BibitemOpen
  \bibfield  {author} {\bibinfo {author} {\bibfnamefont {A.}~\bibnamefont
  {Scheie}}, \bibinfo {author} {\bibfnamefont {P.}~\bibnamefont {Laurell}},
  \bibinfo {author} {\bibfnamefont {B.}~\bibnamefont {Lake}}, \bibinfo {author}
  {\bibfnamefont {S.~E.}\ \bibnamefont {Nagler}}, \bibinfo {author}
  {\bibfnamefont {M.~B.}\ \bibnamefont {Stone}}, \bibinfo {author}
  {\bibfnamefont {J.-S.}\ \bibnamefont {Caux}},\ and\ \bibinfo {author}
  {\bibfnamefont {D.~A.}\ \bibnamefont {Tennant}},\ }\bibfield  {title}
  {\bibinfo {title} {Quantum wake dynamics in heisenberg antiferromagnetic
  chains},\ }\bibfield  {journal} {\bibinfo  {journal} {Nature Communications}\
  }\textbf {\bibinfo {volume} {13}},\ \href
  {https://doi.org/10.1038/s41467-022-33571-8} {10.1038/s41467-022-33571-8}
  (\bibinfo {year} {2022})\BibitemShut {NoStop}%
\bibitem [{\citenamefont {Scheie}\ \emph {et~al.}(2021)\citenamefont {Scheie},
  \citenamefont {Laurell}, \citenamefont {Samarakoon}, \citenamefont {Lake},
  \citenamefont {Nagler}, \citenamefont {Granroth}, \citenamefont {Okamoto},
  \citenamefont {Alvarez},\ and\ \citenamefont {Tennant}}]{Scheie2021}%
  \BibitemOpen
  \bibfield  {author} {\bibinfo {author} {\bibfnamefont {A.}~\bibnamefont
  {Scheie}}, \bibinfo {author} {\bibfnamefont {P.}~\bibnamefont {Laurell}},
  \bibinfo {author} {\bibfnamefont {A.~M.}\ \bibnamefont {Samarakoon}},
  \bibinfo {author} {\bibfnamefont {B.}~\bibnamefont {Lake}}, \bibinfo {author}
  {\bibfnamefont {S.~E.}\ \bibnamefont {Nagler}}, \bibinfo {author}
  {\bibfnamefont {G.~E.}\ \bibnamefont {Granroth}}, \bibinfo {author}
  {\bibfnamefont {S.}~\bibnamefont {Okamoto}}, \bibinfo {author} {\bibfnamefont
  {G.}~\bibnamefont {Alvarez}},\ and\ \bibinfo {author} {\bibfnamefont {D.~A.}\
  \bibnamefont {Tennant}},\ }\bibfield  {title} {\bibinfo {title} {Witnessing
  entanglement in quantum magnets using neutron scattering},\ }\bibfield
  {journal} {\bibinfo  {journal} {Physical Review B}\ }\textbf {\bibinfo
  {volume} {103}},\ \href {https://doi.org/10.1103/physrevb.103.224434}
  {10.1103/physrevb.103.224434} (\bibinfo {year} {2021})\BibitemShut {NoStop}%
\bibitem [{\citenamefont {Scheie}\ \emph {et~al.}(2025)\citenamefont {Scheie},
  \citenamefont {Laurell}, \citenamefont {Simeth}, \citenamefont {Dagotto},\
  and\ \citenamefont {Tennant}}]{ScheieReview2025}%
  \BibitemOpen
  \bibfield  {author} {\bibinfo {author} {\bibfnamefont {A.}~\bibnamefont
  {Scheie}}, \bibinfo {author} {\bibfnamefont {P.}~\bibnamefont {Laurell}},
  \bibinfo {author} {\bibfnamefont {W.}~\bibnamefont {Simeth}}, \bibinfo
  {author} {\bibfnamefont {E.}~\bibnamefont {Dagotto}},\ and\ \bibinfo {author}
  {\bibfnamefont {D.~A.}\ \bibnamefont {Tennant}},\ }\bibfield  {title}
  {\bibinfo {title} {Tutorial: Extracting entanglement signatures from neutron
  spectroscopy},\ }\href
  {https://doi.org/https://doi.org/10.1016/j.mtquan.2024.100020} {\bibfield
  {journal} {\bibinfo  {journal} {Materials Today Quantum}\ }\textbf {\bibinfo
  {volume} {5}},\ \bibinfo {pages} {100020} (\bibinfo {year}
  {2025})}\BibitemShut {NoStop}%
\bibitem [{\citenamefont {Mazza}\ \emph {et~al.}(2024)\citenamefont {Mazza}
  \emph {et~al.}}]{Assaad2024}%
  \BibitemOpen
  \bibfield  {author} {\bibinfo {author} {\bibfnamefont {F.}~\bibnamefont
  {Mazza}} \emph {et~al.},\ }\bibfield  {title} {\bibinfo {title} {Quantum
  fisher information in a strange metal},\ }\href@noop {} {\bibfield  {journal}
  {\bibinfo  {journal} {arXiv}\ }\textbf {\bibinfo {volume} {2403}},\ \bibinfo
  {pages} {12779} (\bibinfo {year} {2024})},\ \bibinfo {note} {preprint,
  available at \url{https://arxiv.org/abs/2403.12779}},\ \Eprint
  {https://arxiv.org/abs/2403.12779} {arXiv:2403.12779 [cond-mat.str-el]}
  \BibitemShut {NoStop}%
\bibitem [{\citenamefont {Malla}\ \emph {et~al.}(2024)\citenamefont {Malla},
  \citenamefont {Weichselbaum}, \citenamefont {Wei},\ and\ \citenamefont
  {Konik}}]{Malla2024}%
  \BibitemOpen
  \bibfield  {author} {\bibinfo {author} {\bibfnamefont {R.~K.}\ \bibnamefont
  {Malla}}, \bibinfo {author} {\bibfnamefont {A.}~\bibnamefont {Weichselbaum}},
  \bibinfo {author} {\bibfnamefont {T.-C.}\ \bibnamefont {Wei}},\ and\ \bibinfo
  {author} {\bibfnamefont {R.~M.}\ \bibnamefont {Konik}},\ }\href
  {https://arxiv.org/abs/2310.05870} {\bibinfo {title} {Detecting multipartite
  entanglement patterns using single particle green's functions}} (\bibinfo
  {year} {2024}),\ \Eprint {https://arxiv.org/abs/2310.05870} {arXiv:2310.05870
  [quant-ph]} \BibitemShut {NoStop}%
\bibitem [{\citenamefont {Hales}\ \emph {et~al.}(2023)\citenamefont {Hales},
  \citenamefont {Bajpai}, \citenamefont {Liu}, \citenamefont {Baykusheva},
  \citenamefont {Li}, \citenamefont {Mitrano},\ and\ \citenamefont
  {Wang}}]{Hales2023}%
  \BibitemOpen
  \bibfield  {author} {\bibinfo {author} {\bibfnamefont {J.}~\bibnamefont
  {Hales}}, \bibinfo {author} {\bibfnamefont {U.}~\bibnamefont {Bajpai}},
  \bibinfo {author} {\bibfnamefont {T.}~\bibnamefont {Liu}}, \bibinfo {author}
  {\bibfnamefont {D.~R.}\ \bibnamefont {Baykusheva}}, \bibinfo {author}
  {\bibfnamefont {M.}~\bibnamefont {Li}}, \bibinfo {author} {\bibfnamefont
  {M.}~\bibnamefont {Mitrano}},\ and\ \bibinfo {author} {\bibfnamefont
  {Y.}~\bibnamefont {Wang}},\ }\bibfield  {title} {\bibinfo {title} {Witnessing
  light-driven entanglement using time-resolved resonant inelastic x-ray
  scattering},\ }\bibfield  {journal} {\bibinfo  {journal} {Nature
  Communications}\ }\textbf {\bibinfo {volume} {14}},\ \href
  {https://doi.org/10.1038/s41467-023-38540-3} {10.1038/s41467-023-38540-3}
  (\bibinfo {year} {2023})\BibitemShut {NoStop}%
\bibitem [{\citenamefont {Baykusheva}\ \emph {et~al.}(2023)\citenamefont
  {Baykusheva}, \citenamefont {Kalthoff}, \citenamefont {Hofmann},
  \citenamefont {Claassen}, \citenamefont {Kennes}, \citenamefont {Sentef},\
  and\ \citenamefont {Mitrano}}]{Baykusheva2023}%
  \BibitemOpen
  \bibfield  {author} {\bibinfo {author} {\bibfnamefont {D.~R.}\ \bibnamefont
  {Baykusheva}}, \bibinfo {author} {\bibfnamefont {M.~H.}\ \bibnamefont
  {Kalthoff}}, \bibinfo {author} {\bibfnamefont {D.}~\bibnamefont {Hofmann}},
  \bibinfo {author} {\bibfnamefont {M.}~\bibnamefont {Claassen}}, \bibinfo
  {author} {\bibfnamefont {D.~M.}\ \bibnamefont {Kennes}}, \bibinfo {author}
  {\bibfnamefont {M.~A.}\ \bibnamefont {Sentef}},\ and\ \bibinfo {author}
  {\bibfnamefont {M.}~\bibnamefont {Mitrano}},\ }\bibfield  {title} {\bibinfo
  {title} {Witnessing nonequilibrium entanglement dynamics in a strongly
  correlated fermionic chain},\ }\bibfield  {journal} {\bibinfo  {journal}
  {Physical Review Letters}\ }\textbf {\bibinfo {volume} {130}},\ \href
  {https://doi.org/10.1103/physrevlett.130.106902}
  {10.1103/physrevlett.130.106902} (\bibinfo {year} {2023})\BibitemShut
  {NoStop}%
\bibitem [{\citenamefont {Peierls}(1933)}]{Peierls1933}%
  \BibitemOpen
  \bibfield  {author} {\bibinfo {author} {\bibfnamefont {R.}~\bibnamefont
  {Peierls}},\ }\bibfield  {title} {\bibinfo {title} {Zur theorie des
  diamagnetismus von leitungselektronen},\ }\href
  {https://doi.org/10.1007/bf01342591} {\bibfield  {journal} {\bibinfo
  {journal} {Zeitschrift f\"ur Physik}\ }\textbf {\bibinfo {volume} {80}},\
  \bibinfo {pages} {763–791} (\bibinfo {year} {1933})}\BibitemShut {NoStop}%
\bibitem [{\citenamefont {Giamarchi}(2004)}]{giamarchi2004quantum}%
  \BibitemOpen
  \bibfield  {author} {\bibinfo {author} {\bibfnamefont {T.}~\bibnamefont
  {Giamarchi}},\ }\href {https://books.google.de/books?id=1MwTDAAAQBAJ} {\emph
  {\bibinfo {title} {Quantum Physics in One Dimension}}},\ International Series
  of Monographs on Physics\ (\bibinfo  {publisher} {Clarendon Press},\ \bibinfo
  {year} {2004})\BibitemShut {NoStop}%
\bibitem [{\citenamefont {Cloizeaux}\ and\ \citenamefont
  {Gaudin}(1966)}]{DesCloizeaux1966}%
  \BibitemOpen
  \bibfield  {author} {\bibinfo {author} {\bibfnamefont {J.~D.}\ \bibnamefont
  {Cloizeaux}}\ and\ \bibinfo {author} {\bibfnamefont {M.}~\bibnamefont
  {Gaudin}},\ }\bibfield  {title} {\bibinfo {title} {Anisotropic linear
  magnetic chain},\ }\href {https://doi.org/10.1063/1.1705048} {\bibfield
  {journal} {\bibinfo  {journal} {Journal of Mathematical Physics}\ }\textbf
  {\bibinfo {volume} {7}},\ \bibinfo {pages} {1384} (\bibinfo {year}
  {1966})}\BibitemShut {NoStop}%
\bibitem [{\citenamefont {Meyer}(2022)}]{PhdConstantin}%
  \BibitemOpen
  \bibfield  {author} {\bibinfo {author} {\bibfnamefont {C.}~\bibnamefont
  {Meyer}},\ }\href {https://doi.org/10.53846/goediss-9286} {\emph {\bibinfo
  {title} {Matrix Product State Approaches to Non\--equilibrium\- Spectral
  Quantities of Strongly Correlated Fermions in One Dimension}}}\ (\bibinfo
  {publisher} {University of G\"ottingen},\ \bibinfo {year} {2022})\BibitemShut
  {NoStop}%
\bibitem [{\citenamefont {Kalthoff}\ \emph {et~al.}(2018)\citenamefont
  {Kalthoff}, \citenamefont {Uhrig},\ and\ \citenamefont
  {Freericks}}]{Kalthoff2018}%
  \BibitemOpen
  \bibfield  {author} {\bibinfo {author} {\bibfnamefont {M.~H.}\ \bibnamefont
  {Kalthoff}}, \bibinfo {author} {\bibfnamefont {G.~S.}\ \bibnamefont
  {Uhrig}},\ and\ \bibinfo {author} {\bibfnamefont {J.~K.}\ \bibnamefont
  {Freericks}},\ }\bibfield  {title} {\bibinfo {title} {Emergence of floquet
  behavior for lattice fermions driven by light pulses},\ }\href
  {https://doi.org/10.1103/PhysRevB.98.035138} {\bibfield  {journal} {\bibinfo
  {journal} {Phys. Rev. B}\ }\textbf {\bibinfo {volume} {98}},\ \bibinfo
  {pages} {035138} (\bibinfo {year} {2018})}\BibitemShut {NoStop}%
\bibitem [{\citenamefont {Nghiem}\ and\ \citenamefont
  {Costi}(2017)}]{NghiemPRL2017}%
  \BibitemOpen
  \bibfield  {author} {\bibinfo {author} {\bibfnamefont {H.~T.~M.}\
  \bibnamefont {Nghiem}}\ and\ \bibinfo {author} {\bibfnamefont {T.~A.}\
  \bibnamefont {Costi}},\ }\bibfield  {title} {\bibinfo {title} {Time evolution
  of the kondo resonance in response to a quench},\ }\href
  {https://doi.org/10.1103/PhysRevLett.119.156601} {\bibfield  {journal}
  {\bibinfo  {journal} {Phys. Rev. Lett.}\ }\textbf {\bibinfo {volume} {119}},\
  \bibinfo {pages} {156601} (\bibinfo {year} {2017})}\BibitemShut {NoStop}%
\bibitem [{\citenamefont {Nghiem}\ \emph {et~al.}(2020)\citenamefont {Nghiem},
  \citenamefont {Dang},\ and\ \citenamefont {Costi}}]{Nghiem2020}%
  \BibitemOpen
  \bibfield  {author} {\bibinfo {author} {\bibfnamefont {H.~T.~M.}\
  \bibnamefont {Nghiem}}, \bibinfo {author} {\bibfnamefont {H.~T.}\
  \bibnamefont {Dang}},\ and\ \bibinfo {author} {\bibfnamefont {T.~A.}\
  \bibnamefont {Costi}},\ }\bibfield  {title} {\bibinfo {title} {Time-dependent
  spectral functions of the anderson impurity model in response to a quench
  with application to time-resolved photoemission spectroscopy},\ }\href
  {https://doi.org/10.1103/PhysRevB.101.115117} {\bibfield  {journal} {\bibinfo
   {journal} {Phys. Rev. B}\ }\textbf {\bibinfo {volume} {101}},\ \bibinfo
  {pages} {115117} (\bibinfo {year} {2020})}\BibitemShut {NoStop}%
\bibitem [{\citenamefont {Uhrig}\ \emph {et~al.}(2019)\citenamefont {Uhrig},
  \citenamefont {Kalthoff},\ and\ \citenamefont {Freericks}}]{UhrigPRL2019}%
  \BibitemOpen
  \bibfield  {author} {\bibinfo {author} {\bibfnamefont {G.~S.}\ \bibnamefont
  {Uhrig}}, \bibinfo {author} {\bibfnamefont {M.~H.}\ \bibnamefont
  {Kalthoff}},\ and\ \bibinfo {author} {\bibfnamefont {J.~K.}\ \bibnamefont
  {Freericks}},\ }\bibfield  {title} {\bibinfo {title} {Positivity of the
  spectral densities of retarded floquet green functions},\ }\href
  {https://doi.org/10.1103/PhysRevLett.122.130604} {\bibfield  {journal}
  {\bibinfo  {journal} {Phys. Rev. Lett.}\ }\textbf {\bibinfo {volume} {122}},\
  \bibinfo {pages} {130604} (\bibinfo {year} {2019})}\BibitemShut {NoStop}%
\bibitem [{\citenamefont {Blum}\ \emph {et~al.}(2025)\citenamefont {Blum},
  \citenamefont {Noack},\ and\ \citenamefont {Manmana}}]{Blum2025}%
  \BibitemOpen
  \bibfield  {author} {\bibinfo {author} {\bibfnamefont {T.}~\bibnamefont
  {Blum}}, \bibinfo {author} {\bibfnamefont {R.~M.}\ \bibnamefont {Noack}},\
  and\ \bibinfo {author} {\bibfnamefont {S.~R.}\ \bibnamefont {Manmana}},\
  }\bibfield  {title} {\bibinfo {title} {Time evolution of the local density of
  states of strongly correlated fermions coupled to a nanoprobe},\ }\href
  {https://doi.org/10.1103/PhysRevB.111.035152} {\bibfield  {journal} {\bibinfo
   {journal} {Phys. Rev. B}\ }\textbf {\bibinfo {volume} {111}},\ \bibinfo
  {pages} {035152} (\bibinfo {year} {2025})}\BibitemShut {NoStop}%
\bibitem [{\citenamefont {Manmana}\ \emph {et~al.}(2005)\citenamefont
  {Manmana}, \citenamefont {Muramatsu},\ and\ \citenamefont
  {Noack}}]{Manmana2005}%
  \BibitemOpen
  \bibfield  {author} {\bibinfo {author} {\bibfnamefont {S.~R.}\ \bibnamefont
  {Manmana}}, \bibinfo {author} {\bibfnamefont {A.}~\bibnamefont {Muramatsu}},\
  and\ \bibinfo {author} {\bibfnamefont {R.~M.}\ \bibnamefont {Noack}},\
  }\bibfield  {title} {\bibinfo {title} {Time evolution of one‐dimensional
  quantum many body systems},\ }\href {https://doi.org/10.1063/1.2080353}
  {\bibfield  {journal} {\bibinfo  {journal} {AIP Conference Proceedings}\
  }\textbf {\bibinfo {volume} {789}},\ \bibinfo {pages} {269} (\bibinfo {year}
  {2005})},\ \Eprint
  {https://arxiv.org/abs/https://pubs.aip.org/aip/acp/article-pdf/789/1/269/11444782/269\_1\_online.pdf}
  {https://pubs.aip.org/aip/acp/article-pdf/789/1/269/11444782/269\_1\_online.pdf}
  \BibitemShut {NoStop}%
\bibitem [{\citenamefont {Pereira}\ \emph {et~al.}(2009)\citenamefont
  {Pereira}, \citenamefont {White},\ and\ \citenamefont
  {Affleck}}]{Pereira2009}%
  \BibitemOpen
  \bibfield  {author} {\bibinfo {author} {\bibfnamefont {R.~G.}\ \bibnamefont
  {Pereira}}, \bibinfo {author} {\bibfnamefont {S.~R.}\ \bibnamefont {White}},\
  and\ \bibinfo {author} {\bibfnamefont {I.}~\bibnamefont {Affleck}},\
  }\bibfield  {title} {\bibinfo {title} {Spectral function of spinless fermions
  on a one-dimensional lattice},\ }\bibfield  {journal} {\bibinfo  {journal}
  {Physical Review B}\ }\textbf {\bibinfo {volume} {79}},\ \href
  {https://doi.org/10.1103/physrevb.79.165113} {10.1103/physrevb.79.165113}
  (\bibinfo {year} {2009})\BibitemShut {NoStop}%
\bibitem [{\citenamefont {Ponte}\ \emph
  {et~al.}(2015{\natexlab{b}})\citenamefont {Ponte}, \citenamefont {Chandran},
  \citenamefont {Papić},\ and\ \citenamefont {Abanin}}]{Ponte2015-2}%
  \BibitemOpen
  \bibfield  {author} {\bibinfo {author} {\bibfnamefont {P.}~\bibnamefont
  {Ponte}}, \bibinfo {author} {\bibfnamefont {A.}~\bibnamefont {Chandran}},
  \bibinfo {author} {\bibfnamefont {Z.}~\bibnamefont {Papić}},\ and\ \bibinfo
  {author} {\bibfnamefont {D.~A.}\ \bibnamefont {Abanin}},\ }\bibfield  {title}
  {\bibinfo {title} {Periodically driven ergodic and many-body localized
  quantum systems},\ }\href {https://doi.org/10.1016/j.aop.2014.11.008}
  {\bibfield  {journal} {\bibinfo  {journal} {Annals of Physics}\ }\textbf
  {\bibinfo {volume} {353}},\ \bibinfo {pages} {196–204} (\bibinfo {year}
  {2015}{\natexlab{b}})}\BibitemShut {NoStop}%
\bibitem [{\citenamefont {Moessner}\ and\ \citenamefont
  {Sondhi}(2017)}]{Moessner2017}%
  \BibitemOpen
  \bibfield  {author} {\bibinfo {author} {\bibfnamefont {R.}~\bibnamefont
  {Moessner}}\ and\ \bibinfo {author} {\bibfnamefont {S.~L.}\ \bibnamefont
  {Sondhi}},\ }\bibfield  {title} {\bibinfo {title} {Equilibration and order in
  quantum floquet matter},\ }\href {https://doi.org/10.1038/nphys4106}
  {\bibfield  {journal} {\bibinfo  {journal} {Nature Physics}\ }\textbf
  {\bibinfo {volume} {13}},\ \bibinfo {pages} {424–428} (\bibinfo {year}
  {2017})}\BibitemShut {NoStop}%
\bibitem [{\citenamefont {Bordia}\ \emph {et~al.}(2017)\citenamefont {Bordia},
  \citenamefont {L\"{u}schen}, \citenamefont {Schneider}, \citenamefont
  {Knap},\ and\ \citenamefont {Bloch}}]{Bordia2017}%
  \BibitemOpen
  \bibfield  {author} {\bibinfo {author} {\bibfnamefont {P.}~\bibnamefont
  {Bordia}}, \bibinfo {author} {\bibfnamefont {H.}~\bibnamefont {L\"{u}schen}},
  \bibinfo {author} {\bibfnamefont {U.}~\bibnamefont {Schneider}}, \bibinfo
  {author} {\bibfnamefont {M.}~\bibnamefont {Knap}},\ and\ \bibinfo {author}
  {\bibfnamefont {I.}~\bibnamefont {Bloch}},\ }\bibfield  {title} {\bibinfo
  {title} {Periodically driving a many-body localized quantum system},\ }\href
  {https://doi.org/10.1038/nphys4020} {\bibfield  {journal} {\bibinfo
  {journal} {Nature Physics}\ }\textbf {\bibinfo {volume} {13}},\ \bibinfo
  {pages} {460–464} (\bibinfo {year} {2017})}\BibitemShut {NoStop}%
\bibitem [{\citenamefont {Abanin}\ \emph {et~al.}(2017)\citenamefont {Abanin},
  \citenamefont {De~Roeck}, \citenamefont {Ho},\ and\ \citenamefont
  {Huveneers}}]{Abanin2017}%
  \BibitemOpen
  \bibfield  {author} {\bibinfo {author} {\bibfnamefont {D.~A.}\ \bibnamefont
  {Abanin}}, \bibinfo {author} {\bibfnamefont {W.}~\bibnamefont {De~Roeck}},
  \bibinfo {author} {\bibfnamefont {W.~W.}\ \bibnamefont {Ho}},\ and\ \bibinfo
  {author} {\bibfnamefont {F.}~\bibnamefont {Huveneers}},\ }\bibfield  {title}
  {\bibinfo {title} {Effective hamiltonians, prethermalization, and slow energy
  absorption in periodically driven many-body systems},\ }\bibfield  {journal}
  {\bibinfo  {journal} {Physical Review B}\ }\textbf {\bibinfo {volume} {95}},\
  \href {https://doi.org/10.1103/physrevb.95.014112}
  {10.1103/physrevb.95.014112} (\bibinfo {year} {2017})\BibitemShut {NoStop}%
\bibitem [{\citenamefont {Park}(2014)}]{Park2014}%
  \BibitemOpen
  \bibfield  {author} {\bibinfo {author} {\bibfnamefont {S.~T.}\ \bibnamefont
  {Park}},\ }\bibfield  {title} {\bibinfo {title} {Interference in
  floquet-volkov transitions},\ }\bibfield  {journal} {\bibinfo  {journal}
  {Physical Review A}\ }\textbf {\bibinfo {volume} {90}},\ \href
  {https://doi.org/10.1103/physreva.90.013420} {10.1103/physreva.90.013420}
  (\bibinfo {year} {2014})\BibitemShut {NoStop}%
\bibitem [{\citenamefont {Lieb}\ and\ \citenamefont
  {Robinson}(1972)}]{Lieb1972}%
  \BibitemOpen
  \bibfield  {author} {\bibinfo {author} {\bibfnamefont {E.~H.}\ \bibnamefont
  {Lieb}}\ and\ \bibinfo {author} {\bibfnamefont {D.~W.}\ \bibnamefont
  {Robinson}},\ }\bibfield  {title} {\bibinfo {title} {The finite group
  velocity of quantum spin systems},\ }\href
  {https://doi.org/10.1007/bf01645779} {\bibfield  {journal} {\bibinfo
  {journal} {Communications in Mathematical Physics}\ }\textbf {\bibinfo
  {volume} {28}},\ \bibinfo {pages} {251–257} (\bibinfo {year}
  {1972})}\BibitemShut {NoStop}%
\bibitem [{\citenamefont {Calabrese}\ and\ \citenamefont
  {Cardy}(2006)}]{Calabrese2006}%
  \BibitemOpen
  \bibfield  {author} {\bibinfo {author} {\bibfnamefont {P.}~\bibnamefont
  {Calabrese}}\ and\ \bibinfo {author} {\bibfnamefont {J.}~\bibnamefont
  {Cardy}},\ }\bibfield  {title} {\bibinfo {title} {Time dependence of
  correlation functions following a quantum quench},\ }\href
  {https://doi.org/10.1103/PhysRevLett.96.136801} {\bibfield  {journal}
  {\bibinfo  {journal} {Phys. Rev. Lett.}\ }\textbf {\bibinfo {volume} {96}},\
  \bibinfo {pages} {136801} (\bibinfo {year} {2006})}\BibitemShut {NoStop}%
\bibitem [{\citenamefont {Calabrese}\ and\ \citenamefont
  {Cardy}(2007)}]{Calabrese2007}%
  \BibitemOpen
  \bibfield  {author} {\bibinfo {author} {\bibfnamefont {P.}~\bibnamefont
  {Calabrese}}\ and\ \bibinfo {author} {\bibfnamefont {J.}~\bibnamefont
  {Cardy}},\ }\bibfield  {title} {\bibinfo {title} {Quantum quenches in
  extended systems},\ }\href {https://doi.org/10.1088/1742-5468/2007/06/p06008}
  {\bibfield  {journal} {\bibinfo  {journal} {Journal of Statistical Mechanics:
  Theory and Experiment}\ }\textbf {\bibinfo {volume} {2007}},\ \bibinfo
  {pages} {P06008–P06008} (\bibinfo {year} {2007})}\BibitemShut {NoStop}%
\bibitem [{\citenamefont {Bravyi}\ \emph {et~al.}(2006)\citenamefont {Bravyi},
  \citenamefont {Hastings},\ and\ \citenamefont {Verstraete}}]{Bravyi2006}%
  \BibitemOpen
  \bibfield  {author} {\bibinfo {author} {\bibfnamefont {S.}~\bibnamefont
  {Bravyi}}, \bibinfo {author} {\bibfnamefont {M.~B.}\ \bibnamefont
  {Hastings}},\ and\ \bibinfo {author} {\bibfnamefont {F.}~\bibnamefont
  {Verstraete}},\ }\bibfield  {title} {\bibinfo {title} {Lieb-robinson bounds
  and the generation of correlations and topological quantum order},\ }\href
  {https://doi.org/10.1103/PhysRevLett.97.050401} {\bibfield  {journal}
  {\bibinfo  {journal} {Phys. Rev. Lett.}\ }\textbf {\bibinfo {volume} {97}},\
  \bibinfo {pages} {050401} (\bibinfo {year} {2006})}\BibitemShut {NoStop}%
\bibitem [{\citenamefont {L\"{a}uchli}\ and\ \citenamefont
  {Kollath}(2008)}]{Laeuchli2008}%
  \BibitemOpen
  \bibfield  {author} {\bibinfo {author} {\bibfnamefont {A.~M.}\ \bibnamefont
  {L\"{a}uchli}}\ and\ \bibinfo {author} {\bibfnamefont {C.}~\bibnamefont
  {Kollath}},\ }\bibfield  {title} {\bibinfo {title} {Spreading of correlations
  and entanglement after a quench in the one-dimensional bose–hubbard
  model},\ }\href {https://doi.org/10.1088/1742-5468/2008/05/p05018} {\bibfield
   {journal} {\bibinfo  {journal} {Journal of Statistical Mechanics: Theory and
  Experiment}\ }\textbf {\bibinfo {volume} {2008}},\ \bibinfo {pages} {P05018}
  (\bibinfo {year} {2008})}\BibitemShut {NoStop}%
\bibitem [{\citenamefont {Manmana}\ \emph {et~al.}(2009)\citenamefont
  {Manmana}, \citenamefont {Wessel}, \citenamefont {Noack},\ and\ \citenamefont
  {Muramatsu}}]{Manmana2009}%
  \BibitemOpen
  \bibfield  {author} {\bibinfo {author} {\bibfnamefont {S.~R.}\ \bibnamefont
  {Manmana}}, \bibinfo {author} {\bibfnamefont {S.}~\bibnamefont {Wessel}},
  \bibinfo {author} {\bibfnamefont {R.~M.}\ \bibnamefont {Noack}},\ and\
  \bibinfo {author} {\bibfnamefont {A.}~\bibnamefont {Muramatsu}},\ }\bibfield
  {title} {\bibinfo {title} {Time evolution of correlations in strongly
  interacting fermions after a quantum quench},\ }\href
  {https://doi.org/10.1103/PhysRevB.79.155104} {\bibfield  {journal} {\bibinfo
  {journal} {Phys. Rev. B}\ }\textbf {\bibinfo {volume} {79}},\ \bibinfo
  {pages} {155104} (\bibinfo {year} {2009})}\BibitemShut {NoStop}%
\bibitem [{\citenamefont {Cheneau}\ \emph {et~al.}(2012)\citenamefont
  {Cheneau}, \citenamefont {Barmettler}, \citenamefont {Poletti}, \citenamefont
  {Endres}, \citenamefont {Schauß}, \citenamefont {Fukuhara}, \citenamefont
  {Gross}, \citenamefont {Bloch}, \citenamefont {Kollath},\ and\ \citenamefont
  {Kuhr}}]{Cheneau2012}%
  \BibitemOpen
  \bibfield  {author} {\bibinfo {author} {\bibfnamefont {M.}~\bibnamefont
  {Cheneau}}, \bibinfo {author} {\bibfnamefont {P.}~\bibnamefont {Barmettler}},
  \bibinfo {author} {\bibfnamefont {D.}~\bibnamefont {Poletti}}, \bibinfo
  {author} {\bibfnamefont {M.}~\bibnamefont {Endres}}, \bibinfo {author}
  {\bibfnamefont {P.}~\bibnamefont {Schauß}}, \bibinfo {author} {\bibfnamefont
  {T.}~\bibnamefont {Fukuhara}}, \bibinfo {author} {\bibfnamefont
  {C.}~\bibnamefont {Gross}}, \bibinfo {author} {\bibfnamefont
  {I.}~\bibnamefont {Bloch}}, \bibinfo {author} {\bibfnamefont
  {C.}~\bibnamefont {Kollath}},\ and\ \bibinfo {author} {\bibfnamefont
  {S.}~\bibnamefont {Kuhr}},\ }\bibfield  {title} {\bibinfo {title}
  {Light-cone-like spreading of correlations in a quantum many-body system},\
  }\href {https://doi.org/10.1038/nature10748} {\bibfield  {journal} {\bibinfo
  {journal} {Nature}\ }\textbf {\bibinfo {volume} {481}},\ \bibinfo {pages}
  {484–487} (\bibinfo {year} {2012})}\BibitemShut {NoStop}%
\bibitem [{\citenamefont {Kalthoff}\ \emph {et~al.}(2019)\citenamefont
  {Kalthoff}, \citenamefont {Kennes},\ and\ \citenamefont
  {Sentef}}]{Kalthoff2019}%
  \BibitemOpen
  \bibfield  {author} {\bibinfo {author} {\bibfnamefont {M.~H.}\ \bibnamefont
  {Kalthoff}}, \bibinfo {author} {\bibfnamefont {D.~M.}\ \bibnamefont
  {Kennes}},\ and\ \bibinfo {author} {\bibfnamefont {M.~A.}\ \bibnamefont
  {Sentef}},\ }\bibfield  {title} {\bibinfo {title} {Floquet-engineered
  light-cone spreading of correlations in a driven quantum chain},\ }\bibfield
  {journal} {\bibinfo  {journal} {Physical Review B}\ }\textbf {\bibinfo
  {volume} {100}},\ \href {https://doi.org/10.1103/physrevb.100.165125}
  {10.1103/physrevb.100.165125} (\bibinfo {year} {2019})\BibitemShut {NoStop}%
\bibitem [{\citenamefont {Fleckenstein}\ and\ \citenamefont
  {Bukov}(2021{\natexlab{b}})}]{Fleckenstein2021_Noise}%
  \BibitemOpen
  \bibfield  {author} {\bibinfo {author} {\bibfnamefont {C.}~\bibnamefont
  {Fleckenstein}}\ and\ \bibinfo {author} {\bibfnamefont {M.}~\bibnamefont
  {Bukov}},\ }\bibfield  {title} {\bibinfo {title} {Thermalization and
  prethermalization in periodically kicked quantum spin chains},\ }\bibfield
  {journal} {\bibinfo  {journal} {Physical Review B}\ }\textbf {\bibinfo
  {volume} {103}},\ \href {https://doi.org/10.1103/physrevb.103.144307}
  {10.1103/physrevb.103.144307} (\bibinfo {year}
  {2021}{\natexlab{b}})\BibitemShut {NoStop}%
\bibitem [{\citenamefont {Dagotto}(1994)}]{Dagotto1994}%
  \BibitemOpen
  \bibfield  {author} {\bibinfo {author} {\bibfnamefont {E.}~\bibnamefont
  {Dagotto}},\ }\bibfield  {title} {\bibinfo {title} {Correlated electrons in
  high-temperature superconductors},\ }\href@noop {} {\bibfield  {journal}
  {\bibinfo  {journal} {Rev. Mod. Phys.}\ }\textbf {\bibinfo {volume} {66}},\
  \bibinfo {pages} {763} (\bibinfo {year} {1994})}\BibitemShut {NoStop}%
\bibitem [{\citenamefont {Gadge}\ and\ \citenamefont
  {Manmana}(2025)}]{gadge_2025_15115513}%
  \BibitemOpen
  \bibfield  {author} {\bibinfo {author} {\bibfnamefont {K.}~\bibnamefont
  {Gadge}}\ and\ \bibinfo {author} {\bibfnamefont {S.~R.}\ \bibnamefont
  {Manmana}},\ }\bibfield  {title} {\bibinfo {title} {Stability of floquet
  sidebands and quantum coherence in 1d strongly interacting spinless
  fermions},\ }\href {https://doi.org/10.5281/zenodo.15115513}
  {10.5281/zenodo.15115513} (\bibinfo {year} {2025})\BibitemShut {NoStop}%
\bibitem [{\citenamefont {Benthien}\ \emph {et~al.}(2004)\citenamefont
  {Benthien}, \citenamefont {Gebhard},\ and\ \citenamefont
  {Jeckelmann}}]{Benthien2004}%
  \BibitemOpen
  \bibfield  {author} {\bibinfo {author} {\bibfnamefont {H.}~\bibnamefont
  {Benthien}}, \bibinfo {author} {\bibfnamefont {F.}~\bibnamefont {Gebhard}},\
  and\ \bibinfo {author} {\bibfnamefont {E.}~\bibnamefont {Jeckelmann}},\
  }\bibfield  {title} {\bibinfo {title} {Spectral function of the
  one-dimensional hubbard model away from half filling},\ }\href
  {https://doi.org/10.1103/PhysRevLett.92.256401} {\bibfield  {journal}
  {\bibinfo  {journal} {Phys. Rev. Lett.}\ }\textbf {\bibinfo {volume} {92}},\
  \bibinfo {pages} {256401} (\bibinfo {year} {2004})}\BibitemShut {NoStop}%
\bibitem [{\citenamefont {Tiegel}(2016)}]{Tiegel2016}%
  \BibitemOpen
  \bibfield  {author} {\bibinfo {author} {\bibfnamefont {A.~C.}\ \bibnamefont
  {Tiegel}},\ }\emph {\bibinfo {title} {Finite-temperature dynamics of
  low-dimensional quantum systems with DMRG methods}},\ \href
  {https://doi.org/10.53846/goediss-5809} {Ph.D. thesis},\ \bibinfo  {school}
  {University Goettingen Repository} (\bibinfo {year} {2016})\BibitemShut
  {NoStop}%
\bibitem [{\citenamefont {K\"ohler}\ \emph {et~al.}(2020)\citenamefont
  {K\"ohler}, \citenamefont {Paeckel}, \citenamefont {Meyer},\ and\
  \citenamefont {Manmana}}]{Khler2020}%
  \BibitemOpen
  \bibfield  {author} {\bibinfo {author} {\bibfnamefont {T.}~\bibnamefont
  {K\"ohler}}, \bibinfo {author} {\bibfnamefont {S.}~\bibnamefont {Paeckel}},
  \bibinfo {author} {\bibfnamefont {C.}~\bibnamefont {Meyer}},\ and\ \bibinfo
  {author} {\bibfnamefont {S.~R.}\ \bibnamefont {Manmana}},\ }\bibfield
  {title} {\bibinfo {title} {Formation of spatial patterns by spin-selective
  excitations of interacting fermions},\ }\bibfield  {journal} {\bibinfo
  {journal} {Physical Review B}\ }\textbf {\bibinfo {volume} {102}},\ \href
  {https://doi.org/10.1103/physrevb.102.235166} {10.1103/physrevb.102.235166}
  (\bibinfo {year} {2020})\BibitemShut {NoStop}%
\bibitem [{\citenamefont {Paeckel}\ and\ \citenamefont {Köhler}()}]{SymMPS}%
  \BibitemOpen
  \bibfield  {author} {\bibinfo {author} {\bibfnamefont {S.}~\bibnamefont
  {Paeckel}}\ and\ \bibinfo {author} {\bibfnamefont {T.}~\bibnamefont
  {Köhler}},\ }\href@noop {} {\bibinfo {title} {Symmps toolkit ---
  symmps.eu}},\ \bibinfo {howpublished} {\url{https://www.symmps.eu}},\
  \bibinfo {note} {[Accessed 23-07-2024]}\BibitemShut {NoStop}%
\bibitem [{\citenamefont {Alvermann}\ and\ \citenamefont
  {Fehske}(2011)}]{ALVERMANN2011}%
  \BibitemOpen
  \bibfield  {author} {\bibinfo {author} {\bibfnamefont {A.}~\bibnamefont
  {Alvermann}}\ and\ \bibinfo {author} {\bibfnamefont {H.}~\bibnamefont
  {Fehske}},\ }\bibfield  {title} {\bibinfo {title} {High-order commutator-free
  exponential time-propagation of driven quantum systems},\ }\href
  {https://doi.org/https://doi.org/10.1016/j.jcp.2011.04.006} {\bibfield
  {journal} {\bibinfo  {journal} {Journal of Computational Physics}\ }\textbf
  {\bibinfo {volume} {230}},\ \bibinfo {pages} {5930} (\bibinfo {year}
  {2011})}\BibitemShut {NoStop}%
\bibitem [{\citenamefont {Bloomfield}(2000)}]{Bloomfield2000}%
  \BibitemOpen
  \bibfield  {author} {\bibinfo {author} {\bibfnamefont {P.}~\bibnamefont
  {Bloomfield}},\ }\href@noop {} {\emph {\bibinfo {title} {Fourier analysis of
  time series}}},\ Wiley Series in Probability and Statistics\ (\bibinfo
  {publisher} {John Wiley and Sons},\ \bibinfo {address} {Nashville, TN},\
  \bibinfo {year} {2000})\BibitemShut {NoStop}%
\end{thebibliography}%
\end{document}